\documentclass[useAMS,usenatbib]{mn2e}

\usepackage{graphicx}
\usepackage{multirow}

\title[Environmental dependence of the S0 offset from the TFR]{S0 galaxies in the Coma cluster: Environmental dependence of the S0 offset from the Tully--Fisher relation}
\author[T. D. Rawle, John R. Lucey, Russell J. Smith and J. T. C. G. Head]{T. D. Rawle$^{1}$\thanks{E-mail:
tim.rawle@sciops.esa.int}, John R. Lucey$^{2}$, Russell J. Smith$^{2}$ and J. T. C. G. Head$^{2}$\\\
$^{1}$ESAC, ESA,  Box 78, Villanueva de la Ca\~{n}ada, 28691 Madrid, Spain\\
$^{2}$Department of Physics, Durham University, South Road, Durham, DH1 3LE}
\begin{document}

\date{Published in MNRAS, 2013, 433, 2667}

\pagerange{\pageref{firstpage}--\pageref{lastpage}} \pubyear{2011}

\maketitle

\label{firstpage}

\begin{abstract}
We present deep GMOS long-slit spectroscopy of 15 Coma cluster S0 galaxies, and extract kinematic properties along the major axis to several times the disc scale-length. Supplementing our dataset with previously published data, we create a combined sample of 29 Coma S0s, as well as a comparison sample of 38 Coma spirals. Using photometry from SDSS and 2MASS, we construct the Tully--Fisher relation (TFR; luminosity versus maximum rotational velocity) for S0 galaxies. At fixed rotational velocity, the Coma S0 galaxies are on average fainter than Coma spirals by $1.10\pm0.18$, $0.86\pm0.19$ and $0.83\pm0.19$ mag in the $g$, $i$ and $K_{\rm s}$ bands respectively. The typical S0 offsets remain unchanged when calculated relative to large field-galaxy spiral samples. The observed offsets are consistent with a simple star formation model in which S0s are identical to spirals until abrupt quenching occurs at some intermediate redshift. The offsets form a continuous distribution tracing the time since the cessation of star formation, and exhibit a strong correlation ($>6\,\sigma$) with residuals from the optical colour--magnitude relation. Typically, S0s which are fainter than average for their rotational velocity are also redder than average for their luminosity. The S0 TFR offset is also correlated with both the projected cluster-centric radius and the $\Sigma$ (projected) local density parameter. Since current local environment is correlated with time of accretion into the cluster, our results support a scenario in which transformation of spirals to S0s is triggered by cluster infall.
\end{abstract}


\section{Introduction}
\label{sec:intro}

Galaxies are often separated into the categories `late-type' and `early-type', where the latter comprises both ellipticals and S0 (lenticular) galaxies. S0s observed to the half-light radius are typically dominated by the bulge, which is akin to a low-luminosity elliptical galaxy  \citep{tho06-510,moo06-583,mor08-341}.

Although the stellar content of an S0 galaxy is broadly similar to a quiescent elliptical galaxy, the structure and kinematics are generally comparable to spiral galaxies \citep{bed06-1912}. Turn off the star formation in a spiral galaxy, and within approximately 1--2 Gyr, an S0-like object is formed. With their spheroidal bulge and flat, mostly gas-free disc, S0s have long been postulated as a transitional stage between spiral and elliptical galaxies. Possible mechanisms for this transformation include minor mergers, slow encounters, harassment, or some combination of these (\citealt*{dre83-664}; \citealt{nei99-2666}). However, as S0s make up the plurality of galaxies in rich local clusters \citep[e.g.][]{dre80-351,dre87-42}, many evolutionary mechanisms have been proposed in which the star formation in a spiral has been abruptly truncated by processes specifically related to cluster in-fall (\citealt{gun72-1}; review in \citealt{bos06-517}). Theories include starvation (removal of hot gas reservoirs via interaction with the intracluster medium; e.g. \citealt{lar80-692,mcc08-593}), tidally-induced starbursts \citep{bek99-15} and ram-pressure stripping of cold gas \citep{gun72-1,aba99-947}. However, if S0s are the direct descendants of objects analogous to local spirals, then for a common environment, the luminosity distribution of S0s would be systematically fainter than spirals. Instead, \citet{bur05-246} and \citet{san05-581} both showed that the typical surface brightness of S0s is greater than for spirals, and \citet{dre80-351} suggested that the bulge luminosity of S0s is too large to evolve from spirals by disc fading alone \citep[also][]{cor13-1382}. The total luminosity and relative bulge mass of S0s could both increase if they were formed via minor mergers, harassment or tidally-induced central starbursts \citep{chr04-192,wil09-298}.

At redshifts up to $z\sim1$, \citet{bun10-1969} observed passive red spirals, likely to be the progenitors of the young S0s observed in nearby clusters. The existence of such galaxies suggests that transformations in colour and in morphology occur at different epochs. For clusters at $z\sim0.5$ ($\sim5$ Gyr ago), \citet{mor07-1503} and \citet{gea09-783} also report populations of transitionary objects likely to be in the process of converting from spiral to S0, while \citet{raw12-106} discover cluster galaxies at $z\sim0.3$ ($\sim3$ Gyr ago) that appear to have undergone recent stripping of outer gas and dust. Additionally, \citet{dre97-577} found that the fraction of S0s in rich clusters at $z\sim0.5$ is 2--3 times smaller than in their local analogues, with a corresponding increase in the spiral fraction. These studies all imply that a large proportion of S0s were transformed from spirals over the last 5 Gyr. \citet{pog01-118} measured the stellar populations of a small sample of S0 galaxies in the nearby Coma cluster, finding that 40 per cent had indeed undergone star formation during the last $\sim$5 Gyr.

For spiral galaxies, the Tully--Fisher relation \citep[TFR;][]{tul77-661} is a tight empirical correlation between luminosity and the maximum rotational velocity. With large H{\sc I}-based galaxy surveys (e.g. SFI++, \citealt{mas06-861}) and optical H-alpha rotation curve measurements \citep[e.g.][]{cou07-203} the TFR for thousands of spirals has now been constructed. Such TFR studies regularly use imaging from the Sloan Digital Sky Survey (SDSS) to obtain homogeneous photometry \citep[e.g][]{piz07-945,moc12-296,hal12-2741}. The spiral TFR exhibits a consistent intrinsic scatter throughout the optical bands: $0.43\pm0.03$ mag in SDSS $griz$ \citep[e.g.]{piz07-945}; 0.41 mag in the $I_{\rm C}$ band \citep{tul12-78}.

Several studies have investigated the same relation for local S0s \citep{nei99-2666,bed06-1125}, and found that they have a systematic luminosity offset compared to the spiral population (in the $B$ band: $1.7\pm0.4$ mag fainter compared to the \citealt{tul00-744} TFR). As truncation of star formation would fade the galaxy disc (by failing to replenish the population with young, luminous stars), this was advocated as evidence for disc fading, with the increased scatter in the relation interpreted as variation in the epoch of truncation. S0s form a continuum with other red spirals, which are also offset from the Tully--Fisher relation \citep[offset $\sim$ 0.5 mag for high rotational velocity Sa galaxies;][]{cou07-203,piz07-945}. The intrinsic scatter of the S0 TFR is generally larger than for spirals (0.6--0.8 mag; \citealt{bed06-1125}), which may reflect a more varied progenitor spiral population. Alternatively, the scatter may merely show that true S0s are hard to isolate cleanly from red spirals.

The merger hypothesis struggles to explain the TFR offset unless the progenitor population was significantly different to spirals at $z=0$ \citep{nei99-2666}. However, the TFR for S0s is still an area of controversy, with some studies \citep[e.g.][]{hin03-2622,wil09-1665} reporting no discernible luminosity offset from spirals. These apparently support a variety of scenarios to form the heterogeneous set of S0s observed.

In order to investigate further the formation of S0s in the cluster environment, we have undertaken deep Gemini Multi-Object Spectrograph (GMOS) long-slit observations of 15 S0 galaxies in the Coma cluster, probing out to several disc scale lengths. Through a detailed exploration of bulge and disc properties, we aim to constrain the effect of local environment on galaxy evolution and discriminate between possible formation mechanisms.

In this paper, we introduce the observations (Section \ref{sec:obs}) and describe in detail the data reduction and derivation of kinematic properties (Section \ref{sec:reduction}). We combine our data with previous Coma cluster S0 samples, and describe a carefully constructed Coma cluster spiral sample (Section \ref{sec:samples}). In Section \ref{sec:results} we calculate the S0 Tully--Fisher relation, relative to both the Coma spirals and published TFRs which used large samples of spirals. We particularly concentrate on interpreting the environmental dependence of the S0-to-spiral offset. Section \ref{sec:conc} summarises our conclusions.

\section{Observations}
\label{sec:obs}

\subsection{GMOS Coma cluster sample}
\label{sec:sample}

The sample consists of 15 edge-on S0 galaxies in the nearby Coma cluster (Abell 1656; $z=0.0231$, $\langle cz_{cmb}\rangle = 7194$ km s$^{-1}$ from the NASA/IPAC Extragalactic Database\footnote{http://ned.ipac.caltech.edu/} (NED), as listed in Table \ref{tab:s0s}. Each galaxy is a confirmed cluster member and was morphologically classified as an S0 by \citet[][D80]{dre80-565}.

\begin{table*}
\centering
\caption{Observed parameters for the GMOS Coma cluster S0 galaxy sample, observed in 2009 and 2011. GMP ID from \citet*{god83-113}. Position (in decimal degrees), disc position angle (PA), $g$- and $i$-band total apparent magnitude (in AB magnitudes) are from SDSS. $K_{\rm s}$-band total apparent magnitude (in Vega magnitudes) is extracted from 2MASS. Heliocentric velocity ($cz_{\rm hel}$) and observed maximum rotation velocity ($V_{\rm obs}$) measured as described in Section \ref{sec:kinematics}. GMP5160 was observed in both 2009 and 2011; kinematics derive from the latter data.} 
\label{tab:s0s} 
\begin{tabular}{@{}llccrrrrccc}
\\ 
\hline
\multicolumn{1}{c}{GMP ID} & \multicolumn{1}{c}{NGC} & RA & Dec & \multicolumn{1}{c}{PA} &\multicolumn{1}{c}{$m_g$} & \multicolumn{1}{c}{$m_i$} & \multicolumn{1}{c}{$m_{K_s}$} &\multicolumn{1}{c}{$cz_{\rm hel}$} & \multicolumn{1}{c}{$V_{\rm obs}$} & \multicolumn{1}{c}{obs} \\
& \multicolumn{1}{c}{\#} & J2000 & J2000 & deg & \multicolumn{1}{c}{mag} & \multicolumn{1}{c}{mag} & \multicolumn{1}{c}{mag} & \multicolumn{1}{c}{km s$^{-1}$} & \multicolumn{1}{c}{km s$^{-1}$} & \multicolumn{1}{c}{year} \\
\hline
gmp1176 & 4931 & 195.75365 & 28.03247 & 78 & 13.796 $\pm$ 0.002 & 12.677 $\pm$ 0.002 & 10.31 $\pm$ 0.03 & 5364 & 192 $\pm$ 5 & 09 \\ 
gmp1504 &  & 195.58967 & 28.23077 & 59 & 15.092 $\pm$ 0.002 & 13.964 $\pm$ 0.002 & 11.57 $\pm$ 0.05 & 5555 & 155 $\pm$ 7 & 09 \\ 
gmp1853 &  & 195.44590 & 28.09500 & 87 & 14.877 $\pm$ 0.002 & 13.700 $\pm$ 0.002 & 11.20 $\pm$ 0.04 & 5836 & 190 $\pm$ 5 & 09 \\ 
gmp2219 &  & 195.30120 & 27.60448 & 132 & 16.096 $\pm$ 0.003 & 14.979 $\pm$ 0.003 & 12.70 $\pm$ 0.08 & 7577 & 110 $\pm$ 3 & 09 \\ 
gmp2584 &  & 195.14822 & 28.14612 & 169 & 15.502 $\pm$ 0.003 & 14.377 $\pm$ 0.003 & 11.97 $\pm$ 0.06 & 5437 & 136 $\pm$ 3 & 09 \\ 
gmp2795 & 4895 & 195.07470 & 28.20240 & 154 & 13.864 $\pm$ 0.002 & 12.673 $\pm$ 0.002 & 10.14 $\pm$ 0.03 & 8513 & 201 $\pm$ 10 & 11 \\ 
gmp2815 & 4894 & 195.06883 & 27.96751 & 32 & 15.478 $\pm$ 0.003 & 14.469 $\pm$ 0.003 & 11.88 $\pm$ 0.07 & 4664 & 85 $\pm$ 2 & 09 \\ 
gmp2956 &  & 195.02293 & 27.80759 & 8 & 15.484 $\pm$ 0.003 & 14.360 $\pm$ 0.003 & 11.87 $\pm$ 0.05 & 6549 & 140 $\pm$ 10 & 11 \\ 
gmp3423 &  & 194.87253 & 27.85016 & 154 & 15.249 $\pm$ 0.002 & 14.017 $\pm$ 0.002 & 11.50 $\pm$ 0.04 & 6895 & 213 $\pm$ 9 & 11 \\ 
gmp3561 & 4865 & 194.83283 & 28.08430 & 115 & 14.367 $\pm$ 0.002 & 13.144 $\pm$ 0.002 & 10.48 $\pm$ 0.03 & 4651 & 203 $\pm$ 7 & 11 \\ 
gmp3997 &  & 194.70303 & 27.81043 & 75 & 14.760 $\pm$ 0.002 & 13.575 $\pm$ 0.002 & 11.12 $\pm$ 0.04 & 5886 & 172 $\pm$ 10 & 11 \\ 
gmp4664 &  & 194.44710 & 27.83331 & 90 & 15.600 $\pm$ 0.003 & 14.421 $\pm$ 0.003 & 12.03 $\pm$ 0.05 & 6046 & 155 $\pm$ 9 & 11 \\ 
gmp4679 &  & 194.44232 & 27.75703 & 114 & 15.473 $\pm$ 0.003 & 14.412 $\pm$ 0.003 & 12.09 $\pm$ 0.07 & 6147 & 100 $\pm$ 13 & 11 \\ 
gmp4907 &  & 194.35767 & 27.54613 & 142 & 15.401 $\pm$ 0.003 & 14.220 $\pm$ 0.002 & 11.84 $\pm$ 0.05 & 5638 & 145 $\pm$ 6 & 11 \\ 
gmp5160 &  & 194.23574 & 28.62338 & 100 & 15.444 $\pm$ 0.003 & 14.300 $\pm$ 0.002 & 12.05 $\pm$ 0.06 & 6566 & 138 $\pm$ 3 & 09/11 \\
\hline 
\end{tabular}
\end{table*}

\begin{figure*}
\centering
\includegraphics[width=170mm]{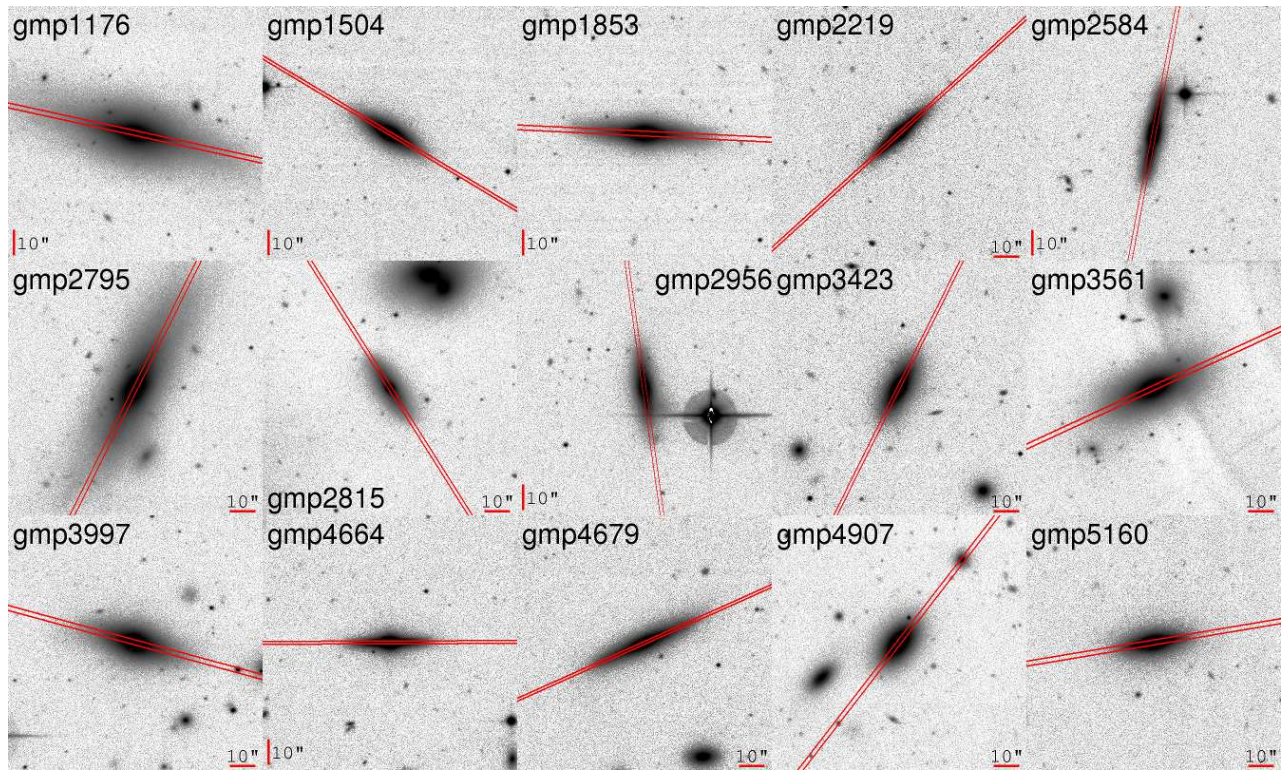}
\caption{CFHT MegaCam $g$-band thumbnails ($100\times100$ arcsec) for the GMOS S0 sample (north up, east left). The orientation of the major axis long-slit is marked.}
\label{fig:thumbs}
\end{figure*}

Although the sample is not complete in any rigorous sense, the galaxies are selected to cover a range in luminosity ($m_i$ = 12.5 -- 15.0) and all lie on the red sequence (see Section \ref{sec:cmr}). They are also located at a range of projected cluster radii (hence a range of local densities), in order to probe the effect of local environment. Each S0 galaxy was examined in deep, optical imaging from MegaCam at the Canada France Hawaii Telescope (CFHT; $g$ band shown in Figure \ref{fig:thumbs}), and visually classified as an `optimum' edge-on S0 by confirming a break in the smooth surface brightness profile. Many of the galaxies appear to have a prominent bulge and an extended disc with a $g$-band surface brightness $\mu_g$ $>$ 22 mag arcsec$^{-2}$ (see Section \ref{sec:decomp} for the final bulge/disc decomposition).

Abundant ancillary data is available for each galaxy. In this paper, we use total magnitudes in $g$ and $i$ bands ($\lambda_{\rm obs}$ $=$ 469, 748 nm) from the SDSS DR9 \citep{ahn12-21} {\it photoObj} catalogue\footnote{accessed via http://skyserver.sdss3.org/CasJobs/} ({\sc cModelMag} parameter), and $K_{\rm s}$ band (2.16 \micron{}) from the 2MASS extended source catalogue\footnote{accessed via http://irsa.ipac.caltech.edu/} \citep[][{\sc k\_m\_ext} parameter]{jar00-2498}. Apparent magnitudes are listed in Table \ref{tab:s0s}, and to facilitate comparison with previous studies, we quote AB mags for SDSS photometry and Vega mags for 2MASS.

\subsection{GMOS observations}
\label{sec:gmos}

\begin{table}
\centering
\caption{GMOS instrument configuration.} 
\label{tab:gmos} 
\begin{tabular}{@{}lcc}
\\
\hline
& 2009 & 2011\\
\hline
Mode & \multicolumn{2}{c}{Long-slit} \\
Grating & B1200 & B600 \\
Slit width & 2 arcsec & 1.5 arcsec \\
Slit length & \multicolumn{2}{c}{330 arcsec} \\
CCD binning & \multicolumn{2}{c}{4 $\times$ 4} \\
Wavelength range & 4060 -- 5522 \AA & 3600 -- 6200 \AA \\
Spectral resolution & 5.1 \AA{} FWHM & 8.5 \AA{} FWHM \\
 & ($\sigma$ $\sim$ 136 km s$^{-1}$) & ($\sigma$ $\sim$ 226 km s$^{-1}$) \\
Spectral sampling & $\sim$0.95 \AA{} pixel$^{-1}$ & $\sim$1.5 \AA{} pixel$^{-1}$ \\
Spatial sampling & \multicolumn{2}{c}{0.2908 arcsec pixel$^{-1}$} \\
\hline 
\end{tabular}
\end{table} 

This investigation uses long-slit spectroscopy from GMOS \citep{hoo04-425} on the Gemini-North telescope, Mauna Kea (programs: GN-2009A-Q-52, PI: Lucey; GN-2011A-Q-50, PI: Rawle). In 2009, GMOS was operated with the B1200 grating and a 2 arcsec wide slit, resulting in a spectral resolution of 5.1 \AA{} FWHM. In 2011, we used the newly available B600 grating to increase the scheduling likelihood, and decreased the slit width to 1.5 arcsec, resulting in a spectral resolution of 8.5 \AA{} FWHM. The full instrument configurations are summarised in Table \ref{tab:gmos}.

Successful exploration of the disc-dominated region of these S0 galaxies, relies on probing to a $g$-band surface brightness $\mu_g$ $\sim$23 mag arcsec$^{-2}$. Derivation of reliable kinematics requires a signal-to-noise ratio (S/N) $\ga$ 20 \AA$^{-1}$, while stellar population gradients (to be presented in a future paper) require S/N $\ga$ 30 \AA$^{-1}$. To achieve this, we devoted three hours of observing time to each target, with four exposures of 2380 s (9520 s in total) on-source. For each observation, the slit was centred on the S0 bulge and orientated along the disc major axis, as shown in Figure \ref{fig:thumbs}.

The galaxies were observed during dark time with the seeing $<$ 1.0 arcsec (FWHM) for all exposures, and better than 0.8 arcsec in many cases. In 2009, GMP5160 was observed in particularly cloudy conditions, and was re-observed in 2011, providing a convenient consistency check between the different instrument configurations of the two campaigns.

\section{Data reduction and analysis}
\label{sec:reduction}

\subsection{Initial reduction}
\label{sec:initial}

Initial data reduction uses the standard Gemini GMOS {\sc pyraf} routines. Variance frames are calculated from the quadrature sum of the Poisson noise in the raw detector counts and the read noise, and propagated through the same pipeline routines as the science frames. We have assumed that the noise associated with bias and flat field frames is negligible.

\subsection{GMOS scattered light}

Visual inspection of the raw detector frames shows that unexposed pixels in the slit bridges and beyond the slit ends do not have zero counts. The cause of this phenomenon is scattered light in the instrument, most likely due to the classically ruled grating \citep[see][]{nor06-815}, which is not accounted for in the standard reduction routines. The scattered light appears as a featureless constant offset which artificially enhances the continuum level, thus decreasing the measured absorption line strengths. The effect is increasingly significant as the galaxy surface brightness decreases ($\sim$50 percent of the measured flux in the outer regions of the galaxy), so scattered light tends to spuriously strengthen index gradients. For the kinematic analysis of this study, the scattered light correction is inconsequential, but we include it here for completeness.

\begin{figure}
\centering
\includegraphics[viewport=0mm 0mm 70mm 175mm,height=84mm,angle=270,clip]{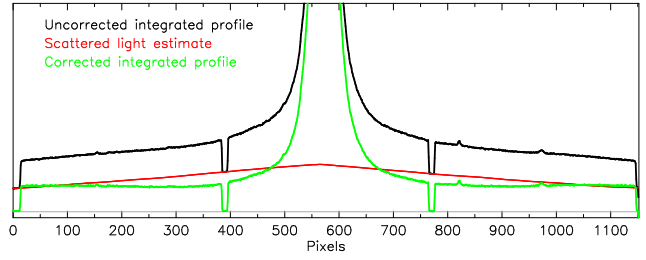}
\caption{The integrated brightness profiles of the uncorrected (black) and scattered-light-corrected (green) frames for GMP1176. The red line shows the scattered light correction profile, as interpolated from the unexposed slit bridges and ends.}
\label{fig:scatlight}
\end{figure}

To quantify the scattered light in each raw exposure, the flux is measured in the unexposed regions (slit bridges and ends). For each wavelength pixel, we interpolate between these regions (using two linear fits) to calculate the scattered light frame. For an example galaxy, Figure \ref{fig:scatlight} shows the interpolated scattered light along the spatial axis, in an arbitrary wavelength slice. The correction removes the inflated wings of the integrated surface brightness profile. During the reduction described above, the scattered light frame is subtracted from the raw image before wavelength calibration.

\subsection{Bulge and disc decomposition}
\label{sec:decomp}

Structural parameters for each galaxy are determined from $g$-band CFHT MegaCam image data, with a PSF FWHM of $\sim$0.7 arcsec and 3.4$\times$ the depth of SDSS. An analytical S\'ersic + exponential component model is fit to thumbnail images using GALFIT (version 3.0.4; \citealt{pen10-2097}). Initial values are generated from the best-fit parameters of a S\'ersic-only model. The search through chi-squared space was extended by perturbing parameters from their `best-fit' positions and refitting, thus improving reliability of the GALFIT-derived characteristics. Parameter uncertainties were estimated from the scatter in Monte-Carlo fitting tests.

The S\'ersic component corresponds to the bulge, parametrized by an effective radius $r_{\rm bul}$, S\'ersic index $n$, and ellipticity, $e_{\rm bul}$. The disc is modelled by the exponential component, parametrized by a scale-length $r_{\rm disc}$ and ellipticity, $e_{\rm disc}$. The disc inclination angle $i_{\rm disc}$ is calculated from the axis ratio via the standard formula
\begin{equation}
cos\,i_{\rm disc} = \sqrt{\frac{(1-e_{\rm disc})^2-q_0^2}{1-q_0^2}}
\end{equation}
for which we assume an intrinsic axis ratio of $q_0$ = 0.22 \citep{dev91-book}. The derived bulge and disc parameters for all GMOS S0s are displayed in the top half of Table \ref{tab:galfit}.

The luminosity profile of each component was computed numerically by integrating along a major axis slit in the best-fit model image. The upper panels of Figures \ref{fig:kin_gmp1176}--\ref{fig:kin_gmp5160}, shows the bulge, disc and total luminosity profiles for each galaxy. Examining these best-fit, two-component decompositions, the relative size of the bulges in this sample vary widely, and can be separated into three broad categories. In most cases (12/15), the decomposed profile exhibits the `classic' S0 form of a dominant bulge at small radii and a significant disc at large radii. In contrast, GMP1176 is better fit by a dominant bulge at all radii, with a small embedded disc which only contributes a small fraction of the light even in the outskirts. The remaining two S0s (GMP3423, 4907) are an intermediate case, strongly resembling the `classic' S0s, with a dominant bulge at small radii, but comparable contribution from bulge and disc components at large radii. The latter two categories may be better fit by a three-component model including a large-scale spherical halo, although we do not attempt such a modelling in the current study.

\begin{table}
\centering
\caption{Bulge and disc parameters from the {\sc Galfit} decomposition described in Section \ref{sec:decomp}. S0 galaxies from the GMOS sample are displayed in the top half of the table. Additional S0s from the \citet{meh00-449} and \citet{hin03-2622} samples (see Section \ref{sec:mores0}) are shown below the dividing line, and were decomposed in exactly the same manner. The fits for GMP1614 and 3761 include an additional small bar component, as described in Section \ref{sec:mores0}.}
\label{tab:galfit} 
\begin{tabular}{@{}lrccccr}
\\
\hline
\multicolumn{1}{c}{GMP ID} & \multicolumn{3}{c}{S\'ersic bulge} & \multicolumn{3}{c}{Exponential disc} \\
 & \multicolumn{1}{c}{$r_{\rm bul}$} & \multicolumn{1}{c}{$n$} & \multicolumn{1}{c}{$e_{\rm bul}$} & \multicolumn{1}{c}{$r_{\rm disc}$} & \multicolumn{1}{c}{$e_{\rm disc}$} & \multicolumn{1}{c}{$i_{\rm disc}$} \\
  & \multicolumn{1}{c}{arcsec} & \multicolumn{1}{c}{} & & \multicolumn{1}{c}{arcsec} & & \multicolumn{1}{c}{deg} \\
\hline
gmp1176 & 13.1 & 6.1 & 0.58 & 3.1 & 0.94 & $\sim$90 \\ 
gmp1504 & 1.6 & 4.3 & 0.22 & 3.9 & 0.71 & 79 \\ 
gmp1853 & 2.3 & 3.0 & 0.37 & 7.1 & 0.82 & $\sim$90 \\ 
gmp2219 & 9.2 & 3.9 & 0.58 & 4.2 & 0.88 & $\sim$90 \\ 
gmp2584 & 2.5 & 4.7 & 0.13 & 5.7 & 0.85 & $\sim$90 \\ 
gmp2795 & 3.5 & 1.4 & 0.38 & 12.3 & 0.72 & 80 \\ 
gmp2815 & 0.8 & 1.4 & 0.16 & 4.4 & 0.66 & 75 \\ 
gmp2956 & 4.0 & 3.4 & 0.50 & 5.0 & 0.82 & $\sim$90 \\ 
gmp3423 & 3.5 & 5.4 & 0.46 & 3.5 & 0.78 & 89 \\ 
gmp3561 & 2.5 & 1.5 & 0.43 & 7.6 & 0.58 & 69 \\ 
gmp3997 & 2.3 & 2.2 & 0.26 & 7.0 & 0.67 & 75 \\ 
gmp4664 & 1.4 & 2.3 & 0.30 & 4.7 & 0.85 & $\sim$90 \\ 
gmp4679 & 32.3 & 9.8 & 0.22 & 7.2 & 0.85 & $\sim$90 \\ 
gmp4907 & 2.0 & 3.8 & 0.39 & 3.2 & 0.48 & 61 \\ 
gmp5160 & 11.4 & 10.0 & 0.25 & 3.8 & 0.78 & 89 \\ 
\hline
gmp0756 & 5.6 & 3.1 & 0.55 & 12.3 & 0.75 & 82 \\ 
gmp1111 & 4.2 & 4.3 & 0.40 & 4.8 & 0.87 & $\sim$90 \\ 
gmp1223 & 5.6 & 3.0 & 0.48 & 6.0 & 0.57 & 67 \\ 
gmp1614 & 1.7 & 1.7 & 0.22 & 6.5 & 0.27 & 45 \\ 
gmp2413 & 1.5 & 1.8 & 0.17 & 6.5 & 0.47 & 61 \\ 
gmp2431 & 4.8 & 4.6 & 0.21 & 8.9 & 0.48 & 61 \\ 
gmp2535 & 4.2 & 2.8 & 0.46 & 6.3 & 0.36 & 52 \\ 
gmp2629 & 2.3 & 2.9 & 0.42 & 7.1 & 0.46 & 60 \\ 
gmp3273 & 1.7 & 1.3 & 0.08 & 8.8 & 0.72 & 80 \\ 
gmp3367 & 1.1 & 2.1 & 0.13 & 4.3 & 0.29 & 46 \\ 
gmp3414 & 1.0 & 1.6 & 0.15 & 5.0 & 0.41 & 56 \\ 
gmp3661 & 1.1 & 1.9 & 0.29 & 5.8 & 0.29 & 47 \\ 
gmp3761 & 4.9 & 4.5 & 0.04 & 6.1 & 0.63 & 72 \\ 
gmp3818 & 3.1 & 2.9 & 0.30 & 5.7 & 0.65 & 74 \\
\hline
\end{tabular}
\end{table} 

\subsection{Spatial binning and kinematic analysis}
\label{sec:kinematics}

We use a simple adaptive-width algorithm in which the data is binned along the slit, imposing the criterion S/N$_{\lambda=4700-5000}$ $\ga$ 20 \AA$^{-1}$ as well as a minimum bin width of 0.5 arcsec. In the outermost bin, we simply use a width to maximise S/N.

The kinematics properties ($cz_{\rm hel}$, $V_{\rm obs}$, $\sigma_{\rm obs}$) are measured using the Penalized Pixel-Fitting ({\sc pPxf}) IDL routine \citep{cap04-138}. {\sc pPxf} extracts kinematics by fitting to a large set of weighted stellar templates, virtually eliminating template mismatch. The non-linear least-squares direct pixel fitting computation is broken into several iterations which find the best-fitting linear combination of the stellar templates. The derived residual spectrum from each iteration forms a penalisation term in the non-linear optimisation. The two central panels of Figures \ref{fig:kin_gmp1176}--\ref{fig:kin_gmp5160} present the observed rotational velocity and velocity dispersion of each GMOS S0 galaxy. The penultimate column of Table \ref{tab:s0s} lists the observed maximum rotational velocity for each galaxy.

\subsection{Derivation of circular velocities}
\label{sec:vc}

We derive the maximum circular velocity from the observed velocity via the prescription described in \citet{nei99-2666}, implemented in numerous S0 studies \citep[e.g.][]{hin03-2622,bed06-1125}. For edge-on S0 galaxies (those with disc inclinations $i$ $\ga$ 60 deg), two steps are required to calculate the circular velocity at a given radius. First, we compute the kinematics in the azimuthal ($\phi$) direction by accounting for line-of-sight integration through the disc. For an exponential disc with a scale height $h_s$ = 0.2$r_{\rm disc}$, \citeauthor{nei99-2666} estimate corrections
\begin{eqnarray}
V_\phi(r) &=& \frac{V_{\rm obs}(r)}{f(r / r_{\rm disc})} \\ 
\sigma_\phi(r)^2 &=& \sigma_{\rm obs}(r)^2 - \frac{1}{2}(V_\phi(r) - V_{\rm obs}(r))^2 
\end{eqnarray}
where
\begin{eqnarray}
f(x) = \frac{exp(-x)}{-0.5772-{\rm ln}x+x-\frac{1}{2}x^2/2!+\frac{1}{3}x^3/3!-...} - x
\end{eqnarray}
and ($V_{\rm obs}$,$\sigma_{\rm obs}$), ($V_\phi$,$\sigma_\phi$) are the observed and corrected azimuthal values respectively. For inclinations $i$ $>$ 70 deg, the error due to assuming $i$ = 90 deg is $\Delta$log$V_\phi$ $\sim$ 0.025. In most of the Coma S0 galaxies presented here, the observed velocity dispersion corresponding to the flat rotation curve region is less than half of the GMOS instrument resolution ($\sim$ 136 or 226 km s$^{-1}$ for 2009/11 respectively), and we were unable to determine $\sigma_\phi$.

The second stage corrects for `asymmetric drift'. Although the net velocity of stars is zero in the vertical and radial directions, the average velocity in the azimuthal direction is not equal to the local circular velocity. The greater the random velocity for individual stars, the larger the lag between net motion and circular velocity. \citet{nei99-2666} correct for this effect using the formula
\begin{eqnarray}
V_{\rm c}(r)^2 = V_\phi(r)^2 + \sigma_\phi(r)^2 \left( 2 \frac{r}{r_{\rm disc}} -1 \right)
\end{eqnarray}
The maximum circular velocity is calculated by taking the mean value of all $V_c$ data points lying on the flat portion of the rotation curve (generally 1.5 $<$ $r$/$r_{\rm disc}$ $<$ 2.5). For the approximations in the asymmetric drift correction to hold, all points also have to conform to the constraint $V_\phi$/$\sigma_\phi$ $>$ 2.5. For the galaxies with no computed $\sigma_\phi$, we were unable to correct for asymmetric drift and, when quoting $V_{\rm c}$ for these galaxies (final column of Table \ref{tab:tfr}), we have given $V_\phi$. However, as $V_{\rm c}$ $\approx$ $V_\phi$ if $\sigma_{\rm obs}$ $\to$ 0, this should not lead to a significant under-estimation. To test this, we used the uncorrected $\sigma_{\rm obs}$ to account for the asymmetric drift (which over-estimates the correction). The difference between this $V_{\rm c}$ and $V_\phi$ is $<10$ per cent of the uncertainty on the values, so will not have a significant effect on any conclusions.

For each galaxy in the Coma S0 sample, $V_{\rm c}$ is presented in Table \ref{tab:tfr}. For GMP5160, we show the velocity derived from the 2011 data ($V_{\rm c}$ $=$ 184 $\pm$ 5 km s$^{-1}$). From the 2009 observation, $V_{\rm c}$ $=$ 170 $\pm$ 13 km s$^{-1}$, as poor S/N severely limits the radial extent of the data.

\subsection{Absolute magnitude derivation}
\label{sec:abmag}

Absolute magnitudes are calculated in each band ($X$ $=$ $g$, $i$, $K_{\rm s}$) using 
\begin{equation}
\label{eq:abmag}
M_X = m_X - A_{i,X} - A_{g,X} - A_{k,X} - \mu_{\rm Coma}
\end{equation}
where the internal extinction $A_{i,X}$ $=$ $\gamma_X$log($a$/$b$), Galactic extinction $A_{g,X}$ and k-correction $A_{k,X}$ are derived in each band from the values listed in Table \ref{tab:mtom}. The axis ratio is estimated via
\begin{equation}
{\rm log}(a/b) = \frac{1}{\sqrt{0.96 i_{\rm disc}^2+0.04}}
\end{equation}

For the S0 sample, we can assume that the internal extinction for S0s is zero, i.e. $\gamma_X$ = 0. Cluster S0 galaxies, including those in the Coma sample (see Section \ref{sec:cmr}), are observed to form a tight red sequence in optical colours, with less than 0.02 mag of the scatter unaccounted for by stellar population differences \citep{smi09-1690}. Such homogeneity is unlikely to occur unless the internal extinction is very small in all S0s. Indeed, dusty early-type galaxies are generally rare in clusters \citep{kav12-49} and only two of our S0 sample have even a hint of a dust lane (GMP3273, 3818). The Herschel Reference Survey has recently shown that local S0s are $\sim$10--100 times less dusty than spirals \citep{smi12-123}. As spiral galaxies in the Coma cluster require a mean extinction correction of only $\sim$0.3 mag in the $g$ band and $\sim$0.1 mag at $K_{\rm s}$ (Section \ref{sec:spirals}), the internal extinction correction for S0 galaxies is likely to be very small. We note many other recent studies also assume a zero internal extinction for S0s \citep[e.g.][]{bed06-1125,dav11-968}.

We adopt the distance modulus $\mu_{\rm Coma}$ = 35.05 mag from NED (Virgo+NED velocity field model). Due to this assumption of an equal line-of-sight depth for all galaxies, the error on the magnitude, $\delta M_X$, includes an additional 0.03 mag uncertainty for S0s, added in quadrature with the measurement error $\delta m_X$.

\begin{table*}
\centering
\caption{Parameters used to derive the absolute magnitude in each band. $V_{\rm c}$ is the circular velocity, log(a/b) is the axial ratio, $E(B-V)$ is the extinction coefficient and $z$ is the redshift. Note that the internal extinction correction, parametrized by $\gamma_{X}$ is different for the S0 and spiral samples.} 
\label{tab:mtom} 
\begin{tabular}{@{}ccccccc}
\\
\hline
 & $g$ & & $i$ & & $K_{\rm s}$ & \\
\hline
$\gamma_{X}$ (S0s) & 0. & & 0. & & 0. & \\
\multirow{2}{*}{$\gamma_{X}$ (spirals)} & \multirow{2}{*}{1.51 + 2.46(log($V_{\rm c}$) -- 2.5)} & \multirow{2}{*}{(1)} & \multirow{2}{*}{1.00 + 1.71(log($V_{\rm c}$) -- 2.5)} & \multirow{2}{*}{(1)} & log(a/b) $>$ 0.5: 1.1 + 0.13/log(a/b) & \multirow{2}{*}{(4)}  \\
 & & & & & log(a/b) $<$ 0.5: 0.26 & \\
$A_{g,X}$ & 3.793$\times$$E(B-V)$ & (2) & 2.086$\times$$E(B-V)$ & (2) & 0.367$\times$$E(B-V)$ & (4) \\
$A_{k,X}$ & 0.01 & (3) & 0.00 & (3) & 1.52$z$ & (4) \\
\hline 
\end{tabular}

(1) \citet{hal12-2741}, (2) \citet{sch98-525}, (3) \citet{bla07-743}, (4) \citet{mas03-158}\\
\end{table*} 

\section{Additional Coma samples}
\label{sec:samples}

\subsection{Extended Coma S0 sample}
\label{sec:mores0}

We supplement our GMOS Coma S0 sample with data from two previous studies, \citet[][M00]{meh00-449} and \citet[][H03]{hin03-2622}. For each additional S0, SDSS and 2MASS total magnitudes are obtained from the catalogues described above. GMP3414 does not have a counterpart in the $K_{\rm s}$ band.

\begin{figure*}
\centering
\includegraphics[width=170mm]{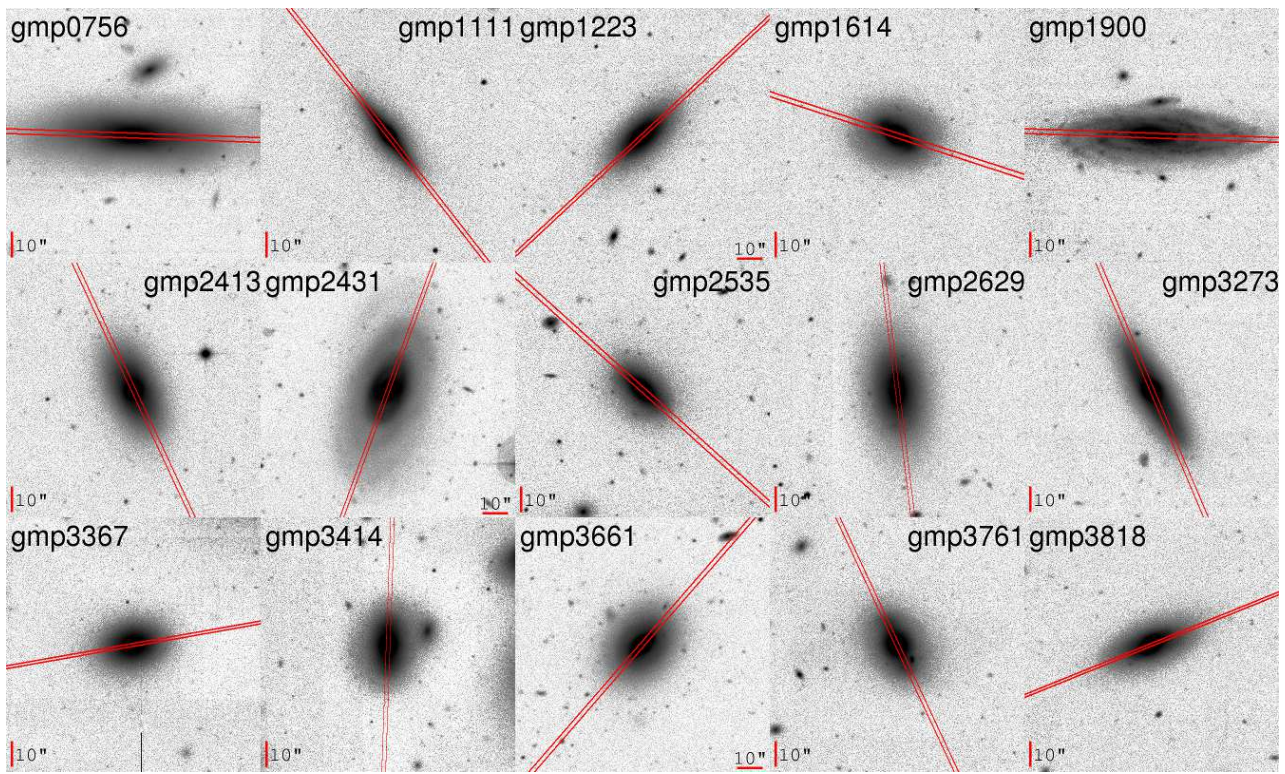}
\caption{CFHT MegaCam $g$-band thumbnails ($100\times100$ arcsec) for the additional 14 S0 galaxies (north up, east left), and GMP1900 which is thrown out of the S0 sample with an obviously spiral-like disc. The orientation of the major axis long-slit is marked.}
\label{fig:thumbs+}
\end{figure*}

Although these galaxies are all typed S0 by M00 and H03, four are not strictly classified S0 according to D80: GMP1614=SB0, GMP2431=Sa, GMP3761=SB0, GMP3818=S0/a. Four further galaxies are not included in D80, while the NED `homogenized' classification gives GMP0756=S0, GMP1111=S0, GMP1900=Sab, GMP3273=S0/a. Visual inspection of the MegaCam imaging (displayed in Figure \ref{fig:thumbs+}) agrees with these types. GMP1900 (from H03) is clearly a spiral with clumpy structure in the disc, and is removed from our S0 sample. We concur with the NED classification of GMP2431, which appears closer in morphology to an S0/a than the D80 Sa type, exhibiting very little structure and only incredibly weak arms. We retain the three S0/a galaxies (GMP2431, 3273, 3818) in the S0 sample.

The bulge/disc decomposition parameters are derived from MegaCam optical imaging using the method described in Section \ref{sec:decomp}, and are displayed below the GMOS sample in Table \ref{tab:galfit}. As with the GMOS S0s, the majority (12/14) exhibit the `classic' S0 structure. GMP2535 has an equal bulge and disc contributions at large radii, while GMP1614 is bulge dominated throughout. As GMP1614 is typed as SB0 by D80, we attempt to fit an additional bar component, finding that the profile is best fit by a 'classic+bar' solution, with the bulge dominating in the centre and the disc at large radii. The bar is subdominant at all radii, but contributes enough flux to remove the extended bulge. We note that although the addition of a bar component significantly reduces the measured effective radius of the bulge ($>4\times$ smaller), the disc scale length is more stable ($<50$ per cent decrease), and the corrected circular velocity remains virtually unchanged.

The disc inclination, displayed in the final column of Table \ref{tab:galfit}, shows that the M00 and H03 S0s are generally not as edge-on as the GMOS sample. Observed parameters for the M00 (six unique S0s) and H03 (eight unique S0s) samples are shown in Table \ref{tab:s0s+}. The following two subsections describe differences between the derivation of the Tully--Fisher parameters for S0s in the M00, H03 and our GMOS sample. The final Tully--Fisher parameters for the full S0 sample (29 Coma S0s) are presented in Table \ref{tab:tfr}.

\begin{table*}
\centering
\caption{Additional Coma cluster S0 galaxies from \citet[][M00]{meh00-449} and \citet[][H03]{hin03-2622}. ID, position and photometric parameters as in Table \ref{tab:s0s}. For M00, $cz$ and $V_{\rm obs}$ are measured using the same method as for the GMOS sample. For the H03 sample, raw spectra are unavailable so $cz$ and $V_{\rm c}$ are taken directly from the paper (see Section \ref{sec:h03})} 
\label{tab:s0s+} 
\begin{tabular}{@{}llccrrrrccc}
\\ 
\hline
\multicolumn{1}{c}{GMP ID} & \multicolumn{1}{c}{NGC} & RA & Dec & \multicolumn{1}{c}{PA} &\multicolumn{1}{c}{$m_g$} & \multicolumn{1}{c}{$m_i$} & \multicolumn{1}{c}{$m_{K_s}$} &\multicolumn{1}{c}{$cz_{\rm hel}$} & \multicolumn{1}{c}{$V_{\rm obs}$} & \\
& \multicolumn{1}{c}{\#} & (J2000) & (J2000) & (deg) & \multicolumn{1}{c}{mag} & \multicolumn{1}{c}{mag} & \multicolumn{1}{c}{mag} & \multicolumn{1}{c}{(km s$^{-1}$)} & \multicolumn{1}{c}{(km s$^{-1}$)} & \\
\hline
gmp0756 & 4944 & 195.95812 & 28.18573 & 88 & 13.436 $\pm$ 0.002 & 12.349 $\pm$ 0.002 & 10.00 $\pm$ 0.03 & 6989 & 206 $\pm$ 10 & M00 \\ 
gmp1111 &  & 195.79057 & 28.58352 & 38 & 14.985 $\pm$ 0.002 & 13.821 $\pm$ 0.002 & 11.38 $\pm$ 0.04 & 6922 & -- & H03 \\ 
gmp1223 &  & 195.73572 & 28.07039 & 133 & 15.348 $\pm$ 0.003 & 14.282 $\pm$ 0.002 & 12.13 $\pm$ 0.09 & 7759 & -- & H03 \\ 
gmp1614 &  & 195.53608 & 28.38711 & 72 & 14.896 $\pm$ 0.002 & 13.650 $\pm$ 0.002 & 11.13 $\pm$ 0.05 & 7605 & -- & H03 \\ 
gmp2413 &  & 195.21701 & 28.36613 & 25 & 14.408 $\pm$ 0.002 & 13.255 $\pm$ 0.002 & 10.82 $\pm$ 0.03 & 7665 & 197 $\pm$ 25 & M00 \\ 
gmp2431 &  & 195.20801 & 27.40578 & 160 & 14.603 $\pm$ 0.002 & 13.572 $\pm$ 0.002 & 11.45 $\pm$ 0.06 & 6569 & -- & H03 \\ 
gmp2535 &  & 195.17020 & 27.99660 & 48 & 15.211 $\pm$ 0.002 & 14.043 $\pm$ 0.002 & 11.66 $\pm$ 0.06 & 7112 & 90 $\pm$ 16 & M00 \\ 
gmp2629 & 4896 & 195.12820 & 28.34636 & 7 & 14.353 $\pm$ 0.002 & 13.196 $\pm$ 0.002 & 10.76 $\pm$ 0.03 & 6012 & 177 $\pm$ 28 & M00 \\ 
gmp3273 &  & 194.91300 & 28.89552 & 23 & 14.508 $\pm$ 0.002 & 13.212 $\pm$ 0.002 & 10.51 $\pm$ 0.03 & 6210 & -- & H03 \\ 
gmp3367 & 4873 & 194.88663 & 27.98360 & 100 & 14.804 $\pm$ 0.002 & 13.585 $\pm$ 0.002 & 11.25 $\pm$ 0.04 & 5818 & -- & H03 \\ 
gmp3414 & 4871 & 194.87484 & 27.95643 & 178 & 14.758 $\pm$ 0.002 & 13.522 $\pm$ 0.002 & \multicolumn{1}{c}{--} & 6729 & 115 $\pm$ 27 & M00 \\ 
gmp3661 &  & 194.80710 & 27.40257 & 139 & 15.173 $\pm$ 0.002 & 14.009 $\pm$ 0.002 & 11.44 $\pm$ 0.05 & 5675 & 126 $\pm$ 24 & M00 \\ 
gmp3761 &  & 194.77509 & 27.99670 & 25 & 14.985 $\pm$ 0.002 & 13.724 $\pm$ 0.002 & 11.38 $\pm$ 0.05 & 7678 & -- & H03 \\ 
gmp3818 &  & 194.75754 & 28.22536 & 112 & 14.820 $\pm$ 0.002 & 13.610 $\pm$ 0.002 & 11.04 $\pm$ 0.04 & 8017 & -- & H03 \\
\hline
\end{tabular}
\end{table*} 

\begin{table*}
\centering
\caption{Final Tully--Fisher parameters for all 29 Coma cluster S0 galaxies in the GMOS, M00 and H03 samples. $\Delta d_{CC}$ is the projected cluster-centric distance from the nominal cluster centre (RA$=$194.9660, Dec$=$27.96849), and $\Sigma$ is the density parameter described in Section \ref{sec:full}. $E(B-V)$, $g$- and $i$-band photometry from SDSS (AB mags); $K_{\rm s}$ band from 2MASS (Vega mags). Absolute magnitude in each band $M_X$ are corrected for Galactic extinction, internal extinction and k-correction as described in Section \ref{sec:abmag}. $\delta M_X$ includes 0.03 mag uncertainty from the assumption that every galaxy is at the same line-of-sight depth (the cluster mean). Maximum circular velocity ($V_{\rm c}$ km s$^{-1}$) as derived in Sections \ref{sec:vc}, \ref{sec:m00} and \ref{sec:h03} (GMOS, M00 and H03 respectively).} 
\label{tab:tfr} 
\begin{tabular}{@{}lccccrrrl}
\\ 
\hline
\multicolumn{1}{c}{GMP ID} & & \multicolumn{1}{c}{$\Delta d_{CC}$} & \multicolumn{1}{c}{$\Sigma$} & \multicolumn{1}{c}{$E(B-V)$} & \multicolumn{1}{c}{$M_g$} & \multicolumn{1}{c}{$M_i$} & \multicolumn{1}{c}{$M_{K_s}$} & \multicolumn{1}{c}{$V_{\rm c}$} \\
& & \multicolumn{1}{c}{Mpc} & \multicolumn{1}{c}{Mpc$^{-2}$} & & \multicolumn{1}{c}{mag} & \multicolumn{1}{c}{mag} & \multicolumn{1}{c}{mag} & \multicolumn{1}{c}{km s$^{-1}$} \\
\hline
gmp0756 & M00 & 1.49 & 52 & 0.008 & --21.65 $\pm$ 0.03 & --22.72 $\pm$ 0.03 & --25.09 $\pm$ 0.15 & 314 $\pm$ 19 \\ 
gmp1111 & H03 & 1.57 & 32 & 0.009 & --20.11 $\pm$ 0.03 & --21.25 $\pm$ 0.03 & --23.71 $\pm$ 0.15 & 208 $\pm$ 12 \\ 
gmp1176 & 09 & 1.16 & 49 & 0.011 & --21.30 $\pm$ 0.03 & --22.40 $\pm$ 0.03 & --24.77 $\pm$ 0.15 & 266 $\pm$ 12 \\ 
gmp1223 & H03 & 1.14 & 55 & 0.009 & --19.75 $\pm$ 0.03 & --20.79 $\pm$ 0.03 & --22.97 $\pm$ 0.17 & 107 $\pm$ 9 \\ 
gmp1504 & 09 & 1.01 & 67 & 0.009 & --20.00 $\pm$ 0.03 & --21.11 $\pm$ 0.03 & --23.51 $\pm$ 0.16 & 204 $\pm$ 8 \\ 
gmp1614 & H03 & 1.08 & 31 & 0.009 & --20.20 $\pm$ 0.03 & --21.42 $\pm$ 0.03 & --23.96 $\pm$ 0.16 & 256 $\pm$ 50 \\ 
gmp1853 & 09 & 0.73 & 53 & 0.008 & --20.21 $\pm$ 0.03 & --21.37 $\pm$ 0.03 & --23.88 $\pm$ 0.15 & 279 $\pm$ 11 \\ 
gmp2219 & 09 & 0.78 & 29 & 0.008 & --18.99 $\pm$ 0.03 & --20.09 $\pm$ 0.03 & --22.40 $\pm$ 0.17 & 152 $\pm$ 6 \\ 
gmp2413 & M00 & 0.75 & 53 & 0.010 & --20.69 $\pm$ 0.03 & --21.81 $\pm$ 0.03 & --24.27 $\pm$ 0.15 & 294 $\pm$ 24 \\ 
gmp2431 & H03 & 1.00 & 17 & 0.008 & --20.49 $\pm$ 0.03 & --21.50 $\pm$ 0.03 & --23.64 $\pm$ 0.16 & 177 $\pm$ 58 \\ 
gmp2535 & M00 & 0.30 & 196 & 0.011 & --19.89 $\pm$ 0.03 & --21.03 $\pm$ 0.03 & --23.42 $\pm$ 0.16 & 136 $\pm$ 20 \\ 
gmp2584 & 09 & 0.40 & 76 & 0.012 & --19.60 $\pm$ 0.03 & --20.70 $\pm$ 0.03 & --23.11 $\pm$ 0.16 & 185 $\pm$ 4 \\ 
gmp2629 & M00 & 0.67 & 55 & 0.010 & --20.75 $\pm$ 0.03 & --21.88 $\pm$ 0.03 & --24.32 $\pm$ 0.15 & 259 $\pm$ 33 \\ 
gmp2795 & 11 & 0.42 & 103 & 0.012 & --21.24 $\pm$ 0.03 & --22.40 $\pm$ 0.03 & --24.95 $\pm$ 0.15 & 318 $\pm$ 21 \\ 
gmp2815 & 09 & 0.15 & 377 & 0.010 & --19.62 $\pm$ 0.03 & --20.60 $\pm$ 0.03 & --23.19 $\pm$ 0.17 & 121 $\pm$ 7 \\ 
gmp2956 & 11 & 0.28 & 261 & 0.008 & --19.61 $\pm$ 0.03 & --20.71 $\pm$ 0.03 & --23.22 $\pm$ 0.16 & 199 $\pm$ 16 \\ 
gmp3273 & H03 & 1.54 & 13 & 0.011 & --20.59 $\pm$ 0.03 & --21.86 $\pm$ 0.03 & --24.58 $\pm$ 0.15 & 269 $\pm$ 8 \\ 
gmp3367 & H03 & 0.12 & 622 & 0.009 & --20.29 $\pm$ 0.03 & --21.48 $\pm$ 0.03 & --23.83 $\pm$ 0.16 & 375 $\pm$ 50 \\ 
gmp3414 & M00 & 0.13 & 796 & 0.010 & --20.34 $\pm$ 0.03 & --21.55 $\pm$ 0.03 & \multicolumn{1}{c}{--} & 171 $\pm$ 27 \\ 
gmp3423 & 11 & 0.24 & 378 & 0.011 & --19.85 $\pm$ 0.03 & --21.06 $\pm$ 0.03 & --23.59 $\pm$ 0.15 & 307 $\pm$ 25 \\ 
gmp3561 & 11 & 0.27 & 159 & 0.010 & --20.73 $\pm$ 0.03 & --21.93 $\pm$ 0.03 & --24.59 $\pm$ 0.15 & 308 $\pm$ 17 \\ 
gmp3661 & M00 & 0.97 & 21 & 0.008 & --19.92 $\pm$ 0.03 & --21.06 $\pm$ 0.03 & --23.64 $\pm$ 0.16 & 189 $\pm$ 25 \\ 
gmp3761 & H03 & 0.28 & 264 & 0.011 & --20.12 $\pm$ 0.03 & --21.35 $\pm$ 0.03 & --23.71 $\pm$ 0.16 & 267 $\pm$ 40 \\ 
gmp3818 & H03 & 0.52 & 53 & 0.011 & --20.28 $\pm$ 0.03 & --21.46 $\pm$ 0.03 & --24.05 $\pm$ 0.15 & 234 $\pm$ 22 \\ 
gmp3997 & 11 & 0.47 & 492 & 0.013 & --20.35 $\pm$ 0.03 & --21.50 $\pm$ 0.03 & --23.97 $\pm$ 0.16 & 265 $\pm$ 15 \\ 
gmp4664 & 11 & 0.79 & 131 & 0.011 & --19.50 $\pm$ 0.03 & --20.65 $\pm$ 0.03 & --23.06 $\pm$ 0.16 & 215 $\pm$ 10 \\ 
gmp4679 & 11 & 0.84 & 72 & 0.011 & --19.63 $\pm$ 0.03 & --20.66 $\pm$ 0.03 & --22.99 $\pm$ 0.16 & 148 $\pm$ 12 \\ 
gmp4907 & 11 & 1.13 & 116 & 0.009 & --19.69 $\pm$ 0.03 & --20.85 $\pm$ 0.03 & --23.24 $\pm$ 0.16 & 212 $\pm$ 12 \\ 
gmp5160 & 11 & 1.52 & 12 & 0.011 & --19.66 $\pm$ 0.03 & --20.77 $\pm$ 0.03 & --23.04 $\pm$ 0.16 & 184 $\pm$ 5 \\
\hline
\end{tabular}
\end{table*} 

\subsubsection{S0s from \citet{meh00-449}}
\label{sec:m00}

M00 obtained long-slit spectra for a sample of 13 Coma S0s using spectrographs on the 2.5--3.5m-class telescopes of the Michigan-Dartmouth-MIT, McDonald and German-Spanish (at Calar Alto) observatories. Five of these galaxies are also in the GMOS sample (GMP1176, 1853, 2795, 3561, 4679).

We use the spatially-binned observed kinematic properties ($V_{\rm obs}$,$\sigma_{\rm obs}$), provided by the authors in an appendix to their original paper, and recalculate $V_{\rm c}$ using exactly the method described in Section \ref{sec:vc}. For two S0s (GMP3073, 5568), S/N beyond the central bin is too low and the galaxies are removed from our sample. GMP2535 and GMP3414 are also faint and while the spectroscopy does probe disc-dominated radii, the data may not extend quite far enough to fully constrain the maximum rotational velocity. We retain these two S0s in our sample, but keep this concern in mind.

For the galaxies overlapping with the GMOS sample, we can assess the quality of the M00 data directly and also recalculate $V_{\rm c}$ from their observations using our method. Generally, our GMOS data probe further into the disc dominated regime, by as much as twice the observed radius in two cases. This leads to a $2-3\times$ increase in usable velocity data points in the disc and hence smaller uncertainties. For the five galaxies in common, the mean difference in $V_{\rm c}$ is only 7 km s$^{-1}$ (GMP1176: 192 km s$^{-1}$, 187 km s$^{-1}$ for the GMOS and M00 data respectively; GMP1853: 190, 174; GMP2795: 201, 197; GMP3561: 203, 201; GMP4679: 100, 92). These offsets are less than the typical measurement uncertainty of $\sim$14 km s$^{-1}$.

\subsubsection{S0s from \citet{hin03-2622}}
\label{sec:h03}

H03 observed a sample of 15 Coma cluster S0s using the Blue Channel long-slit spectrograph on the 6.5m MMT. With the raw spectra for these observations inaccessible on tape (private communication, J. Hinz), we use the values of $V_{\rm c}$ presented in the original paper. We note that H03 obtains the circular velocity via the same methodology with the same corrections as presented in Section \ref{sec:vc}. From H03 Figure 4, we see that four of the spectra do not reach the required signal-to-noise beyond the central bin (GMP2413, 2495, 4664, 4907) and these S0s are cut from our sample. We also note that while the data for GMP1223 probes the disc-dominated regime, the maximal velocity may not be fully constrained: we retain this S0, but remember this potential shortcoming. Two of the remaining galaxies overlap with the GMOS sample (GMP3423, 3997). Finally, as we noted previously, GMP1900 is a spiral galaxy and is not included in our S0 sample.

The two galaxies with adequate data from both GMOS and H03 offer a useful indicator of the compatibility of the derived velocities from the two studies. H03 presents $V_{\rm c}$ $=$ 317 $\pm$ 31 km s$^{-1}$ for GMP3423 and $V_{\rm c}$ $=$ 267 $\pm$ 19 km s$^{-1}$ for GMP3997. These maximum velocities are well within the errors of our GMOS values, $V_{\rm c}$ $=$ 306 $\pm$ 25 and 265 $\pm$ 15 km s$^{-1}$ respectively, giving us confidence in using the remaining H03 velocity measurements to increase our sample size.

\subsection{Coma cluster spiral sample}
\label{sec:spirals}

\begin{figure*}
\centering
\includegraphics[viewport=15mm 10mm 135mm 300mm,height=187mm,angle=270,clip]{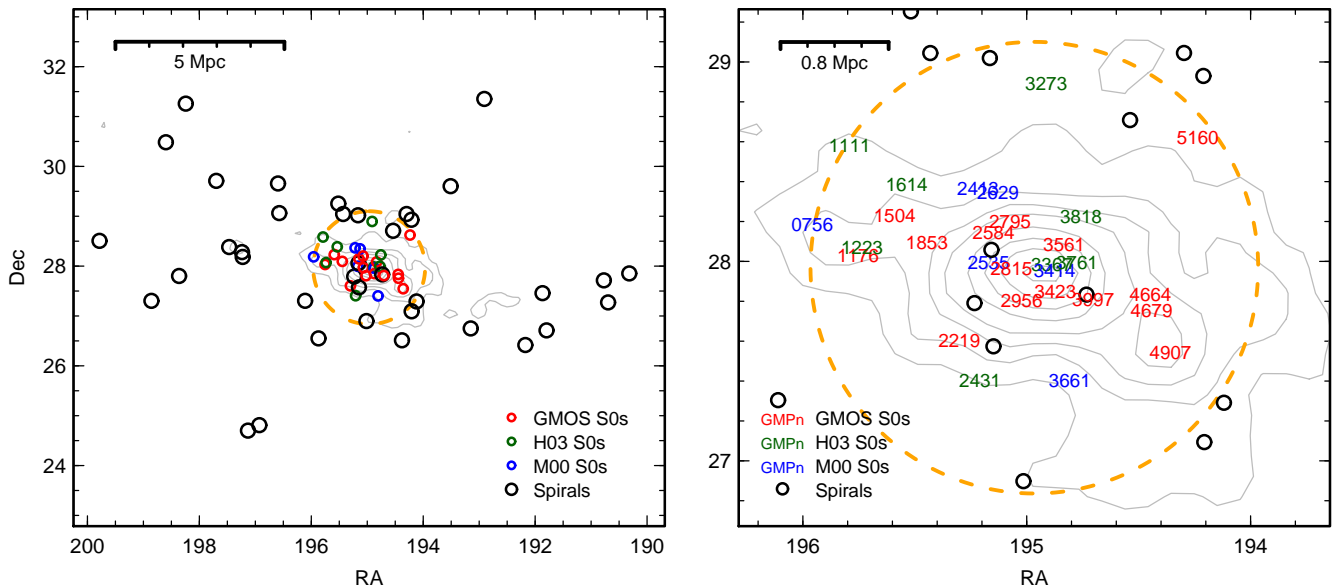}
\caption{Spatial distribution of the S0 (red $=$ GMOS, blue $=$ M00, green $=$ H03) and spiral (black) samples within the Coma cluster. The spiral sample extends to significantly larger radii so two levels of zoom are shown: in both panels the thick dashed orange circle shows a projected cluster-centric radius of 1 deg (1.66 Mpc at the distance of the Coma cluster). Grey contours represent the local galaxy density, as derived from the SDSS sample described in Section \ref{sec:full}.}
\label{fig:field}
\end{figure*}

We wish to compare the S0 population to spiral galaxies in the Coma cluster. The SFI++ Tully--Fisher catalogue \citep{spr07-599} contains H{\sc i} line widths log($W$) and disc inclinations $i_{\rm disc}$ for a sample of nearly 5000 spiral galaxies in the local Universe. This includes 38 Coma cluster members (including GMP1900 from H03), which forms our spiral sample. For each of the Coma spiral galaxy, $g$-, $i$- and $K_{\rm s}$-band total apparent magnitudes are extracted from SDSS and 2MASS in the same manner as for the S0s. Two faint spirals do not have $K_{\rm s}$-band counterparts in the 2MASS XSC. The observed properties for the spiral sample are presented in Table \ref{tab:sps}.

Derivation of the Tully--Fisher parameters ($V_{\rm c}$, $M_X$) for the spiral sample differs in two key respects compared to the S0 analysis: the measurement of the rotational velocities and the internal extinction correction.

For the spirals, rotational velocity is usually derived from H{\sc i} line widths, measured at 50 per cent of the total flux and corrected for instrument effects. For the Coma spirals we adopt the widths from \citet{spr07-599} and assume that the maximum rotational velocity is half of the inclination-corrected line width; this does not attempt to account for turbulent motion intrinsic to the H{\sc i} gas.

Several spirals in the full SFI++ catalogue have both H{\sc i} line width measurements and H$\alpha$ long-slit gas kinematic rotation curves, allowing a direct comparison of $V_{\rm c}$ derived from the two methods. \citet{cat07-334} show that the mean difference $V_{\rm c,HI}-V_{\rm c,H\alpha} = +24\pm4$ km s$^{-1}$, with a systematic dependence on the surface brightness profile (larger differences are associated with brighter bulges). A similar mean difference has also been reported by other authors \citep{cou97-2402,ray97-2046}. While the usually adopted technique to determine $V_{\rm c}$ from H-alpha rotation curve data \citep[see e.g.][]{cou97-2402} is not precisely the
same as the method we have used for our S0 absorption line rotation curves (Section \ref{sec:vc}), these techniques for near edge-on disc galaxies are practically identical.

Correction of the H{\sc i}-derived $V_{\rm c}$ to the optical rotation curve system, would move the spiral TFR systematically towards marginally lower velocities ($\sim$0.03 dex) and flatten the relation in log-space. Pre-empting the discussion in Section \ref{sec:tfr}, we note that our Coma spiral TFR (uncorrected for the above offset) compares well to the TFR reported by \citet{piz07-945}, who also derive $V_{\rm c}$ from H$\alpha$ rotation curves. \citeauthor{piz07-945} also use a different functional form for the rotation curve; their Equation 1 compared to our Section \ref{sec:vc}. Refitting the GMOS S0s with this alternative function, we derive $V_{\rm c}$ with a mean offset of $1\pm9$ km s$^{-1}$, which is comparable to the uncertainty. Generally for edge on disc galaxies, different methods of calculated $V_{\rm c}$ only result in marginal differences and for simplicity we choose not to apply an additional correction to the spiral $V_{\rm c}$ measurements.

The other significant difference between spiral and S0 Tully--Fisher parameters is the internal extinction correction. Whereas we assume that S0s lack significant internal extinction (Section \ref{sec:abmag}), the dusty spirals require an inclination-dependent correction. In the SDSS bands, we use the corrections from \citet{hal12-2741}, which are based on the observed optical internal extinction of spiral galaxies reported by \citet{tul98-2264}. In the $K_{\rm s}$ band, we apply the empirical corrections from \citet{mas03-158}, which are considered the standard for extragalactic infrared photometry \citep[e.g.][]{mas08-1738,jar13-6}. The exact form of these corrections are given in Table \ref{tab:mtom}. The mean internal extinction correction for the spiral sample is $0.34\pm0.10$, $0.22\pm0.07$ and $0.10\pm0.02$ mag in the $g$-, $i$- and $K_{\rm s}$ bands respectively.

Photometric corrections for the spirals (including internal extinction) are applied as described by Equation \ref{eq:abmag}. We use the Coma cluster distance modulus as for the S0s, but adopt an error of 0.12 to allow for the broader line-of-sight distribution of the spirals (c.f. 0.03 mag adopted for the S0s; Section \ref{sec:abmag}).

The derived Tully--Fisher parameters for the Coma cluster spiral sample are presented in Table \ref{tab:tfr_sp} (c.f. Table \ref{tab:tfr} for the S0 equivalent).

\subsection{Local density and the full SDSS Coma sample}
\label{sec:full}

Figure \ref{fig:field} displays the position of our Coma S0 and spiral samples on the sky. We aim to examine the environmental dependence of the S0 Tully--Fisher relation. The simplest approach is to use the projected cluster-centric radius as an indicator of local density, effectively assuming spherical symmetry in the Coma cluster. Line-of-sight distance and differential velocity with respect to the total system are degenerate, so we do not attempt to correct for projection. However, we note that the effect can only decrease the apparent radius (increase apparent local density) as outer members are projected onto inner regions.

We account for the non-spherical morphology of the cluster by deriving the local density for each galaxy via a nearest-neighbour algorithm. Such a density parameter requires a complete selection of galaxies within the Coma cluster. We select all sources in the SDSS DR9 {\it specphot} catalogue\footnote{accessed via http://skyserver.sdss3.org/CasJobs/} (the intersection of the full primary science spectroscopic and photometric catalogues) within $\pm$3500 km s$^{-1}$ of the Coma cluster systemic redshift ($z$ $=$ 0.0231) and with a $r$-band magnitude $M_r$ $<-18$ mag. There are 1113 such galaxies within a 5 deg radius of the nominal cluster centre.

\begin{figure}
\centering
\includegraphics[viewport=15mm 5mm 140mm 153mm,height=90mm,angle=270,clip]{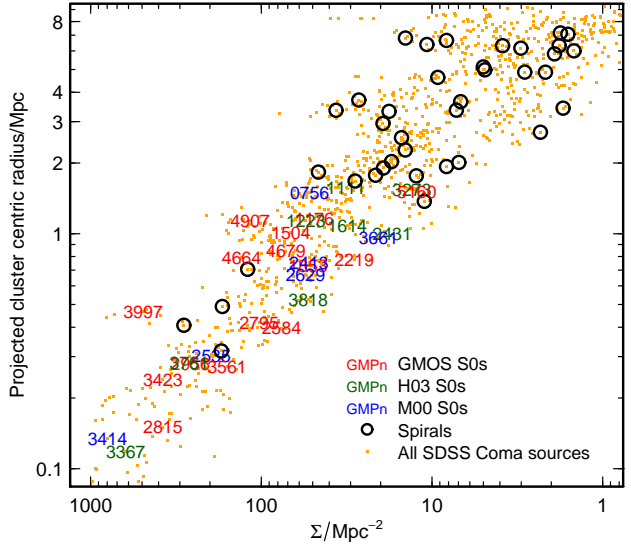}
\caption{Comparison of the projected cluster-centric radius and the projected local galaxy density parameter $\Sigma$, as defined in Section \ref{sec:full}. The full SDSS spectroscopic cluster member sample is shown by orange dots; further symbols as in Figure \ref{fig:field}. The deviations from linearity indicate regions where the galaxy distribution departs from a projected circular symmetry.}
\label{fig:density}
\end{figure}

We adopt the density parameter, $\Sigma$, from \citet{bal06-469}, which is defined as follows

\begin{equation}
\Sigma = \frac{1}{2} \left[ \frac{4}{\pi d^2_4} + \frac{5}{\pi d^2_5} \right]
\end{equation}
where $d_N$ is the projected co-moving distance to the $N$th nearest cluster member. We calculate $\Sigma$ for each galaxy, and display the local density as contours on Figure \ref{fig:field}. Figure \ref{fig:density} compares this $\Sigma$ parameter to the projected cluster-centric radius. The generally good correspondence suggests that the Coma cluster is mostly relaxed, although some significant substructure is revealed, such as the south-west over-density, indicated by a bulge to larger $\Sigma$ at $\sim$1 Mpc.

\section{Results and Discussion}
\label{sec:results}

\subsection{S0 Tully--Fisher relation}
\label{sec:tfr}

The Tully--Fisher relation (TFR) links the luminosity of spirals to their maximum circular velocity. If S0s are quenched spirals, the ageing stellar population would result in a luminosity decrease. In contrast, the galaxy mass (and hence rotational velocity) should remain constant causing an offset between the spiral and S0 TFRs.

We consider the TFR in the $g$, $i$ and $K_{\rm s}$ bands. In each band we derive the best fitting spiral TFR via an orthogonal regression, using the uncertainty in both parameters, as the least biased numerical fitting method available \citep[][and references therein]{hal12-2741}. The S0 TFR is then calculated using the spiral TFR gradient as a reference. The best fit parameters for all the TFRs described in this section are given in Table \ref{tab:tfr_fits}.

\begin{table*}
\centering
\caption{The Coma cluster S0 Tully--Fisher offset in the $g$, $i$ and $K_{\rm s}$ bands (Column 6). Columns 2--5 describe the reference spiral TFR in the form $M_X = a(\log V_{\rm c}-\log V_0)+b$, where $\log V_0=2.220$, and $\sigma_{\rm Sp}$ is the dispersion in the magnitude axis. In each case, the gradient of the Coma cluster S0 TFR (fitted to our S0 sample) is fixed to the reference spiral relation. The S0 offset and dispersion $\sigma_{\rm S0}$ are given in the direction of luminosity. Note that \citet{tul12-78} report $I_{\rm C}$ band results; for Coma spirals $m_i=m_{I_C}+0.43$.}
\label{tab:tfr_fits} 
\begin{tabular}{@{}clccccc}
\\
\hline
\multicolumn{1}{l}{band} & \multicolumn{1}{c}{reference spiral TFR} & \multicolumn{1}{c}{$a$} & \multicolumn{1}{c}{$b$} & \multicolumn{1}{c}{$\sigma_{\rm Sp}$} & \multicolumn{1}{c}{Coma S0 offset} & \multicolumn{1}{c}{$\sigma_{\rm S0}$}  \\
 & & & & mag & mag & mag \\
\hline
\multirow{3}{*}{$g$} & \citet{piz07-945} & $-5.48\pm0.23$ & $-20.69\pm0.04$ & 0.45 & {\bf 1.17$\pm$0.15} & {\bf 0.58} \\ 
& \citet{hal12-2741} & $-8.04\pm0.26$ & $-20.54\pm0.05$ & 0.26 & {\bf 1.32$\pm$0.12} & {\bf 0.84} \\ 
& {\bf Coma cluster (This Study)} & {\bf --6.17$\pm$0.45} & {\bf --20.54$\pm$0.04} & {\bf 0.31} & {\bf 1.10$\pm$0.18} & {\bf 0.64} \\ 
\hline
\multirow{5}{*}{$i$} & \citet{piz07-945} & $-6.32\pm0.22$ & $-21.39\pm0.04$ & 0.42 & {\bf 0.83$\pm$0.14} & {\bf 0.58} \\ 
& \citet{hal12-2741} & $-8.71\pm0.24$ & $-21.50\pm0.05$ & 0.26 & {\bf 1.23$\pm$0.12} & {\bf 0.87} \\ 
& \citet{tul12-78} & $-8.81\pm0.16$ & $-21.15\pm0.04$ & 0.41 & {\bf 1.03$\pm$0.14} & {\bf 0.88}  \\ 
& \citet{tul12-78} [Coma only] & $-6.96\pm0.56$ & $-21.42\pm0.06$ & 0.27 & {\bf 0.94$\pm$0.12} & {\bf 0.68} \\ 
& {\bf Coma cluster (This Study)} & {\bf --7.04$\pm$0.45} & {\bf --21.33$\pm$0.05} & {\bf 0.32} & {\bf 0.86$\pm$0.19} & {\bf 0.69} \\ 
\hline
\multirow{3}{*}{$K_{\rm s}$} & \citet{mas08-1738} & $-7.25\pm0.11$ & $-23.63\pm0.08$ & 0.40 & {\bf 0.80$\pm$0.12}  & {\bf 0.55} \\
& \citet{wil10-1330} & $-8.15\pm0.76$ & $-23.06\pm0.11$ & 0.37 & {\bf 0.35$\pm$0.11} & {\bf 0.54} \\ 
& {\bf Coma cluster (This Study)} & {\bf --7.31$\pm$0.49} & {\bf --23.66$\pm$0.07} & {\bf 0.35} & {\bf 0.83$\pm$0.19} & {\bf 0.67} \\ 
\hline
\end{tabular}
\end{table*} 

\subsubsection{$g$-band TFR} 
\label{sec:tfr_g}

\begin{figure}
\centering
\includegraphics[viewport=15mm 5mm 140mm 153mm,height=90mm,angle=270,clip]{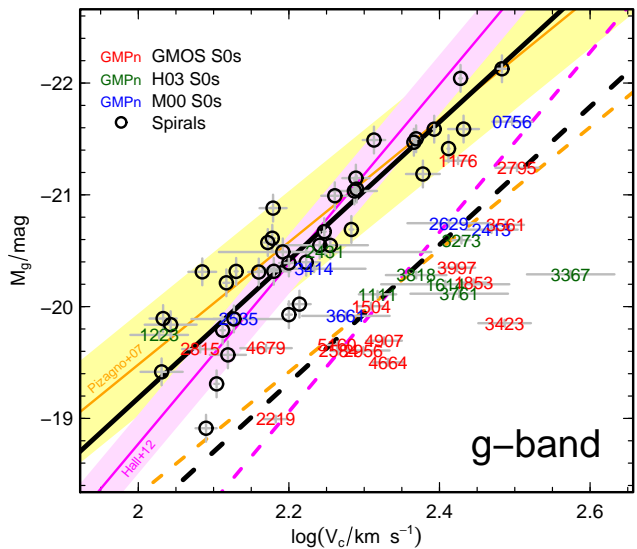}
\caption{The $g$-band Tully--Fisher relation for the Coma cluster. S0s are represented by their GMP ID number, while spirals are shown as black circles. The black solid line shows the best orthogonal regression fit to the Coma spiral TFR (gradient $=$ $-6.17\pm0.45$), while the black dashed line shows the best fit to the S0 sample, fixing the gradient to match the Coma spirals. The mean S0 offset is $1.10\pm0.18$ mag. Additional spiral TFRs are shown by the orange \citep[][gradient $=$ $-5.48\pm0.23$]{piz07-945} and magenta \citep[][gradient $=$ $-8.04\pm0.26$]{hal12-2741} solid lines (yellow and purple shaded regions correspond to 1$\,\sigma$ dispersions). Best fit Coma S0 TFRs fixed to these spiral reference gradients are shown by orange and magenta dashed lines.}
\label{fig:tf_g}
\end{figure}

The $g$-band TFR for the 38 spirals in our Coma cluster sample (Table \ref{tab:tfr_sp}) is shown in Figure \ref{fig:tf_g}. The best fit relation is $M_g$ $\propto$ ($-6.17\pm0.45$)$\,logV_{\rm c}$, with an rms dispersion of 0.31 mag.

We verify the reliability of this TFR by comparing to previous studies of larger samples. Recently, \citet{hal12-2741} presented local spiral TFRs using newly derived photometry from DR7 SDSS images. For a ``best'' sample of 668 spirals, they derive rotational velocities from H{\sc i} linewidths in the \citet{spr07-599} SFI++ catalogue, using exactly the same method as for our spiral sample (Section \ref{sec:spirals}). Using the same orthogonal fitting method, they report a steeper gradient of $-8.04\pm0.26$. Several of our Coma spirals are included in the \citeauthor{hal12-2741} sample, and we find no significant difference in the TF parameters of these individual galaxies (rms dispersion of $\Delta V_{\rm c}$ $<$ 1 km s$^{-1}$). In their Section 7.1, \citeauthor{hal12-2741} suggest that their steep gradient results from a lack of explicit morphological selection: the sample includes a large fraction of early-type spirals (including many S0/a galaxies), particularly at higher luminosity, compared to other studies. In contrast, less than one quarter of our Coma cluster spiral sample is listed as Sa or Sab by NED, and of course includes no S0 or S0/a galaxies.

\citet{piz07-945} present a sample of 200 local spirals with no morphological selection. Photometry is taken from SDSS, while maximum rotational velocity is derived from H$\alpha$ rotation curves, using a long-slit kinematic analysis similar to the method we describe for our S0s in Section \ref{sec:vc}. \citeauthor{piz07-945} use a bivariate fit to their data, deriving a gradient of $-5.48\pm0.23$, and an rms dispersion of 0.45 mag. The marginally shallower gradient can be attributed to the fitting method \citep{hal12-2741}, and the TFR is  compatible with our Coma relation, despite the very different origin of the $V_{\rm c}$ measurement.

We now consider the Coma cluster S0 sample comprising 29 galaxies (Table \ref{tab:tfr}). The S0s are significantly offset from the Coma spiral relation. Assuming the same gradient as the spiral TFR, we calculate the mean S0 offset from the spirals to be $1.10\pm0.18$ mag. At a given rotational velocity, S0s in the Coma cluster are on average fainter than spirals by more than a magnitude. The mean offset is more than three times the rms dispersion of the spirals. For comparison, we also calculate the S0 offset assuming the spiral TFR from the two studies discussed above: $1.32\pm0.12$ and $1.17\pm0.15$ mag from the \citeauthor{hal12-2741} and \citeauthor{piz07-945} TFRs respectively (Table \ref{tab:tfr_fits}). The larger offset from the \citeauthor{hal12-2741} TFR is a consequence of the steeper gradient.

A few individual galaxies in our analysis warrant further comment, starting with the five M00/H03 objects not strictly classified as S0 (see Section \ref{sec:mores0}). The two SB0 (GMP1614, 3761) and two S0/a (GMP3273, 3818) exhibit an offset indistinguishable from the true S0s. In contrast, GMP2431 (designated Sa by D80, but S0/a in NED) is located close to the spiral TFR. Several other S0s have small offsets from the spiral TFR, and most were previously identified as having possibly underestimated maximum velocities due to shallow spectroscopy (GMP1223, 2535, 3414; Sections \ref{sec:m00}, \ref{sec:h03}). For these to be located on the mean S0 relation, $V_{\rm c}$ would need to be larger by 40--70 km s$^{-1}$, which is not implausible given the shape of their observed velocity curves. However, the GMOS S0s with the smallest TFR offsets (GMP1176, 2815) have well constrained $V_{\rm c}$ (Figures \ref{fig:kin_gmp1176}, \ref{fig:kin_gmp2815}), and therefore really are located closer to the spiral TFR.

\subsubsection{$i$-band TFR} 
\label{sec:tfr_i}

\begin{figure}
\centering
\includegraphics[viewport=15mm 5mm 140mm 153mm,height=90mm,angle=270,clip]{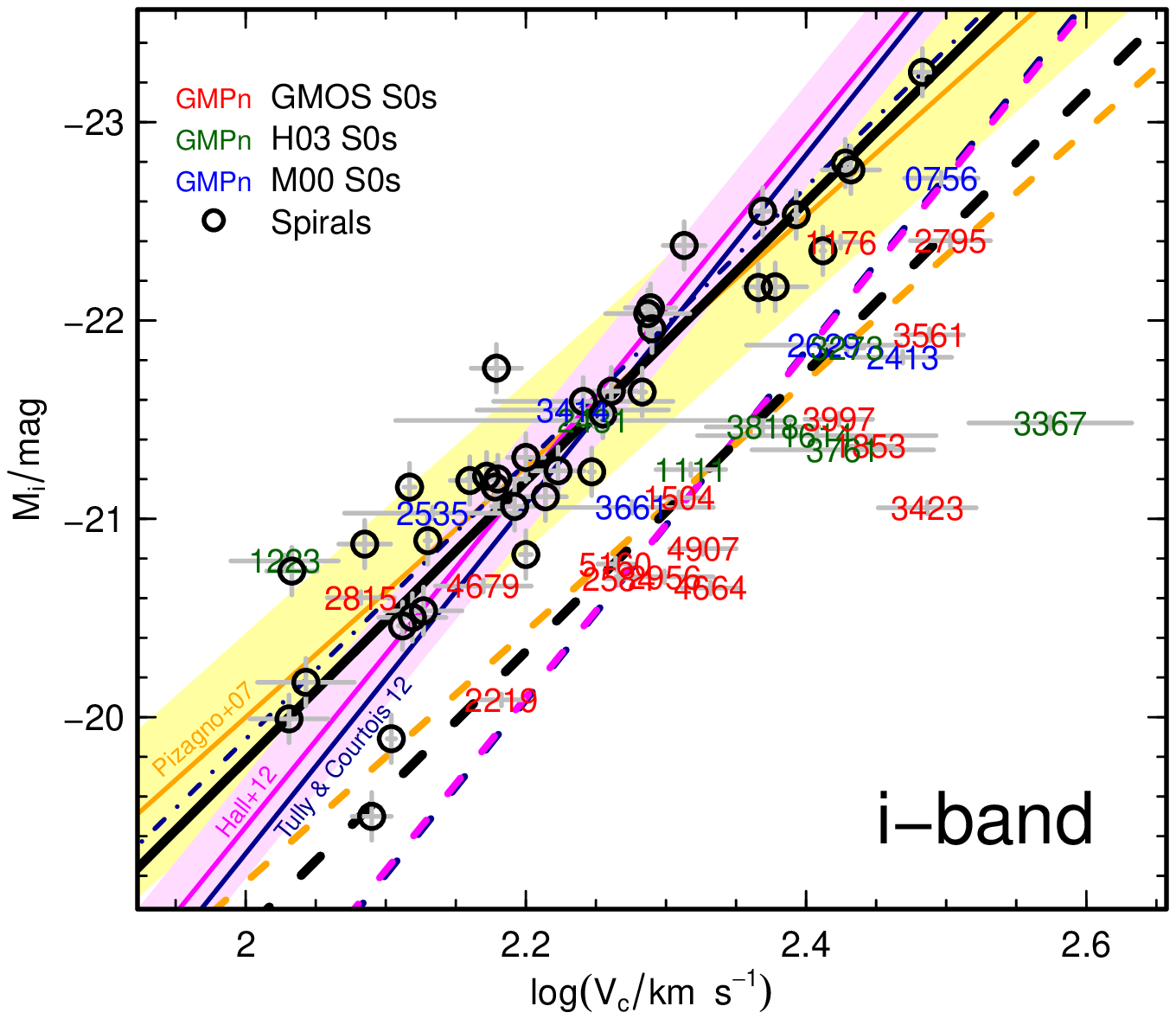}
\caption{The $i$-band Tully--Fisher relation for the Coma cluster. Layout identical to Figure \ref{fig:tf_g}. The bivariate fit to the Coma spiral TFR has a gradient $=$ $-7.04\pm0.45$, and the mean offset for the S0s is $0.86\pm0.19$ mag. The spiral TFR from \citet[][gradient $=$ $-6.32\pm0.22$]{piz07-945}, \citet[][gradient $=$ $-8.71\pm0.24$]{hal12-2741} and \citet[][gradient $=$ $-8.81\pm0.16$]{tul12-78} are shown by the orange, magenta and blue solid lines. The Coma only sample from \citet[][gradient $=$ $-6.96\pm0.56$]{tul12-78} is displayed as a blue dashed-dotted line.}
\label{fig:tf_i}
\end{figure}

Figure \ref{fig:tf_i} presents the SDSS $i$-band TFR. The best fit to the Coma cluster spiral galaxy sample gives $M_i$ $\propto$ ($-7.04\pm0.45$)$\,logV_{\rm c}$, with a dispersion of 0.32 mag.

Our $i$-band TFR is similar to \citet{piz07-945}, who calculate a gradient of $-6.32\pm0.22$ and a dispersion of 0.42 mag, while \citet{hal12-2741} report a steeper gradient ($-8.71\pm0.24$). In addition, we compare to the recent $I_{\rm C}$-band TFR relation presented by \citet{tul12-78}, who use H{\sc i} linewidths of 267 galaxies in 13 clusters, including 23 spirals in the Coma cluster. For the Coma cluster, \citeauthor{tul12-78} derive a TFR gradient of $-6.96\pm0.56$, which is very similar to our value. For their full sample, they derive a gradient of $-8.81\pm0.16$ (rms dispersion of 0.41 mag), which is more compatible with the \citeauthor{hal12-2741} TFR. Our shallower gradient in the $g$ and $i$ bands may result from a variation in the TFR between different clusters \citep[as discussed by][]{ber94-1962}. However, \citet{tul12-78} report that the Coma cluster sample can be drawn from the universal relation, and the variation may simply result from morphological selection bias in different environments.

We derive a mean $i$-band S0 TFR offset of $0.86\pm0.19$ mag, which is twice the rms dispersion of the spiral sample. This offset is smaller (by $\sim$0.2 mag) than for the corresponding $g$-band TFR. The mean Coma S0 offsets from the \citet{piz07-945}, \citet{hal12-2741} and \citet{tul12-78} spiral TFRs are $0.83\pm0.14$, $1.23\pm0.12$ and $1.03\pm0.14$ mag respectively (Table \ref{tab:tfr_fits}).

\subsubsection{$K_{\rm s}$-band TFR} 
\label{sec:tfr_k}

\begin{figure}
\centering
\includegraphics[viewport=15mm 5mm 140mm 153mm,height=90mm,angle=270,clip]{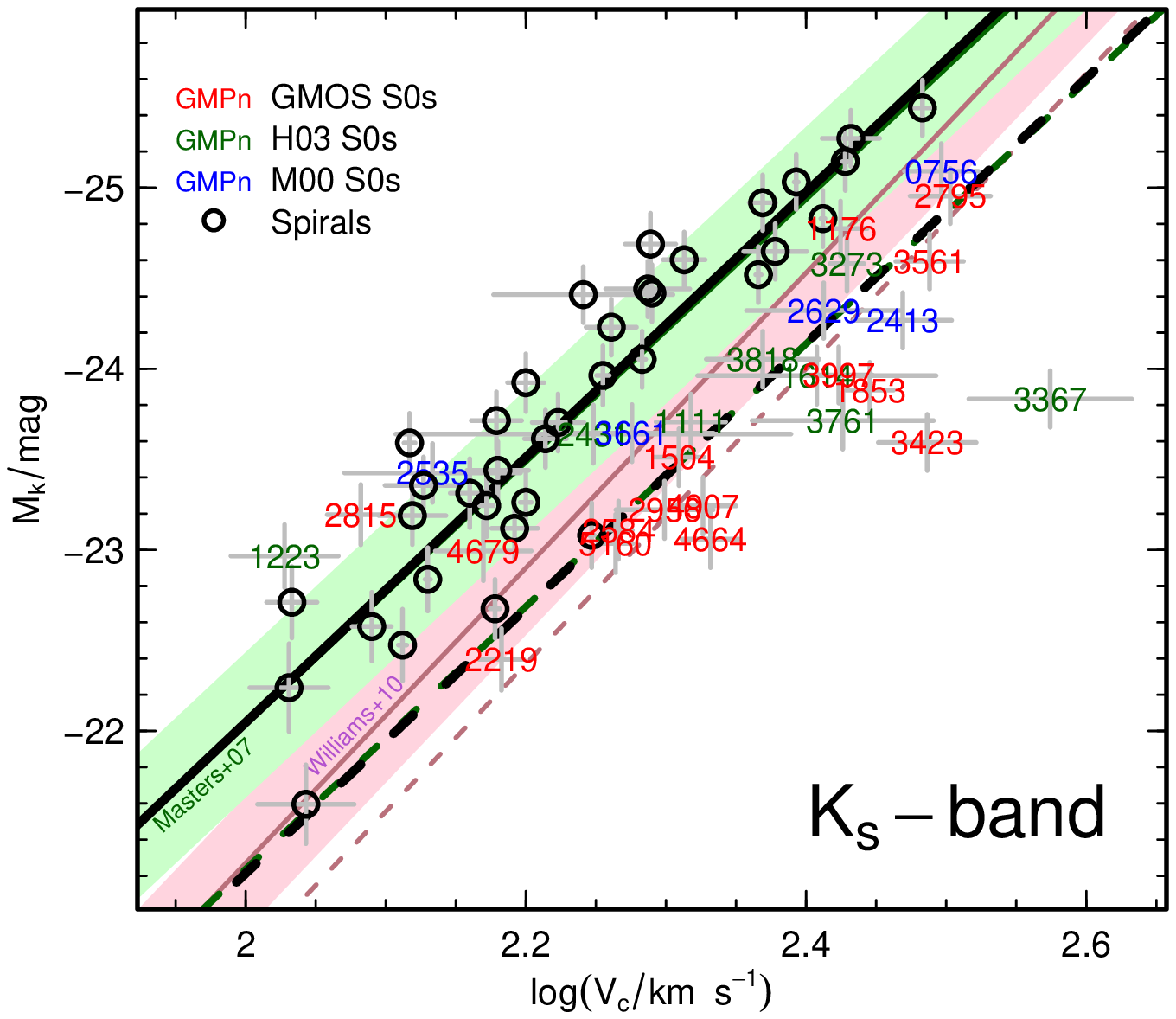}
\caption{The $K_{\rm s}$-band Tully--Fisher relation for the Coma cluster. Layout as in Figure \ref{fig:tf_g}. The fit to the Coma spiral TFR has a gradient $=$ $-7.31\pm0.49$, and the mean offset for the S0 sample is $0.83\pm0.19$ mag. Spiral TFRs from \citet[][gradient $=$ $-7.25\pm0.11$]{mas08-1738} and \citet[][gradient $=$ $-8.15\pm0.76$]{wil10-1330} are shown by the green and mauve solid lines respectively. While the green dashed line indicates the best fit to our Coma S0s using \citeauthor{mas08-1738} as the spiral reference, the mauve dashed line shows the S0 TFR directly from \citet{wil10-1330}.}
\label{fig:tf_k}
\end{figure}

The 2MASS $K_{\rm s}$ data is shallower than SDSS, restricting the S0 and spiral samples to 28 and 36 members respectively (see Tables \ref{tab:tfr} and \ref{tab:tfr_sp}). The best fit to the Coma cluster spiral galaxies, shown in Figure \ref{fig:tf_k}, reveals $M_{K_s}$ $\propto$ ($-7.31\pm0.49$)$\,logV_{\rm c}$, which is marginally steeper than in the $i$ band. The rms dispersion of the spirals in $M_{K_s}$ is 0.35 mag. The mean S0 offset in the $K_{\rm s}$-band is $0.83\pm0.19$ mag.

We compare our near-infrared TFR to the 2MASS Tully--Fisher Survey \citep{mas08-1738}, which once again uses the SFI++ H{\sc i} linewidths \citep{spr07-599}. For their full sample of 888 spiral galaxies, the authors use a bivariate best fit to derive a $K_{\rm s}$-band gradient of $-7.25\pm0.11$, with an rms dispersion of 0.40 mag. Attempting to correct for morphology (i.e. estimating the TFR of an Sc-only sample), they derive a much steeper gradient of $-8.92\pm0.10$. Using the full \citeauthor{mas08-1738} spiral TFR as the reference for our Coma S0s, we calculate an offset of $0.80\pm0.12$ mag (Table \ref{tab:tfr_fits}).

Finally, we compare our $K_{\rm s}$-band S0 TFR to \citet{wil10-1330}. Figure \ref{fig:tf_k} shows that their S0 TFR is very similar to the Coma cluster. However, these authors report a S0-to-spiral TFR offset of only $0.53\pm0.15$ mag. This smaller offset arises because their spiral TFR is significantly displaced from the spiral TFRs for either our Coma cluster sample or that of \citeauthor{mas08-1738} ($\sim$0.6 mag; Table \ref{tab:tfr}). The disagreement does not originate from luminosity, as each study uses 2MASS photometry with similar internal extinction corrections ($\le0.1$ mag). Nor is it due to different velocity systems: although the primary result from \citeauthor{wil10-1330} (plotted in Figure \ref{fig:tf_k}), uses dynamic velocity derived from an NFW model of the dark matter mass distribution for each galaxy, their TFR is not significantly shifted if kinematic circular velocity is employed instead. The mean difference between these two velocities is 10 and 7 km s$^{-1}$ for their spiral and S0 samples respectively. Rather, the difference is due to the exclusive use of earlier types in the \citet{wil10-1330} spiral sample (Sa--Sb only; and including several S0/a). In contrast, Sc--Sd spirals make up a third of our Coma cluster spiral sample. This clear offset of Sa/b spirals from much later-types hints at a continuous trend in TFR offset, rather than discrete populations of spirals and S0s.

\subsubsection{Simple model for the multi-band TFR offsets}
\label{sec:tfr_summary}

Generally, differences between spiral TFRs are due to sample selection (e.g. morphological bias) and population fitting procedures \citep[as discussed in][]{hal12-2741}. The typical S0 TFR offset in the Coma cluster is $\sim$0.8--1.2 mag, depending on the photometric band and selection criteria for the reference spiral TFR (Table \ref{tab:tfr_fits}). The $i$- and $K_{\rm s}$-band fluxes are well correlated, with an colour dispersion of only 0.34 mag for spirals and 0.13 mag for S0s. In the $g$ band, the S0 offset is $\sim$0.2 mag larger than for the redder bands. 

Quantitatively, the size of the mean observed spiral-to-S0 offset, and its dependence on photometric band, are consistent with a simple model: S0s initially have similar star-formation histories (SFHs) to the spirals, but are abruptly quenched at some intermediate redshift. As an example, we consider two variants of an exponentially-decaying SFH, beginning 13 Gyr ago,
and with an $e$-folding time ($\tau$) of 14 Gyr. The first version is allowed to continue forming stars to the present day, while the second variant is cut off 5 Gyr before the present (i.e. $z\approx0.5$). Convolving these SFHs with the \citet{mar05-799} single-burst models, we find that at $z=0$ the quenched variant is 1.15 mag fainter in $g$ than the unquenched version. The equivalent differences in $i$ and $K_{\rm s}$, which are less sensitive to the youngest stars, are 0.79 mag and 0.74 mag respectively. These results are clearly consistent with the TFR offsets we observe, but they are not unique: other combinations of age, $\tau$, and quenching time would also be compatible with the observations.

\subsection{TFR offset correlations}
\label{sec:offset}

In the previous section we report that S0s are offset from the spiral TFR by an average of $\sim$0.8--1.2 mag, in the sense that S0s are fainter than spirals for a given rotational velocity. The existence and extent of the offset for any individual S0 in the Coma cluster is relatively independent of the choice of reference spiral TFR or observed band. Simple models show that the TFR offset reflects a dimming of the stellar population, and can be interpreted as a probe of the time since the S0-to-spiral transformation began.

\subsubsection{Correlation with colour}
\label{sec:cmr}

\begin{figure}
\centering
\includegraphics[viewport=15mm 5mm 140mm 153mm,height=90mm,angle=270,clip]{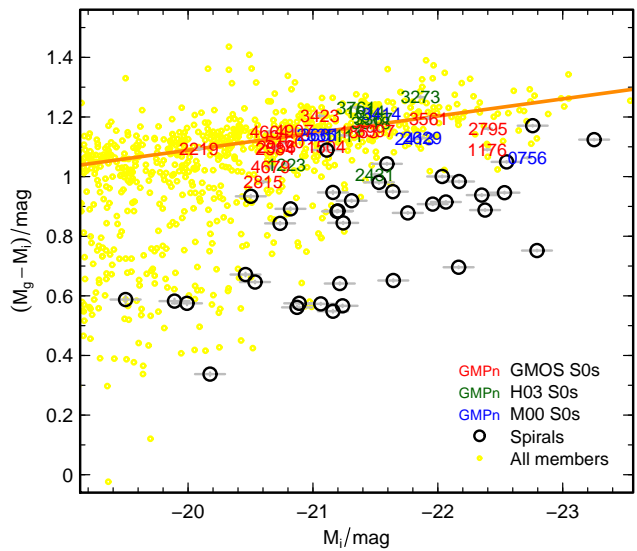}
\caption{The $g-i$ colour--magnitude diagram for the full SDSS spectroscopic Coma cluster member sample (yellow circles) with the strong red sequence (colour--magnitude relation, CMR) highlighted via an orange line. By selection, our S0 galaxies are all located near to the CMR, while the spirals scatter to much bluer colours.}
\label{fig:cmr}
\end{figure}

First, we explore the interplay between the TFR offset and the integrated optical colour of a galaxy. We begin by examining the SDSS $g-i$ colour--magnitude diagram for the full Coma cluster spectroscopic member sample (Figure \ref{fig:cmr}). The cluster members exhibit a strong red sequence (hereafter colour--magnitude relation, CMR) with an rms dispersion of 0.05 mag \citep[e.g.][]{bow92-601,smi12-3167}. By selection, the galaxies within our S0 sample are located in the vicinity of this CMR, and have only a marginally larger rms dispersion (0.07 mag). The spiral sample has bluer $g-i$ colours, and a much larger rms dispersion (0.39 mag). While the two samples have similar luminosity distributions, the colour--magnitude diagram clearly displays the difference in colour. However, the populations do overlap in a `green valley' region, which is often interpreted as the location for the evolutionary intermediary stage between star-forming late-type galaxies and quiescent early-types \citep{fab07-265}. The offset from the CMR is an alternative tracer of the evolutionary progress of a galaxy.

\begin{figure}
\centering
\includegraphics[viewport=15mm 5mm 140mm 153mm,height=90mm,angle=270,clip]{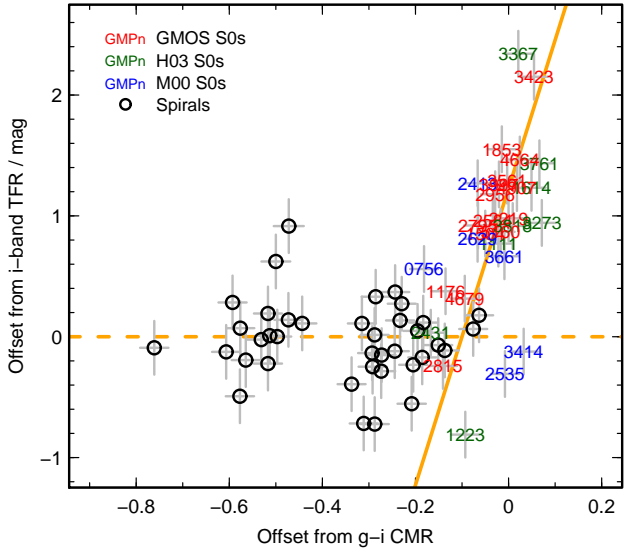}
\caption{Relation between the offsets from the $i$-band TFR and the $g-i$ CMR. The solid orange line shows the best bivariate fit to all S0s (gradient $=12.3\pm1.9$). S0s with larger TFR offsets show generally redder  $g-i$ colours, while those closer to the spiral TFR also exhibit colours similar to the `green valley' spirals.}
\label{fig:tf_i_cmr}
\end{figure}

We compare the $g-i$ CMR offset to the $i$-band TFR offset in Figure \ref{fig:tf_i_cmr}. The two parameters, almost by definition, have complementary interpretive power. The TFR offset is based on the tight spiral relation, offering little distinction within this population, while distributing the S0s over a two magnitude range. In contrast, the CMR offset is based on the tight red sequence for early-types (including S0s), but shows a wide scatter in the spirals. Within the S0 population, there is a highly significant correlation between these offsets. S0s which are {\it fainter} than average for their rotational velocity are also {\it redder} than average for their luminosity. S0s with a colour most similar to `green valley' spirals exhibit the smallest TFR offsets. The best bivariate fit to all S0s produces a gradient of $12.3\pm1.9$ ($>6\,\sigma$ significance), with rms dispersion of 0.57 and 0.06 mag in the two axial directions. The three S0s falling significantly below the general trend (GMP1223, 2535, 3414), were previously identified as examples where the long-slit data may not measure the maximal $V_{\rm c}$.

\begin{figure*}
\centering
\includegraphics[viewport=15mm 15mm 137mm 292mm,height=185mm,angle=270,clip]{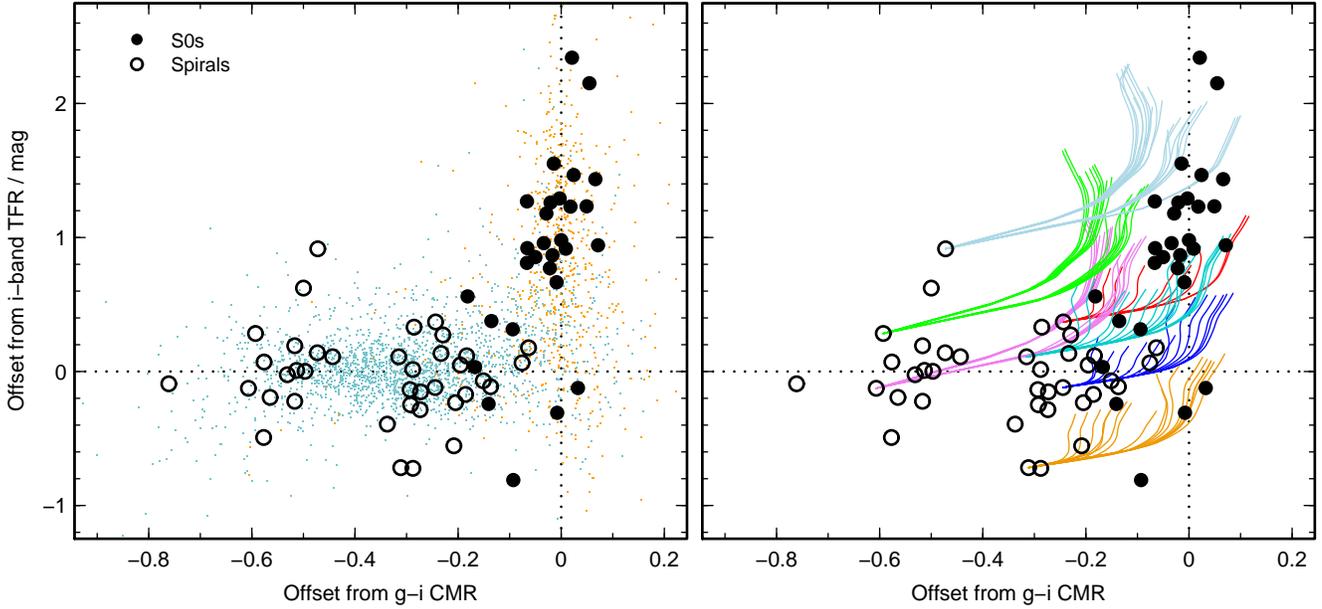}
\caption{Relation between the offsets from the $i$-band TFR and the $g-i$ CMR, as in Figure \ref{fig:tf_i_cmr}, with model predictions over-plotted. {\it Left:} Disc dominated galaxies from semi-analytic models \citep{del07-2}, separated by SFR: non-star-forming galaxies (SFR $<$ 0.1 M$_{\sun}$ yr$^{-1}$) in orange and star-forming in blue. {\it Right:} Single-burst star formation models from Maraston (2005). We select all unquenched model galaxies that match the observed optical colours of seven representative Coma spirals at $z=0$. From the equivalent quenched models, we predict observables assuming a range of different quenching epochs. For example, the green lines show the possible $z=0$ locations of model galaxies based on one example observed spiral, assuming a range of different SFHs, metallicities and quenching times. Hence the lines are not evolutionary tracks, but the result of fading hypothetical 'parent' spirals via a range of models. These models can successfully reproduce the location of most observed Coma S0s.}
\label{fig:tf_cmr_mod}
\end{figure*}

We now explore the interconnection of the TFR and CMR offsets by considering the likely star-formation histories of the S0 and spiral galaxies. Specifically we investigate whether the same `parent' population of spirals can give rise to both the (current) spirals and the (current) S0s in Coma, dependent only on whether or not they experienced instantaneous quenching at some intermediate epoch. We compare to predictions from both complex semi-analytic models, and simple, quenched star-formation history models.

For the semi-analytic predictions, we use catalogues from the models of \citealt{del07-2}, which were based on merger trees from the Millennium Simulation \citep{spr05-629}. We use the virial velocity $V_{\rm vir}$ of the dark matter halo in place of $V_{\rm c}$, and applying a small offset of $\sim$0.25 mag to match the observed zero-point for the spiral TFR. (Note that models which attempt to compute $V_{\rm c}$ self-consistently fail to match the TFR zero point, as discussed by \citealt{bau06-3101}). Figure \ref{fig:tf_cmr_mod} (left panel) shows the model predictions for galaxies with bulge mass-fractions less than 60 per cent, and with $V_{\rm vir} >$ 100 km s$^{-1}$, and divided according to star-formation rate (SFR). The general form of the distributions of star-forming and non-star-forming galaxies is well-matched to the properties of the observed spirals and S0s respectively. As in the observed sample, the region corresponding to small colour residuals and small TFR offsets is populated by a mixture of star-forming and non-star-forming galaxies. However, the colours of the simulated non-star-forming galaxies do not seem to be correlated with their TFR offsets, in contrast to the trend seen in the observed sample. This may be because the simple SFR cut cannot reproduce the morphological spiral/S0 distinction made in the observed samples, particularly for objects in transition between the classes.

The semi-analytic predictions include many physical processes, and perhaps obscure the effect of quenched SFHs on the residuals from both the CMR and TFR. To isolate this effect, we return to the simplified star formation model considered in Section \ref{sec:tfr_summary}. Here, the model library spans a range of formation times, (positive) exponential decline times, metallicities and quenching times in the past 5 Gyr. We generate predictions by convolving the library SFHs with the predictions for the single-burst models from \citet{mar05-799}. For each galaxy in the Coma spiral sample, we find all {\it unquenched} models which at $z=0$ match the observed $g-i$ colour. For each of these models, we extract predictions at $z=0$ from the equivalent {\it quenched} models, for a range of quenching times. The loci displayed in Figure \ref{fig:tf_cmr_mod} (right panel) show the effect of instantaneous quenching at different epochs on a sample of hypothetical `parent' spirals.

Figure \ref{fig:tf_cmr_mod} (right panel) demonstrates that this model can reproduce the location of most of the S0s by fading the predicted progenitors of the current spiral population: the observed spirals and S0s can all originate from the same `parent' spiral population. However, we note that for the reddest S0s with large TFR offsets, the required quenching times are close to the epoch of `formation'. The gradient of the upturn for these loci (as quenching time tends to formation time) may go some way to explain the observed S0 correlation between TFR and CMR offsets (Figure \ref{fig:tf_i_cmr}). We emphasise that these calculations are very simplified (e.g. dust is neglected; we treat the whole galaxy with a single SFH ignoring any old ``bulge''; no starbursts occur at quenching time, etc) and more realistic models would likely have even more freedom to match the observed S0 properties.

\subsubsection{Evidence for environmental triggering}
\label{sec:environment}

\begin{figure*}
\centering
\includegraphics[viewport=15mm 15mm 135mm 292mm,height=185mm,angle=270,clip]{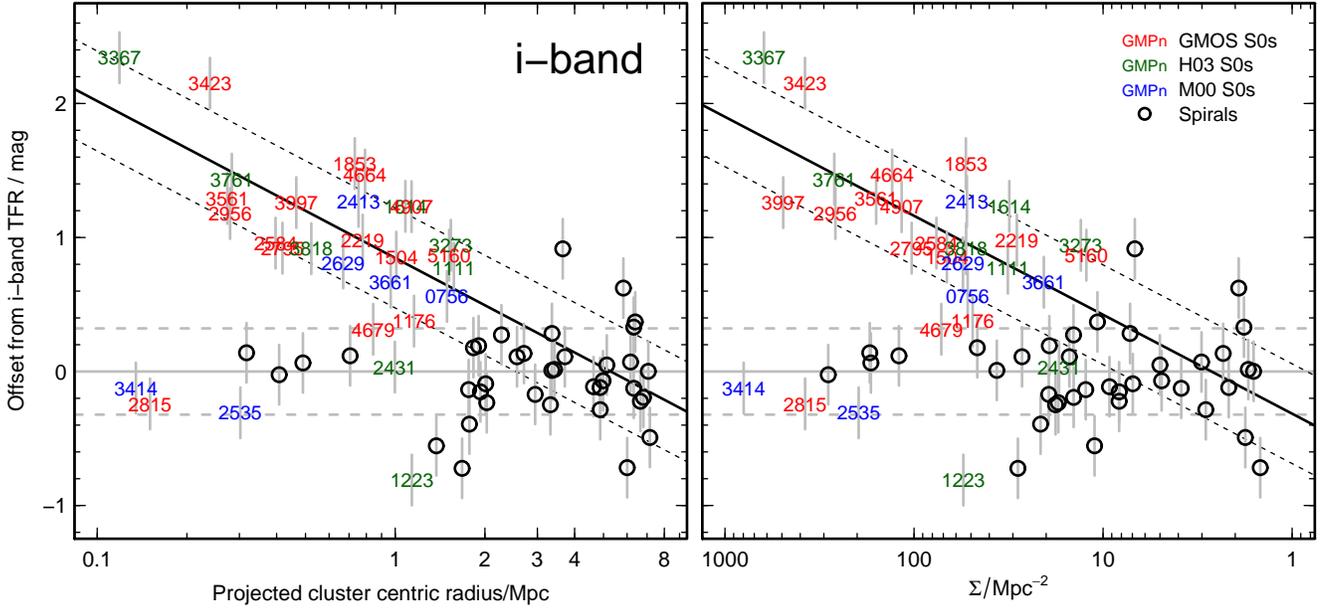}
\caption{S0 offset from the $i$-band spiral TFR versus the projected cluster-centric radius ({\it left}) and the local density parameter $\Sigma$ ({\it right}). The 1$\,\sigma$ dispersion of the spiral galaxies (black circles; by definition, mean population offset $=$ 0) is shown by the grey horizontal dashed lines. The S0s display a strong relationship between TFR offset and environment, with gradients of $-1.17\pm0.27$ and $0.74\pm0.17$ respectively (solid black lines; 1$\,\sigma$ dispersion $=$ 0.37 mag, as black dotted lines), after removing four galaxies highlighted in previous sections (see full explanation in Section \ref{sec:offset}).}
\label{fig:tf_i_offset}
\end{figure*}

The S0 TFR offset is consistent with abrupt quenching of a spiral population. We now explore the offset as a function of local environment. Galaxies at larger projected distance from the cluster center were accreted into the cluster (and its progenitors) at earlier times on average, than galaxies projected near the cluster core \citep{gao04-1,smi12-3167,del12-1277,oma13-arxiv}. Hence cluster-centric radius can be employed as a proxy for infall time, albeit with substantial scatter.

The transformation of spirals into S0s could conceivably occur in small groups before infall into the cluster (pre-processing), where differential velocities are small and galaxy-galaxy interactions dominate the quenching process \citep[e.g.][]{jus10-192}. Alternatively, transformation could occur during in-fall into the cluster, where the cessation of star formation is triggered by the stripping of cold gas through interaction with the increasingly dense cluster medium \citep{gun72-1,aba99-947}. Observational evidence of such stripping in the Coma cluster was reported by \citet{smi10-1417}. Naively, a correlation between local environment and time since quenching began may be expected only if the process occurred entirely within the cluster environment. However, recent evidence \citep{smi12-3167} shows that galaxies at larger projected cluster-centric radii not only entered the cluster later on average than those near the centre, but also entered progenitor groups later as well. Therefore, even if quenching were triggered in smaller groups, radial trends would still be observed within the cluster.

We begin our discussion of the TFR offset with the $i$-band data, which has the higher photometric precision of SDSS and the smaller extinction-correction uncertainties. Figure \ref{fig:tf_i_offset} presents the $i$-band TFR offset as a function of local environment, as traced by both the simple projected cluster-centric radius and the $\Sigma$ density parameter. For the Coma cluster spiral sample, the mean offset is zero by definition, and the individual TFR offsets show no correlation with either density parameter. As emphasised previously, the spiral sample is located at much larger radii (lower densities) than the S0s.

By eye, the TFR offset for the S0 sample appears to have a strong trend with local environment, in the sense that most central S0s (located at highest density) exhibit the largest offset. However, statistically for the full S0 sample, this trend is not significant; the gradients are $-0.34\pm0.37$ and $0.25\pm0.25$ versus log(radius) and $\log\Sigma$ respectively. The four S0s with negative offsets, i.e. GMP1223, 2535, 2815, 3414, dilute the dominant trend that is apparent by eye. Three of these galaxies are the faintest S0s highlighted in Section \ref{sec:cmr}, and may have underestimated $V_{\rm c}$. The fourth (GMP2815) has an extremely low $cz_{\rm hel}$ (4664 km s$^{-1}$), indicating that the galaxy may be in the foreground: a projection-corrected radius parameter would place GMP2815 amongst the spiral population at a larger true cluster-centric radius. If we remove these four galaxies, the remaining 25 galaxies show a significant gradient ($>4\,\sigma$; solid black lines in Figure \ref{fig:tf_i_offset}) of $-1.12\pm0.26$ and $0.70\pm0.16$ versus log(radius) and $\log\Sigma$ respectively. The measured dispersion is 0.36 mag in both cases, with an intrinsic scatter of 0.32 mag. The majority of the spiral galaxies ($\sim60$ per cent, not accounting for projection uncertainties) are also located within the $1\,\sigma$ limits of the S0 trend, indicating that there may indeed be a continuum of objects, or possibly an evolutionary track, linking the two populations.

In the $g$ and $K_{\rm s}$ bands, we observe a similar trend in the S0s (not shown here for brevity). In the $g$ band, the derived best fit gradients are $-1.04\pm0.27$ and $0.64\pm0.17$ (versus projected radius and $\Sigma$ respectively), with 1$\,\sigma$ dispersion of 0.37 and 0.38 mag (0.34 mag intrinsic scatter), after removing the same four outliers as identified above. In the near-infrared $K_{\rm s}$ band, the gradients are $-1.12\pm0.26$ and $0.73\pm0.16$, with 1$\,\sigma$ measured dispersion for either density parameter of 0.36 mag (0.27 mag intrinsic). In all three bands, there is a $>3\,\sigma$ correlation of the S0 TFR offset with the tracers of local density, consistent with each other within the uncertainties. This trend suggests a link between the onset of interaction with the intracluster medium and the cessation of star formation.

\begin{figure}
\centering
\includegraphics[viewport=15mm 5mm 140mm 153mm,height=90mm,angle=270,clip]{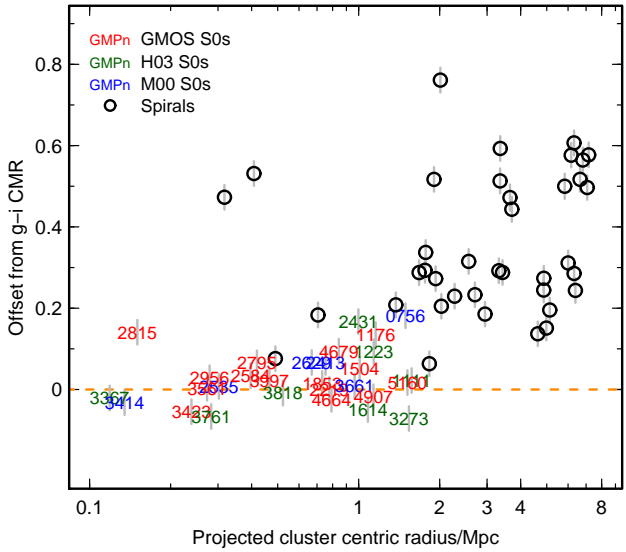}
\caption{Offset from the $g-i$ colour--magnitude relation (CMR), as a function of projected cluster-centric radius. Generally, the the mean CMR offset increases, with increasing scatter, as cluster radius decreases: $\le0.5$ Mpc ($\Sigma\ge100$ Mpc$^{-2}$) the mean offset is 0.00 mag (0.05 mag rms scatter; including the obvious outlier GMP2815); $>0.5$ Mpc ($\Sigma<100$ Mpc$^{-2}$) the mean offset is 0.05 mag (0.08 mag rms).}
\label{fig:cmr_offset}
\end{figure}

For completeness, Figure \ref{fig:cmr_offset} presents the offset from the CMR as a function of local environment, as traced by the simple projected cluster-centric radius. The spiral population alone shows no trend with environment. Two blue spirals (GMP2559, GMP3896) appear to be near the cluster core ($0.3-0.4$ Mpc) but this is likely to be a projection effect. Within the S0 population there is a marginal trend, whereby the S0s at larger cluster radii exhibit an increased scatter (towards higher offsets; bluer colours) compared to S0s in the cluster core. S0s at $\le0.5$ Mpc ($\Sigma\ge100$ Mpc$^{-2}$) have a mean CMR offset of 0.00 mag and an rms scatter of 0.05 mag, whereas those at $>0.5$ Mpc ($\Sigma<100$ Mpc$^{-2}$) exhibit a mean offset of 0.05 mag and a scatter of 0.08 mag. Generally, bluer S0s in the Coma cluster are more likely to be located at larger cluster-centric radii, much like the spiral sample itself, which is simply a confirmation of the morphology--density relation.

\subsubsection{Central age correlation}
\label{sec:age}

A spiral infalling into a rich cluster will encounter the increasing density of the intracluster medium and this may trigger quenching. A pure fading model of S0 transformation would necessarily increase the average stellar population age. If quenching is accompanied by a nuclear starburst, the central age of more recently quenched galaxies would be even younger, strengthening any age--TFR offset correlation for bulge-dominated ages. For a small sample of Fornax cluster galaxies, \citet{bed06-1125} report that central ages are more strongly correlated with the TFR offset than the `global' ages. In this final section, we briefly attempt to constrain the transformation mechanism in Coma via single stellar population ages derived from the VErsatile SPectral Analysis (VESPA) catalogue\footnote{http://www-wfau.roe.ac.uk/vespa/} based on SDSS DR7 spectra \citep{toj09-1}. The catalogue includes all of the Coma cluster members in our S0 and spiral samples, and describes SFH via discrete stellar mass estimates in several stellar age bins. From this information, we calculate the mass-weighted stellar population age
\begin{equation}
\langle Age\rangle_{\rm \mathcal{M}} = \sum\limits_{i}^{} \left( Age_i \mathcal{M}_i \right) /  \sum\limits_{i}^{} \mathcal{M}_i
\end{equation}
where $Age_i$ and $\mathcal{M}_i$ are the age and estimated stellar mass in the $i$-th bin. The 3 arcsec SDSS fibres ensure that for Coma, the mass-weighted stellar population age simply reflects the typical star formation history of the central $\sim$1.4 kpc of the bulge.

\begin{figure}
\centering
\includegraphics[viewport=15mm 5mm 140mm 153mm,height=90mm,angle=270,clip]{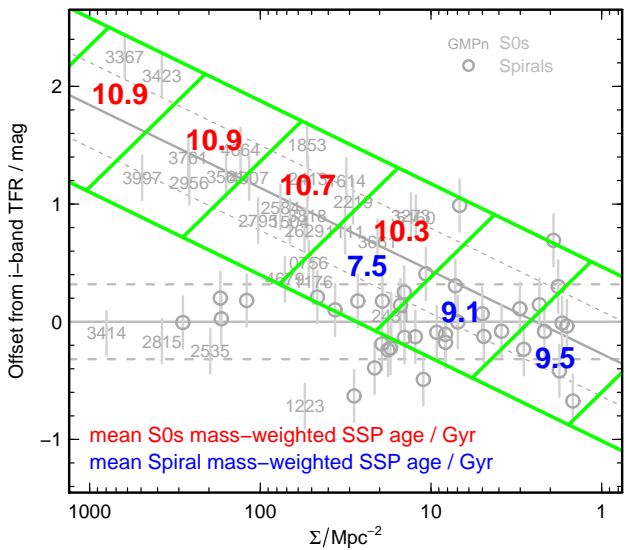}
\caption{The mean mass-weighted simple stellar population (SSP) age (red/upper=S0s, blue/lower=spirals) binned along the correlation between projected local density $\Sigma$ and offset from the $i$-band TFR (the best fit line shown in the right panel of Figure \ref{fig:tf_i_offset}). The points representing S0 and spiral galaxies are displayed in grey to indicate the trend. The S0s show a marginal trend towards older ages with increasing $\Sigma$, while the spirals become younger, possibly indicating a central starburst.}
\label{fig:age}
\end{figure}

In Figure \ref{fig:tf_i_offset} we reported the trend between TFR offset and local projected density for S0s. In bins long this correlation, we now determine the mean ages of the spiral and S0 populations (Figure \ref{fig:age}). The S0 population shows a very marginal trend of increasing age (10.3 to 10.9 Gyr) with increasing local density, in agreement with \citet{smi12-3167}, although the age uncertainty in each bin is $\sim$1 Gyr. Interestingly, the spiral sample exhibits the opposite trend: decreasing age (9.5 to 7.5 Gyr) with increasing local density. This may be indicative of a central starburst in the early stages of the transformation, before quenching in the disc transforms the observed morphology from spiral to S0.

\section{Conclusions}
\label{sec:conc}

In order to characterise early-type disc galaxies in the rich cluster environment we have undertaken deep, long-slit spectroscopy with GMOS of 15 Coma cluster S0s. From absorption line measurements we have determined kinematic properties along the major axis to several times the disc scale length, and hence derived rotation curves. We have supplemented our kinematics measurements with literature data from \citet[][M00]{meh00-449} and \citet[][H03]{hin03-2622}, yielding a combined sample of 29 Coma cluster S0s. Using SDSS and 2MASS photometry we have investigated the Tully-Fisher relation for cluster S0s.

We confirm the existence of a Tully--Fisher offset for S0 galaxies, calculating that, at fixed rotational velocity, S0s are fainter than spirals by an average $1.06\pm0.18$, $0.85\pm0.19$ and $0.86\pm0.18$ mag in the $g$, $i$ and $K_{\rm s}$ bands respectively. The TFR offsets are consistent with a simple star formation model in which S0s initially have similar SFHs to the spirals, but are abruptly quenched at some intermediate redshift. The TFR offset can be interpreted as a tracer of the time since the cessation of star formation, and exhibits a strong correlation ($>6\,\sigma$) with the residual from the optical CMR. Typically, S0s which are fainter than average for their rotational velocity are also redder than average for their luminosity.

The S0s in our study span a wide range of local densities within the Coma cluster, which has allowed environmental trends to be investigated. We find a correlation between the TFR offset and environment, in the sense that S0s located in regions of lower local density are closer to the spiral Tully--Fisher relation. Since current cluster-centric radius is related to time since accretion into the cluster (or its progenitors), the correlation of TFR offset with the radius suggests that the transformation of spirals into S0s is associated with cluster infall. We also observe a decrease in the mean stellar population age of spirals with increasing local density, which may indicate that immediately prior to quenching in the disc, the transformative process includes a burst of star formation in the bulge.

In future papers, we will present absorption line index profiles for the 15 GMOS S0s. The long-slit observations were specifically designed to enable stellar population analysis at radii beyond several disc scale-lengths. Examination of such gradients \citep[e.g.][]{raw10-852} from both bulge and disc components, in conjunction with GALEX ultraviolet colours tracing recent star formation \citep{raw08-2097,smi12-2982}, which will allow further constraints to be put on the S0 transformation process.

\section*{Acknowledgments}

The authors are grateful to Richard Bower for useful discussion while preparing this manuscript.

TDR is supported by a Research Fellowship at the European Space Agency (ESA). JRL and RJS are supported by STFC Rolling Grant PP/C501568/1 ``Extragalactic Astronomy and Cosmology at Durham 2008--2013". JTCGH is supported by an STFC studentship.

Based on observations obtained at the Gemini Observatory, which is operated by the Association of Universities for Research in Astronomy, Inc., under a cooperative agreement with the NSF on behalf of the Gemini partnership: the National Science Foundation (United States), the National Research Council (Canada), CONICYT (Chile), the Australian Research Council (Australia), MinistŽrio da Cincia, Tecnologia e Inova‹o (Brazil) and Ministerio de Ciencia, Tecnolog'a e Innovaci—n Productiva (Argentina). At the time of observation, the Gemini partnership also included the Science and Technology Facilities Council (United Kingdom), who supported TDR as a visiting astronomer at Gemini North in March 2009.

This publication also makes use of data products from the Two Micron All Sky Survey, which is a joint project of the University of Massachusetts and the Infrared Processing and Analysis Center/California Institute of Technology, funded by NASA and the National Science Foundation. Funding for SDSS-III has been provided by the Alfred P. Sloan Foundation, the Participating Institutions, the National Science Foundation, and the U.S. Department of Energy Office of Science. The SDSS-III web site is http://www.sdss3.org/. Also based on observations obtained at the Canada-France-Hawaii Telescope (CFHT) which is operated by the National Research Council of Canada, the Institut National des Sciences de l'Univers of the Centre National de la Recherche Scientifique of France, and the University of Hawaii. The Millennium Simulation databases used in this paper and the web application providing online access to them were constructed as part of the activities of the German Astrophysical Virtual Observatory. This research has made use of the NASA/IPAC Extragalactic Database (NED) which is operated by the Jet Propulsion Laboratory, California Institute of Technology, under contract with the National Aeronautics and Space Administration.


\appendix

\section{Photometric and kinematic profiles for the GMOS S0 galaxies}
\label{app:profiles}

\begin{figure*}
\centering
\includegraphics[height=160mm,angle=270,clip]{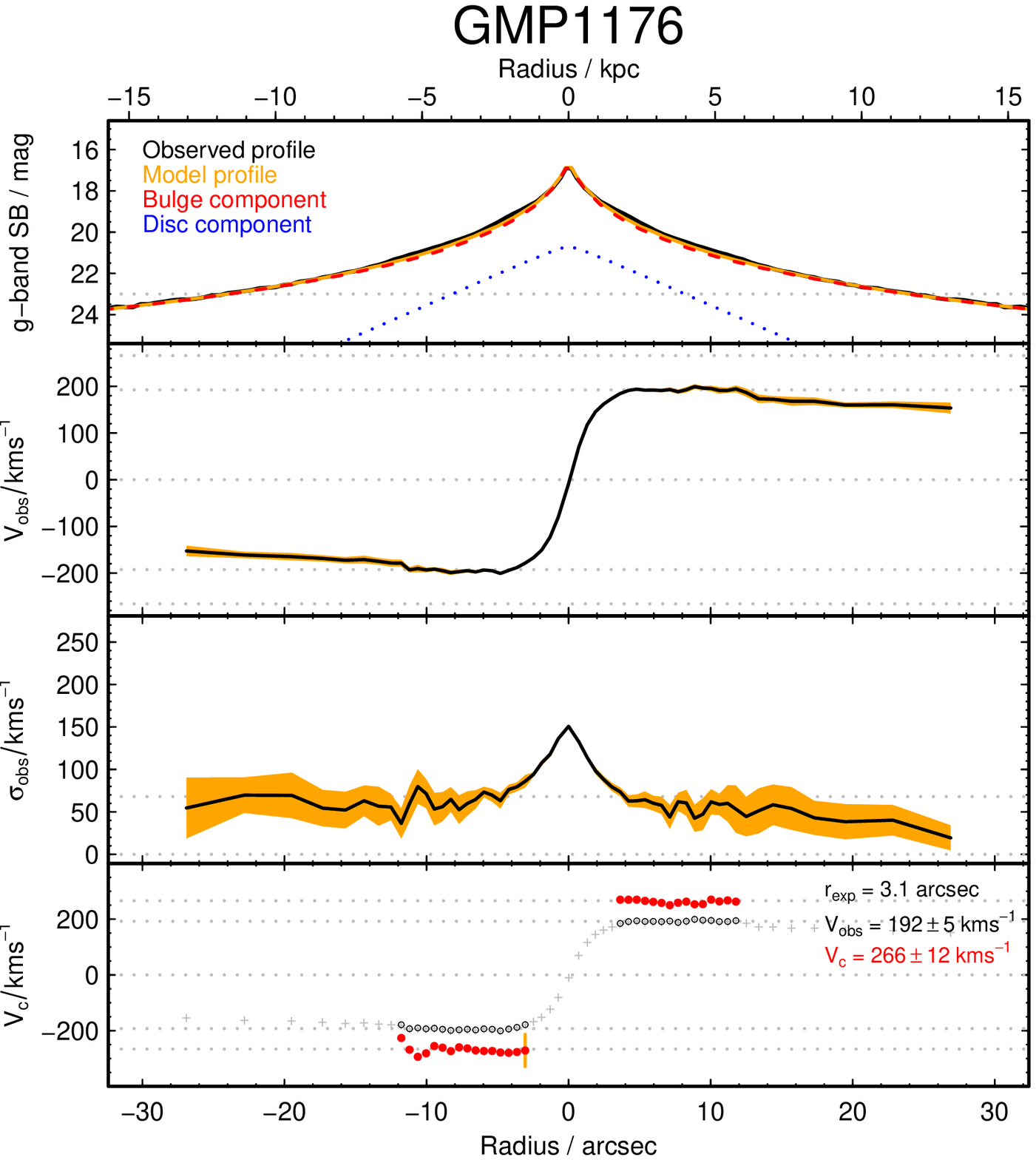}
\caption[Radial profiles for GMP1176]{Photometric and kinematic profiles for GMP1176. All x-axes are delineated in arcsec, apart from the very top axis which indicates the corresponding physical scale in kpc (1 arcsec $=$ 0.48 kpc at the distance of Coma). {\bf Upper Panel:} Observed $g$-band surface brightness profile (black solid line). {\sc Galfit} two-component decomposition model surface brightness also shown for the Sersic bulge (red dashes), exponential disc (blue dots) and total profile (orange solid). {\bf Central panels:} Observed rotational velocity ($V_{\rm obs}$) and velocity dispersion ($\sigma_{\rm obs}$). 1$\,\sigma$ errors are indicated by orange shading. {\bf Lowest panel:} Open black circles show observed velocities, and filled red circles show the circular rotation velocity in the flat turnover region of the radial profile ($V_c$, see text for details). The maximum values of V$_{\rm obs}$ and $V_c$ are given in the top right of the panel, along with the exponential disc scale length.}
\label{fig:kin_gmp1176}
\end{figure*}

\begin{figure*}
\centering
\includegraphics[height=160mm,angle=270,clip]{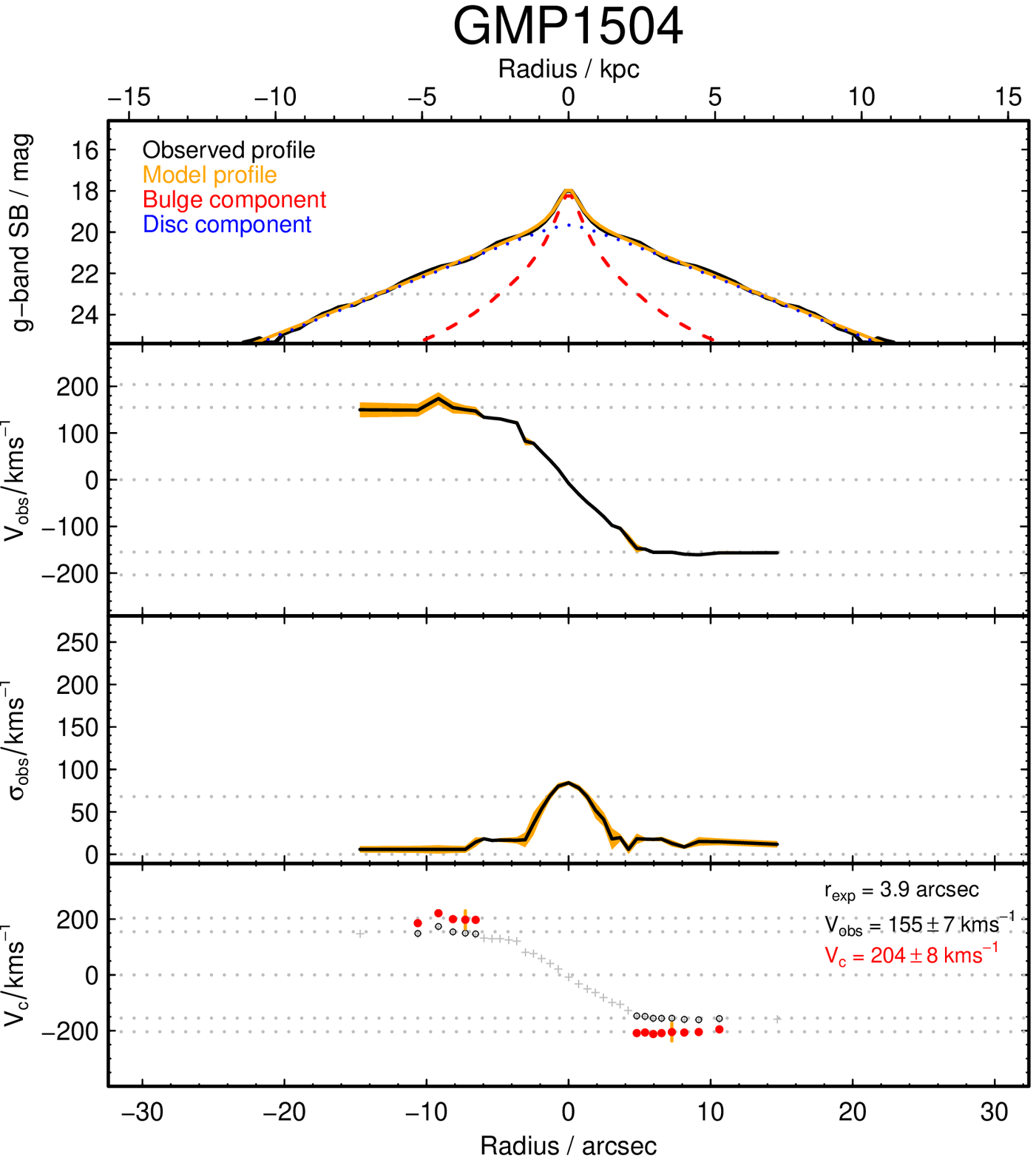}
\caption[Radial profiles for GMP1504]{Photometric and kinematic profiles for GMP1504. Layout as in Figure \ref{fig:kin_gmp1176}.}
\label{fig:kin_gmp1504}
\end{figure*}

\begin{figure*}
\centering
\includegraphics[height=160mm,angle=270,clip]{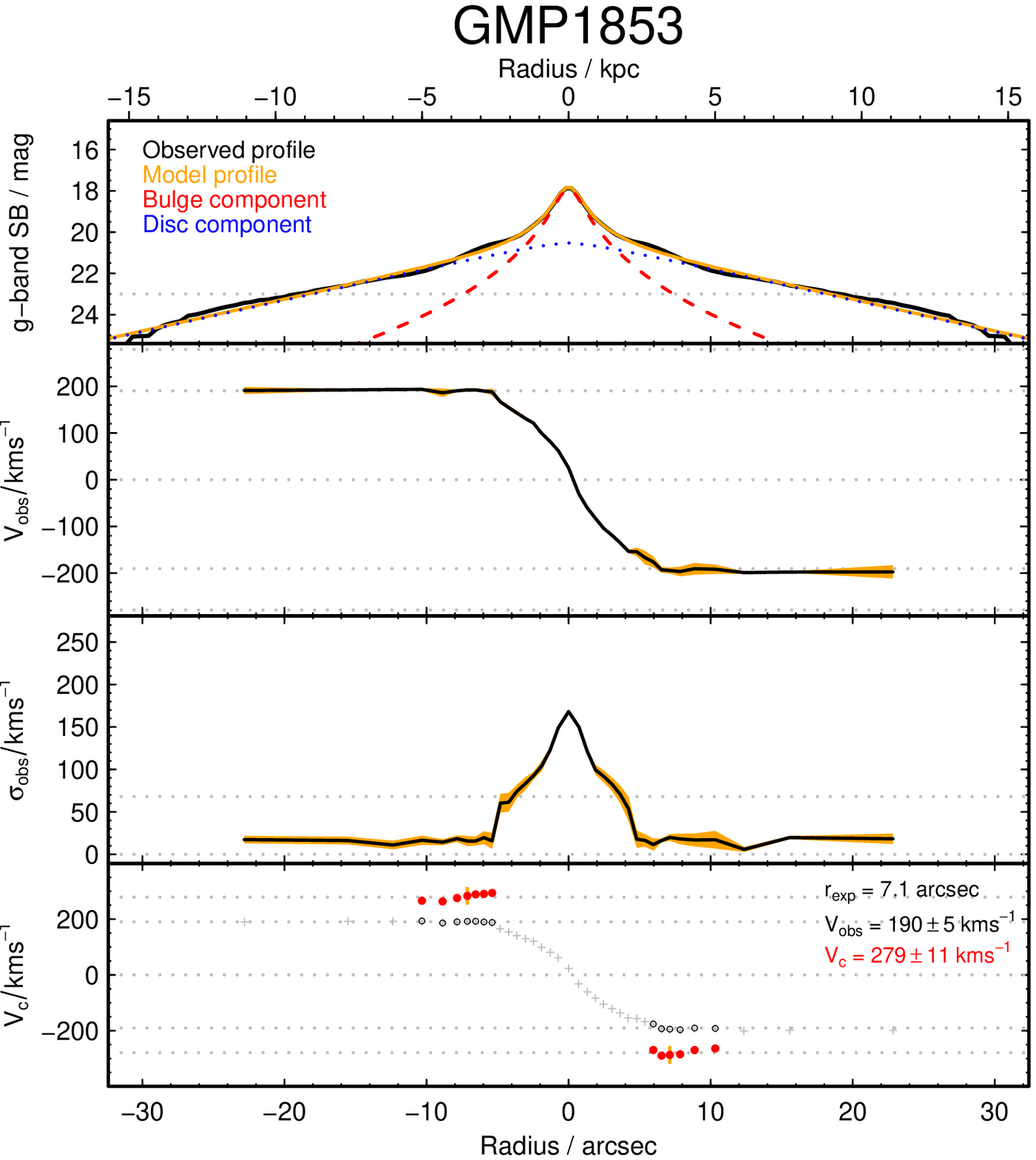}
\caption[Radial profiles for GMP1853]{Photometric and kinematic profiles for GMP1853. Layout as in Figure \ref{fig:kin_gmp1176}.}
\label{fig:kin_gmp1853}
\end{figure*}

\pagebreak

\begin{figure*}
\centering
\includegraphics[height=160mm,angle=270,clip]{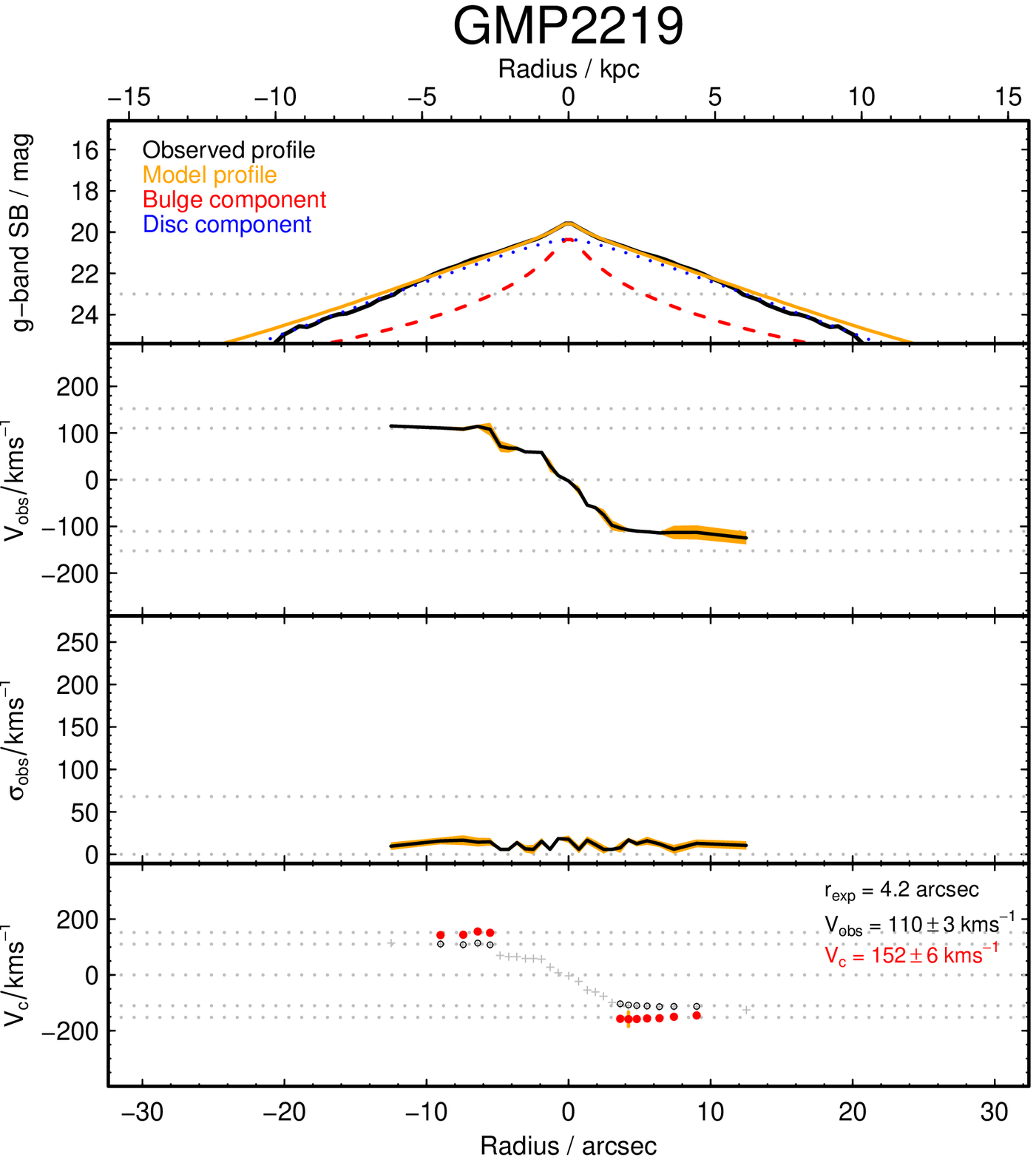}
\caption[Radial profiles for GMP2219]{Photometric and kinematic profiles for GMP2219. Layout as in Figure \ref{fig:kin_gmp1176}.}
\label{fig:kin_gmp2219}
\end{figure*}

\begin{figure*}
\centering
\includegraphics[height=160mm,angle=270,clip]{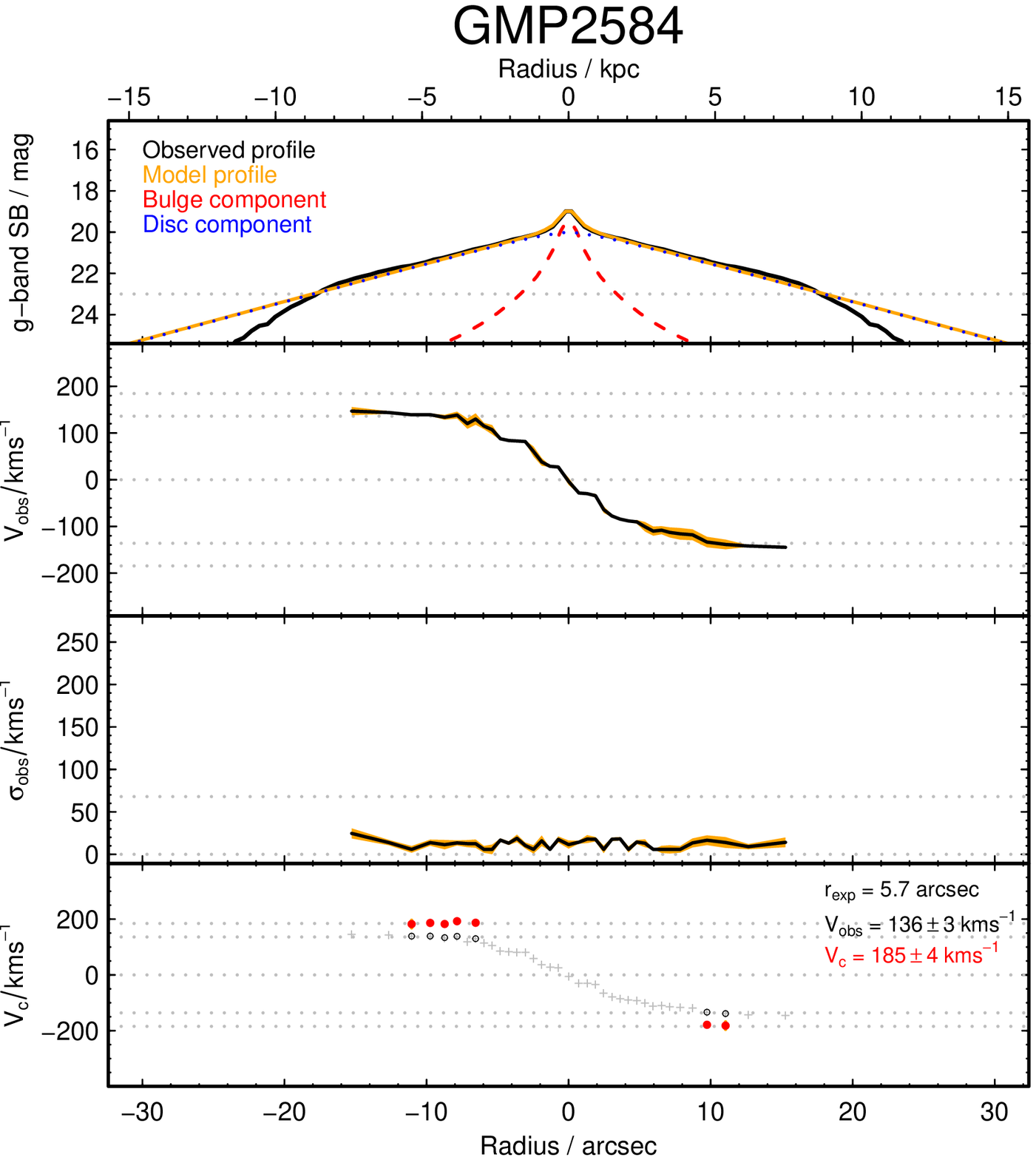}
\caption[Radial profiles for GMP2584]{Photometric and kinematic profiles for GMP2584. Layout as in Figure \ref{fig:kin_gmp1176}.}
\label{fig:kin_gmp2584}
\end{figure*}

\begin{figure*}
\centering
\includegraphics[height=160mm,angle=270,clip]{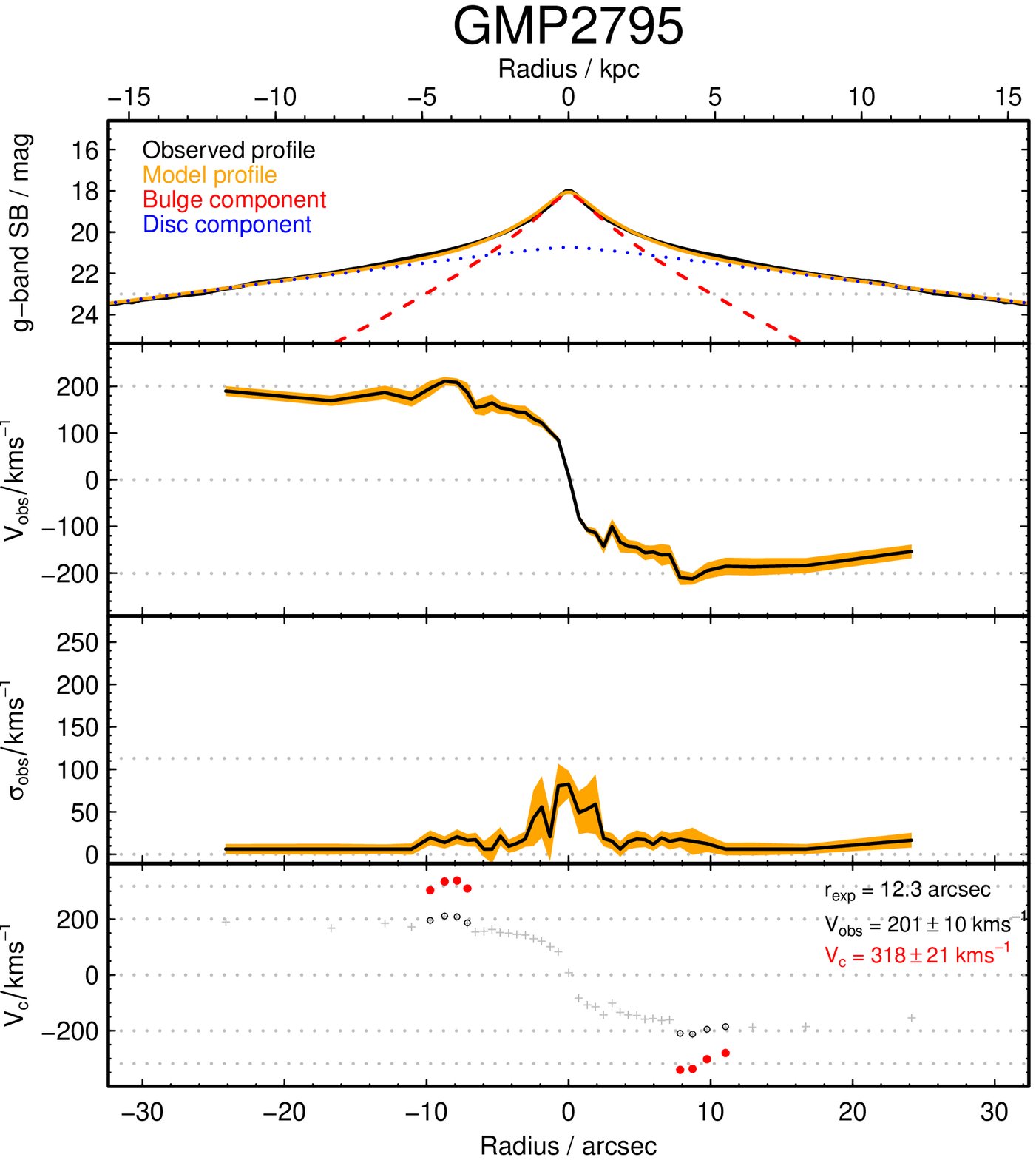}
\caption[Radial profiles for GMP2795]{Photometric and kinematic profiles for GMP2795. Layout as in Figure \ref{fig:kin_gmp1176}.}
\label{fig:kin_gmp2795}
\end{figure*}

\pagebreak

\begin{figure*}
\centering
\includegraphics[height=160mm,angle=270,clip]{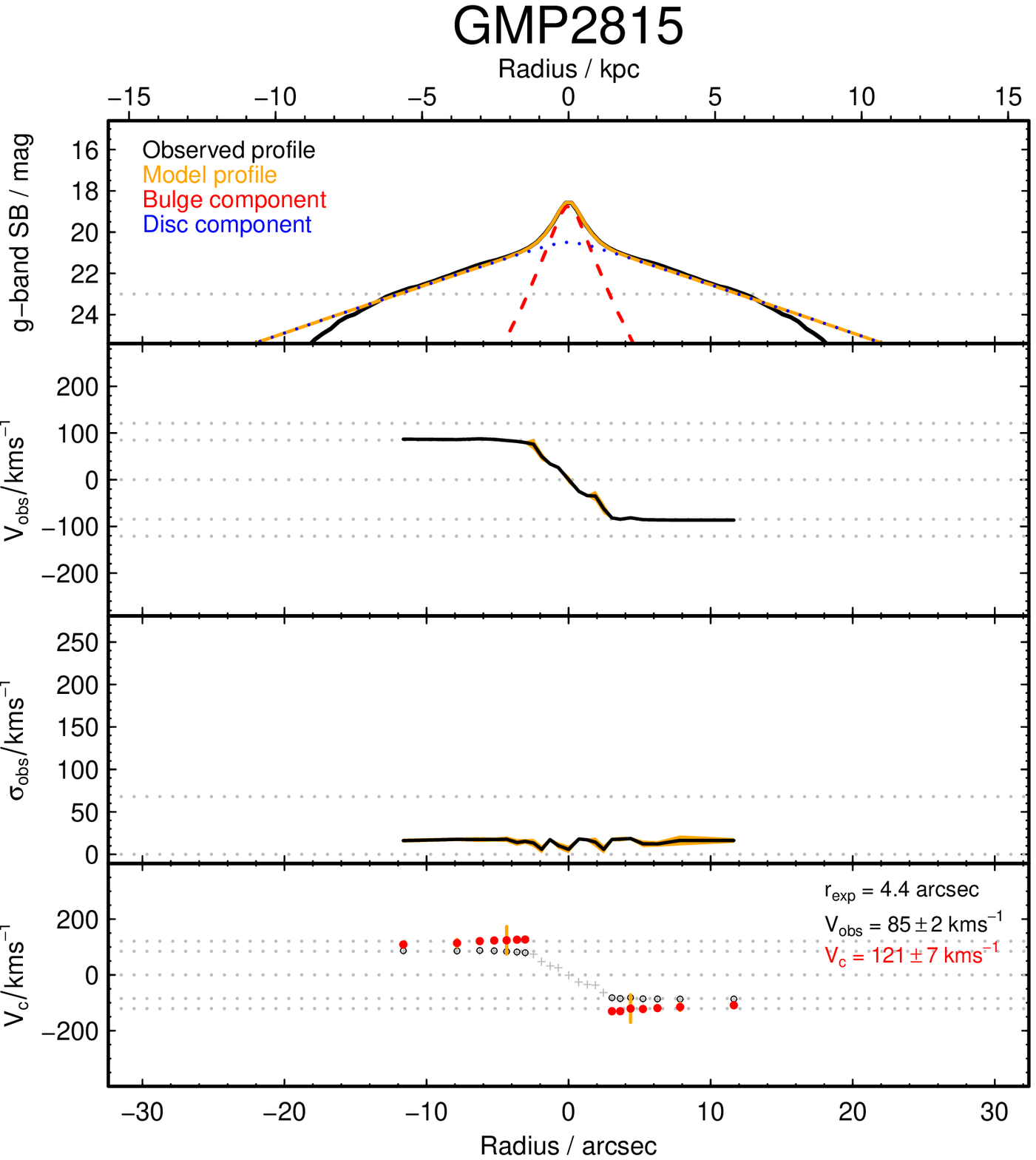}
\caption[Radial profiles for GMP2815]{Photometric and kinematic profiles for GMP2815. Layout as in Figure \ref{fig:kin_gmp1176}.}
\label{fig:kin_gmp2815}
\end{figure*}

\begin{figure*}
\centering
\includegraphics[height=160mm,angle=270,clip]{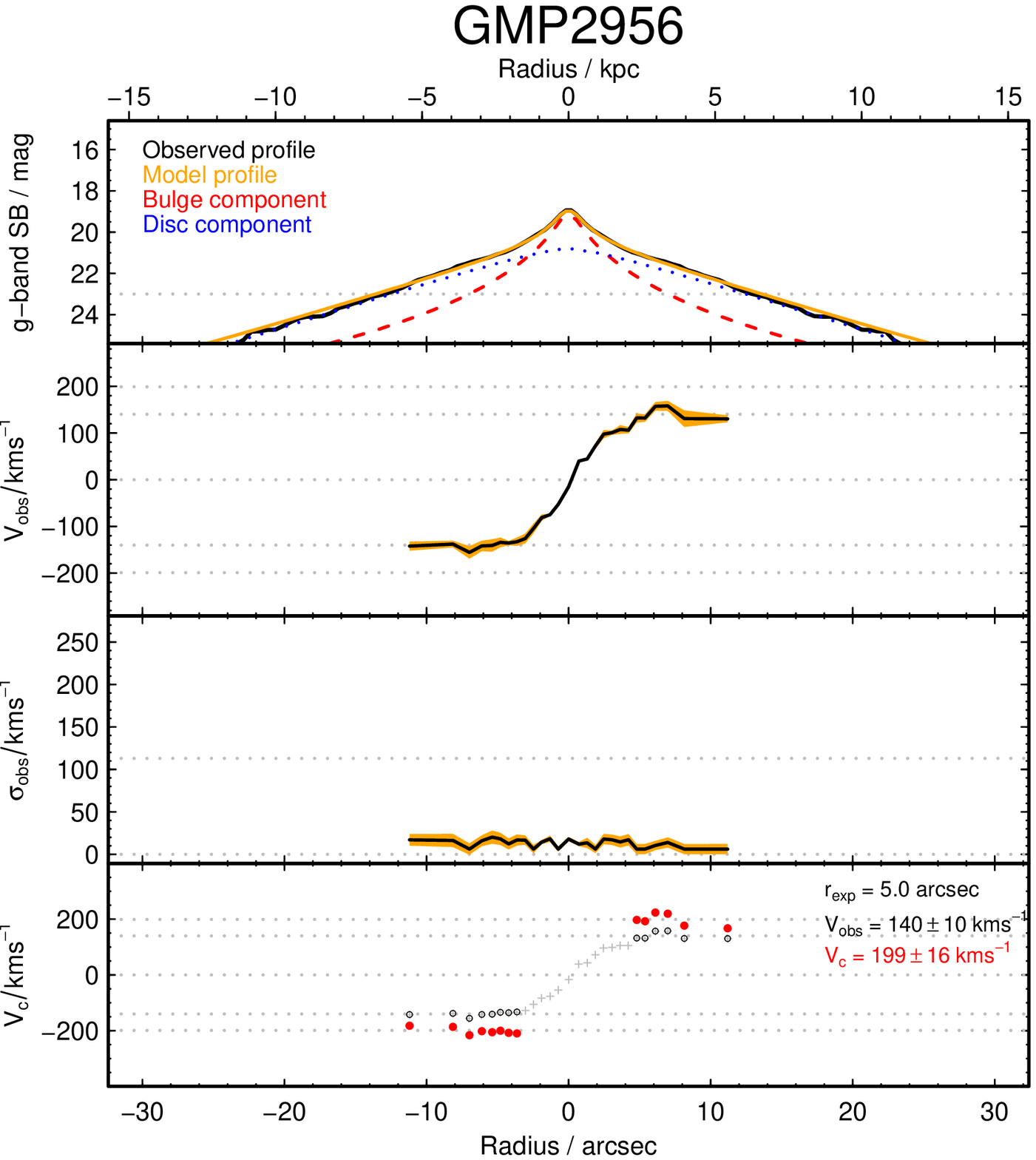}
\caption[Radial profiles for GMP2956]{Photometric and kinematic profiles for GMP2956. Layout as in Figure \ref{fig:kin_gmp1176}.}
\label{fig:kin_gmp2956}
\end{figure*}

\begin{figure*}
\centering
\includegraphics[height=160mm,angle=270,clip]{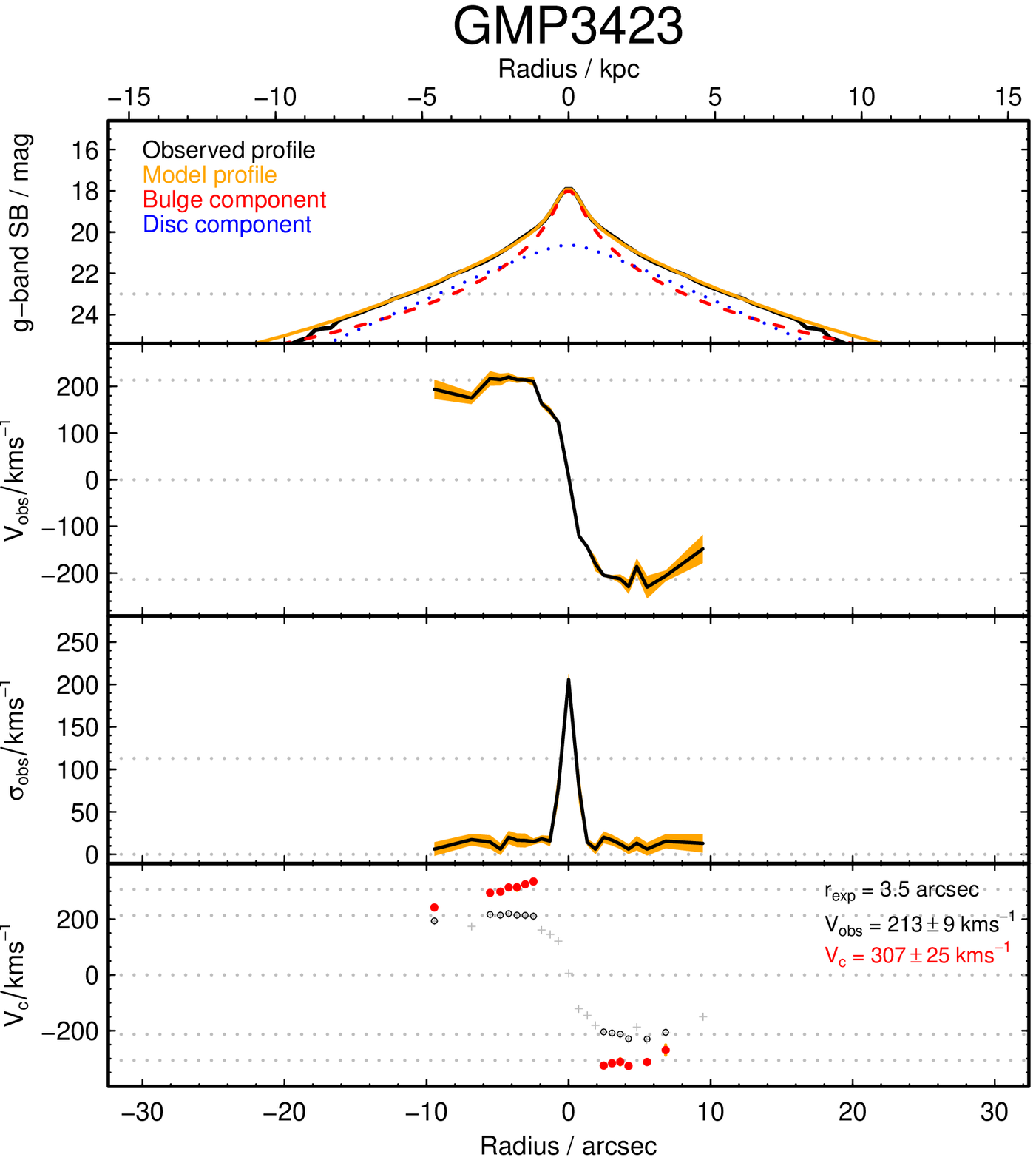}
\caption[Radial profiles for GMP3423]{Photometric and kinematic profiles for GMP3423. Layout as in Figure \ref{fig:kin_gmp1176}.}
\label{fig:kin_gmp3423}
\end{figure*}

\pagebreak

\begin{figure*}
\centering
\includegraphics[height=160mm,angle=270,clip]{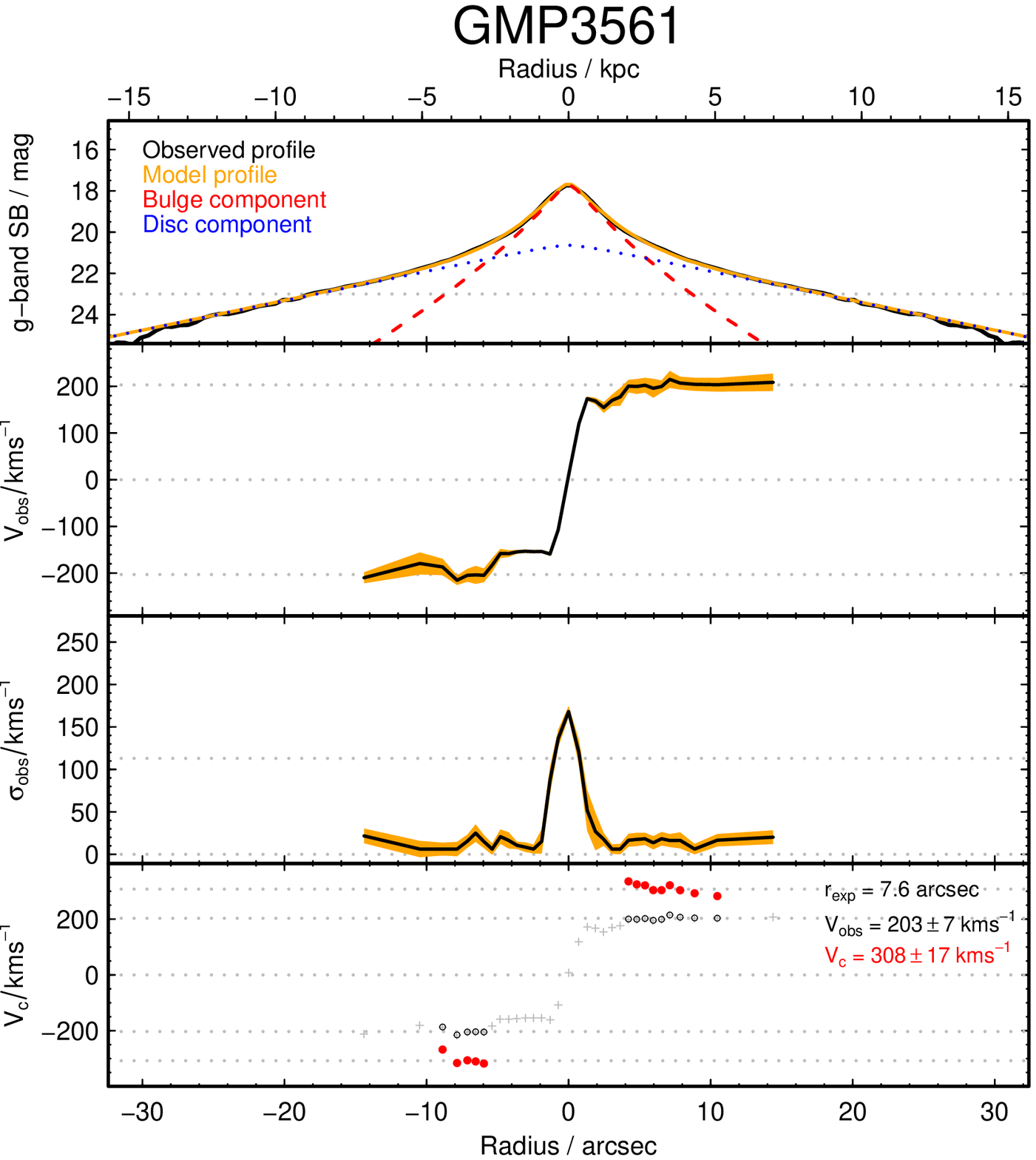}
\caption[Radial profiles for GMP3561]{Photometric and kinematic profiles for GMP3561. Layout as in Figure \ref{fig:kin_gmp1176}.}
\label{fig:kin_gmp3561}
\end{figure*}

\begin{figure*}
\centering
\includegraphics[height=160mm,angle=270,clip]{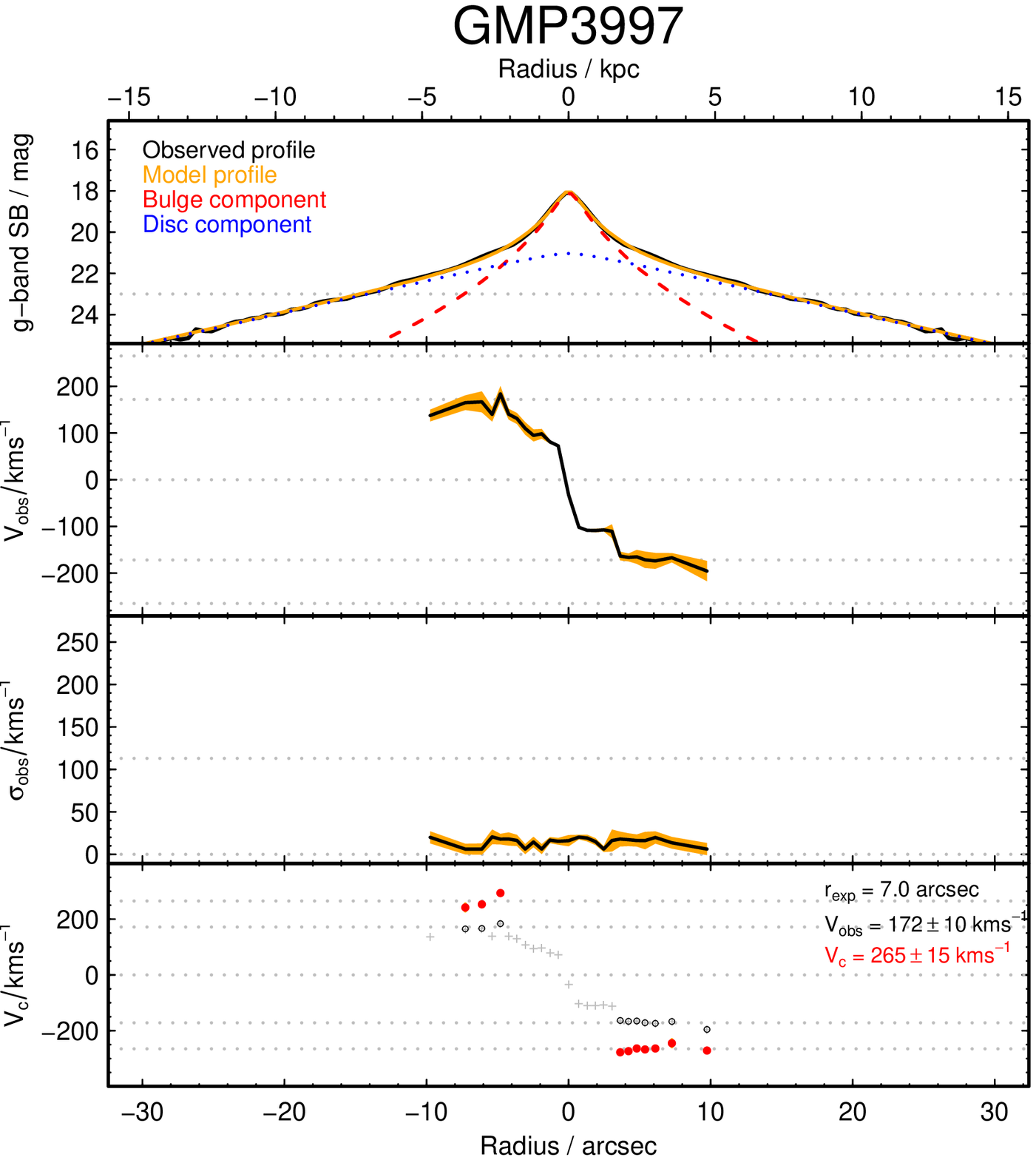}
\caption[Radial profiles for GMP3997]{Photometric and kinematic profiles for GMP3997. Layout as in Figure \ref{fig:kin_gmp1176}.}
\label{fig:kin_gmp3997}
\end{figure*}

\begin{figure*}
\centering
\includegraphics[height=160mm,angle=270,clip]{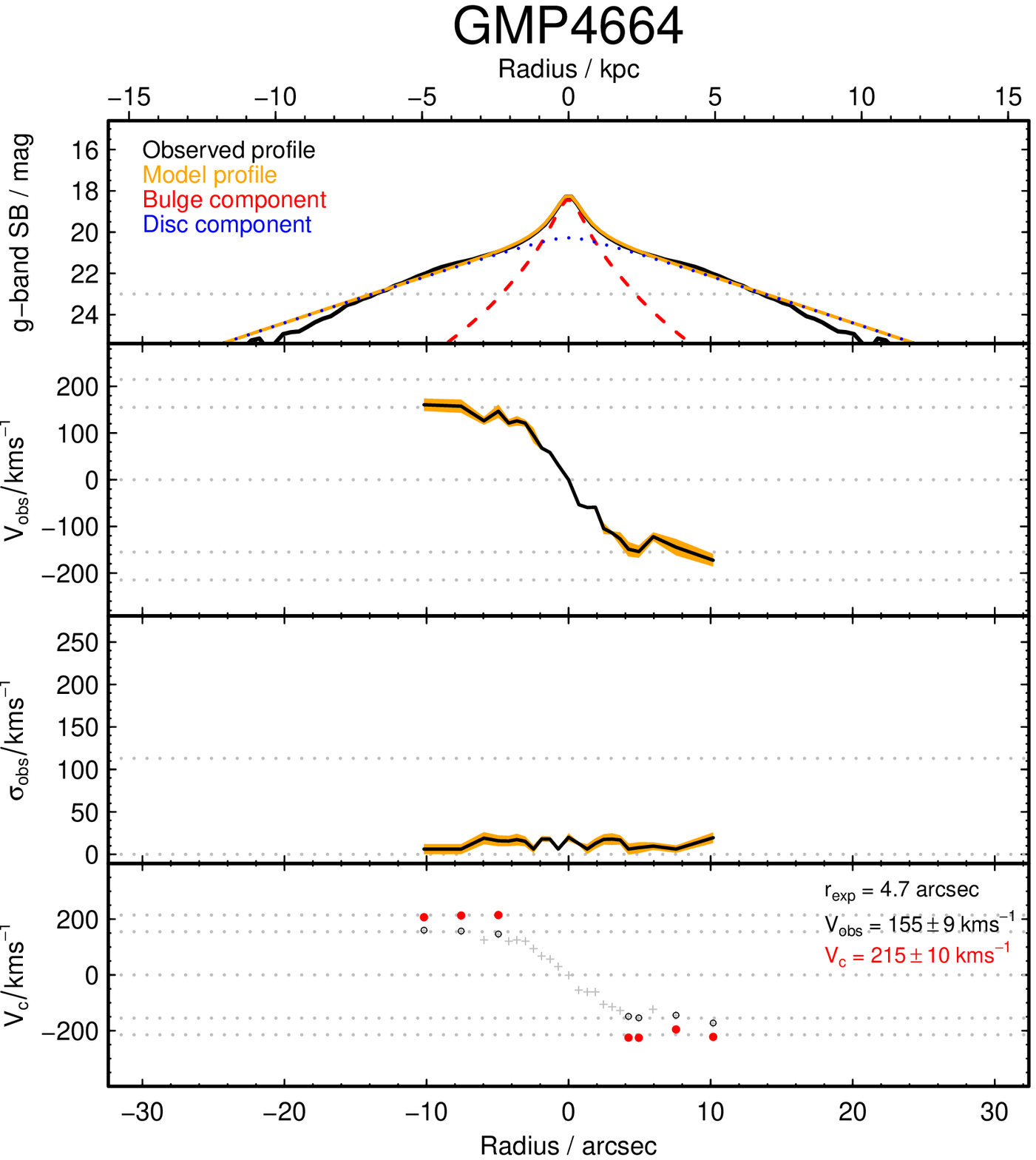}
\caption[Radial profiles for GMP4664]{Photometric and kinematic profiles for GMP4664. Layout as in Figure \ref{fig:kin_gmp1176}.}
\label{fig:kin_gmp4664}
\end{figure*}

\pagebreak

\begin{figure*}
\centering
\includegraphics[height=160mm,angle=270,clip]{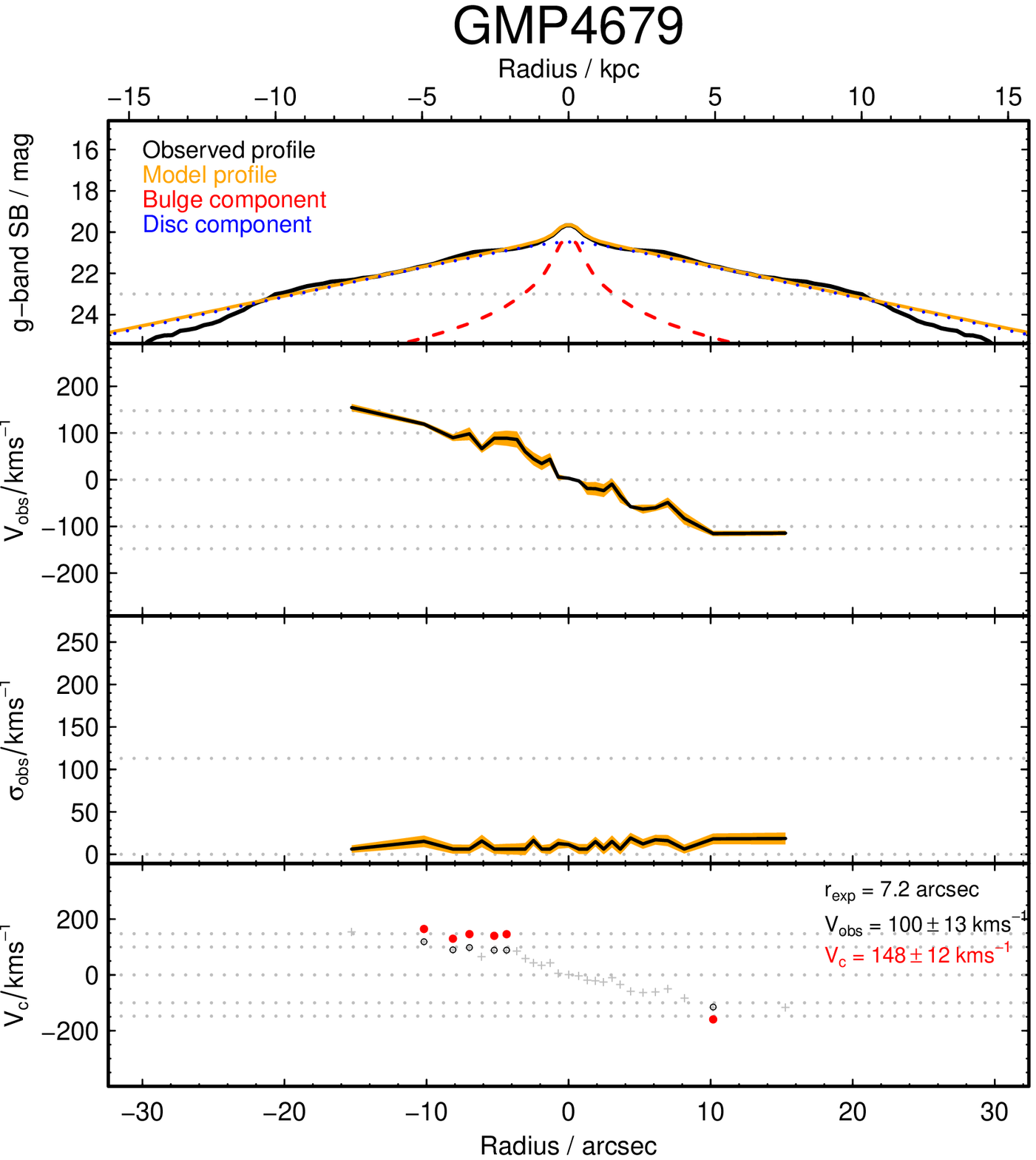}
\caption[Radial profiles for GMP4679]{Photometric and kinematic profiles for GMP4679. Layout as in Figure \ref{fig:kin_gmp1176}.}
\label{fig:kin_gmp4679}
\end{figure*}

\begin{figure*}
\centering
\includegraphics[height=160mm,angle=270,clip]{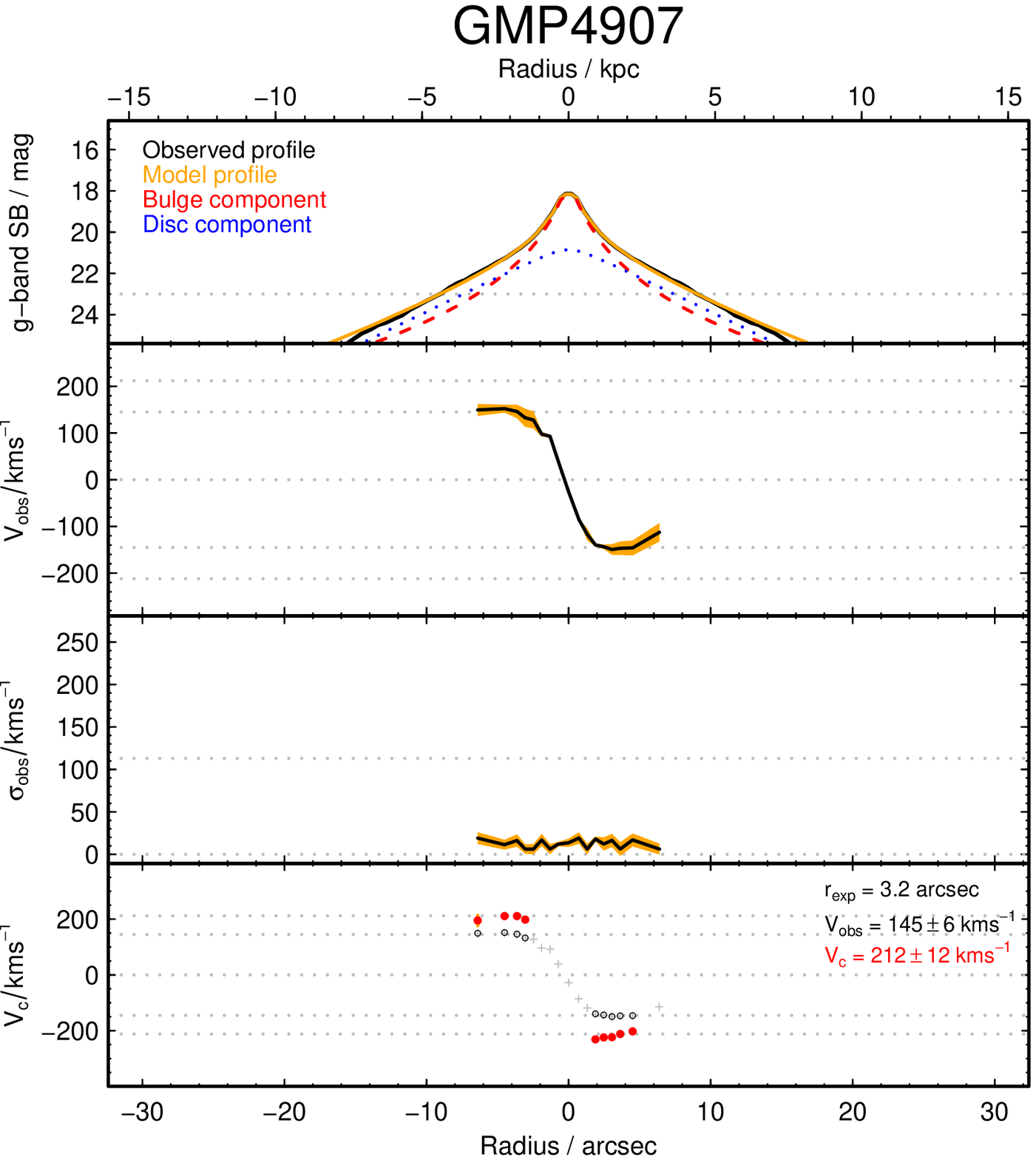}
\caption[Radial profiles for GMP4907]{Photometric and kinematic profiles for GMP4907. Layout as in Figure \ref{fig:kin_gmp1176}.}
\label{fig:kin_gmp4907}
\end{figure*}

\begin{figure*}
\centering
\includegraphics[height=160mm,angle=270,clip]{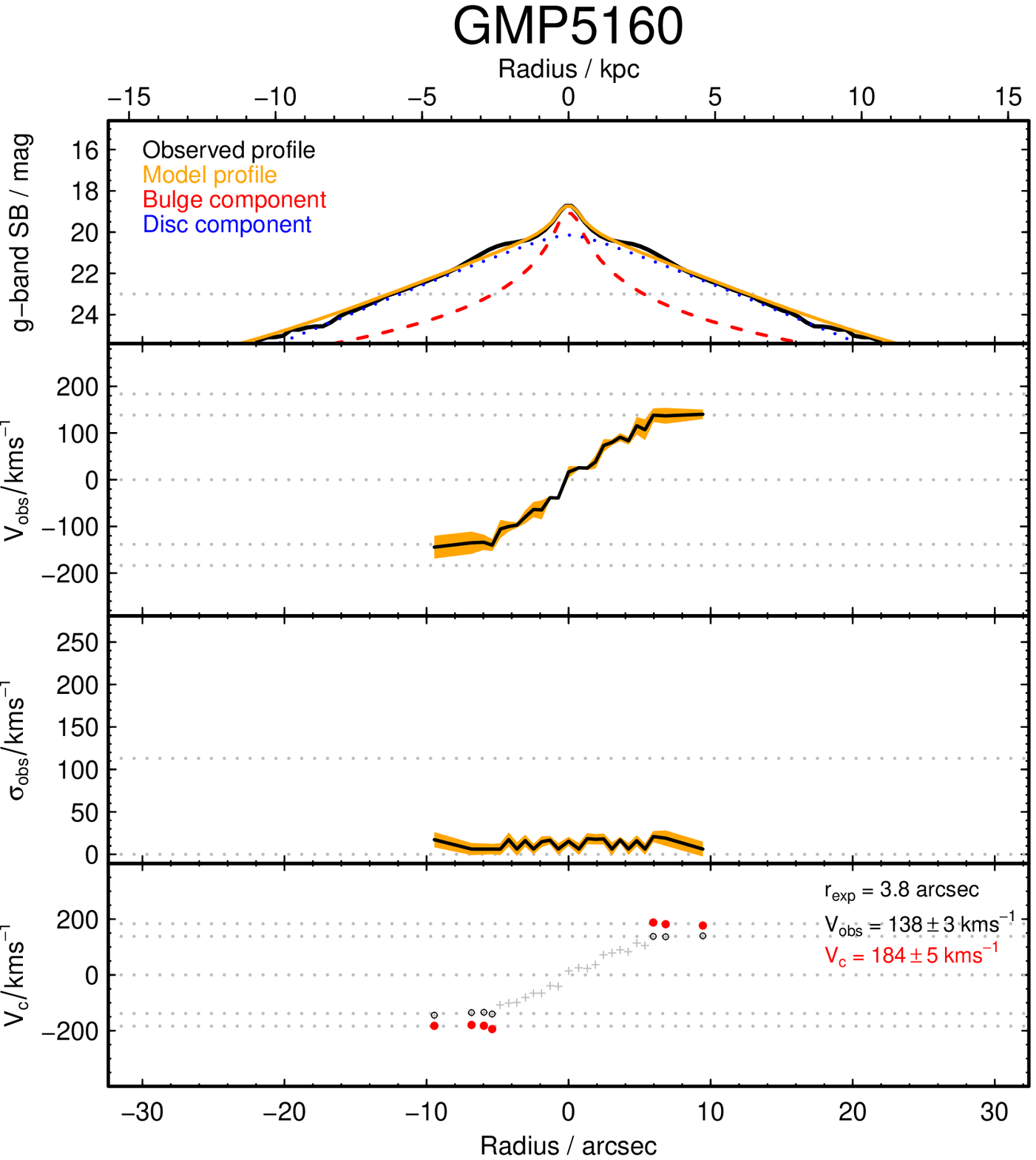}
\caption[Radial profiles for GMP5160]{Photometric and kinematic profiles for GMP5160. Layout as in Figure \ref{fig:kin_gmp1176}.}
\label{fig:kin_gmp5160}
\end{figure*}

\pagebreak

\section{Coma cluster spiral galaxy sample}
\label{app:sp}

\begin{table*}
\centering
\caption{Observed quantities for the 38* Coma spiral galaxies sample introduced in Section \ref{sec:spirals}. Position and optical photometric from SDSS (AB mags). $K_{\rm s}$ band from 2MASS (Vega mags). CMB frame velocity ($cz_{\rm CMB}$ km s$^{-1}$), line width and inclination from \citet{spr07-599}.} 
\label{tab:sps} 
\begin{tabular}{@{}crccrrrccc}
\\ 
\hline
\multicolumn{1}{c}{GMP ID} & \multicolumn{1}{c}{U/AGC} & RA & Dec & \multicolumn{1}{c}{$m_g$} & \multicolumn{1}{c}{$m_i$} & \multicolumn{1}{c}{$m_{K_s}$} &\multicolumn{1}{c}{$cz_{\rm CMB}$} & \multicolumn{1}{c}{logW} & \multicolumn{1}{c}{$i$} \\
& \multicolumn{1}{c}{\#} & (J2000) & (J2000) & \multicolumn{1}{c}{mag} & \multicolumn{1}{c}{mag} & \multicolumn{1}{c}{mag} & \multicolumn{1}{c}{(km s$^{-1}$)} & \multicolumn{1}{c}{} & \multicolumn{1}{c}{deg} \\
\hline
-- & 7845 & 190.31816 & 27.85315 & 15.128 $\pm$ 0.003 & 14.391 $\pm$ 0.003 & 12.40 $\pm$ 0.09 & 8014 & 2.431 & 78.0 \\ 
-- & 7877 & 190.69666 & 27.27193 & 15.689 $\pm$ 0.004 & 14.577 $\pm$ 0.003 & 11.97 $\pm$ 0.07 & 6186 & 2.501 & 84.0 \\ 
-- & 7890 & 190.77231 & 27.71406 & 14.750 $\pm$ 0.002 & 14.095 $\pm$ 0.003 & 12.02 $\pm$ 0.08 & 7802 & 2.493 & 46.8 \\ 
-- & 7955 & 191.79675 & 26.71083 & 15.113 $\pm$ 0.003 & 13.901 $\pm$ 0.003 & 11.27 $\pm$ 0.07 & 7034 & 2.556 & 81.0 \\ 
-- & 221022 & 191.86821 & 27.45778 & 14.772 $\pm$ 0.002 & 13.620 $\pm$ 0.002 & 10.74 $\pm$ 0.04 & 6885 & 2.542 & 56.7 \\ 
-- & 221033 & 192.17530 & 26.41730 & 15.108 $\pm$ 0.003 & 14.013 $\pm$ 0.003 & 11.31 $\pm$ 0.05 & 7137 & 2.501 & 76.2 \\ 
-- & 8004 & 192.90820 & 31.35276 & 14.773 $\pm$ 0.004 & 14.089 $\pm$ 0.003 & 12.51 $\pm$ 0.06 & 6445 & 2.479 & 68.1 \\ 
-- & 8013 & 193.15121 & 26.74988 & 14.918 $\pm$ 0.003 & 14.147 $\pm$ 0.003 & 12.16 $\pm$ 0.10 & 8157 & 2.548 & 78.0 \\ 
-- & 8025 & 193.51033 & 29.60361 & 14.391 $\pm$ 0.002 & 13.111 $\pm$ 0.002 & 10.20 $\pm$ 0.03 & 6583 & 2.694 & 86.3 \\ 
gmp5422 & 221130 & 194.11907 & 27.29127 & 15.261 $\pm$ 0.003 & 14.366 $\pm$ 0.003 & 12.42 $\pm$ 0.13 & 7796 & 2.334 & 46.9 \\ 
gmp5234 & 221147 & 194.20689 & 27.09391 & 15.400 $\pm$ 0.003 & 14.165 $\pm$ 0.002 & 11.58 $\pm$ 0.05 & 7117 & 2.515 & 69.7 \\ 
gmp5197 & 221149 & 194.21089 & 28.92983 & 15.082 $\pm$ 0.003 & 14.058 $\pm$ 0.002 & 11.76 $\pm$ 0.05 & 8297 & 2.481 & 70.3 \\ 
gmp5006 & 8069 & 194.29749 & 29.04508 & 14.432 $\pm$ 0.002 & 13.273 $\pm$ 0.002 & 10.75 $\pm$ 0.03 & 7927 & 2.588 & 67.1 \\ 
-- & 221174 & 194.38002 & 26.51215 & 15.096 $\pm$ 0.003 & 14.078 $\pm$ 0.002 & 11.51 $\pm$ 0.05 & 7533 & 2.524 & 72.7 \\ 
gmp4437 & 221206 & 194.53844 & 28.70856 & 15.159 $\pm$ 0.004 & 14.077 $\pm$ 0.004 & 11.62 $\pm$ 0.06 & 7881 & 2.418 & 73.3 \\ 
gmp3896 & 8096 & 194.73308 & 27.83337 & 14.755 $\pm$ 0.003 & 13.842 $\pm$ 0.003 & 11.01 $\pm$ 0.04 & 7803 & 2.562 & 87.1 \\ 
gmp2987 & 8108 & 195.01470 & 26.89811 & 14.219 $\pm$ 0.002 & 13.089 $\pm$ 0.002 & 10.63 $\pm$ 0.03 & 6173 & 2.614 & 80.1 \\ 
gmp2582 & 221402 & 195.14865 & 27.57422 & 15.801 $\pm$ 0.003 & 14.734 $\pm$ 0.003 & 12.02 $\pm$ 0.06 & 5389 & 2.420 & 74.0 \\ 
gmp2559 & 221406 & 195.15775 & 28.05797 & 15.437 $\pm$ 0.003 & 14.674 $\pm$ 0.003 & 11.84 $\pm$ 0.05 & 7896 & 2.428 & 68.1 \\ 
gmp2544 & 8118 & 195.16479 & 29.01941 & 14.381 $\pm$ 0.002 & 13.326 $\pm$ 0.002 & 10.76 $\pm$ 0.05 & 7551 & 2.591 & 63.7 \\ 
gmp2374 & 8128 & 195.23359 & 27.79085 & 13.610 $\pm$ 0.002 & 12.378 $\pm$ 0.002 & 9.84 $\pm$ 0.04 & 8251 & 2.733 & 34.5 \\ 
gmp1900 & 8140 & 195.43072 & 29.04466 & 14.286 $\pm$ 0.002 & 13.332 $\pm$ 0.002 & 10.72 $\pm$ 0.04 & 7357 & 2.667 & 77.5 \\ 
gmp1657 & 221460 & 195.51750 & 29.25345 & 14.590 $\pm$ 0.002 & 13.344 $\pm$ 0.002 & 10.59 $\pm$ 0.03 & 7573 & 2.679 & 78.5 \\ 
-- & 8161 & 195.87117 & 26.55050 & 14.750 $\pm$ 0.002 & 13.651 $\pm$ 0.002 & 11.13 $\pm$ 0.04 & 6945 & 2.584 & 65.2 \\ 
gmp0455 & 230051 & 196.11060 & 27.30431 & 15.328 $\pm$ 0.003 & 14.933 $\pm$ 0.004 & 13.53 $\pm$ 0.16 & 5766 & 2.344 & 46.2 \\ 
-- & 8194 & 196.57206 & 29.06318 & 13.942 $\pm$ 0.002 & 12.746 $\pm$ 0.002 & 10.25 $\pm$ 0.03 & 7309 & 2.670 & 61.0 \\ 
-- & 8195 & 196.59462 & 29.65753 & 16.496 $\pm$ 0.007 & 15.756 $\pm$ 0.010 & 12.66 $\pm$ 0.12 & 7296 & 2.391 & 81.2 \\ 
-- & 8209 & 196.92840 & 24.81060 & 14.346 $\pm$ 0.002 & 13.390 $\pm$ 0.002 & 11.41 $\pm$ 0.05 & 6599 & 2.480 & 45.4 \\ 
-- & 8220 & 197.13158 & 24.70076 & 14.537 $\pm$ 0.002 & 13.275 $\pm$ 0.002 & 10.41 $\pm$ 0.04 & 7398 & 2.713 & 82.5 \\ 
-- & 8229 & 197.22579 & 28.18390 & 14.205 $\pm$ 0.002 & 13.170 $\pm$ 0.002 & 10.46 $\pm$ 0.08 & 6248 & 2.590 & 56.7 \\ 
-- & 230117 & 197.23822 & 28.28051 & 15.761 $\pm$ 0.004 & 15.124 $\pm$ 0.005 & 12.90 $\pm$ 0.19 & 6108 & 2.332 & 54.5 \\ 
-- & 8244 & 197.46713 & 28.38244 & 15.505 $\pm$ 0.004 & 14.730 $\pm$ 0.004 & 12.71 $\pm$ 0.13 & 7358 & 2.413 & 66.8 \\ 
-- & 230139 & 197.69853 & 29.70990 & 14.912 $\pm$ 0.003 & 13.954 $\pm$ 0.003 & 11.82 $\pm$ 0.12 & 6619 & 2.461 & 48.6 \\ 
-- & 8294 & 198.24282 & 31.25862 & 14.894 $\pm$ 0.003 & 14.262 $\pm$ 0.004 & \multicolumn{1}{c}{--} & 6320 & 2.386 & 53.0 \\ 
-- & 8300 & 198.36229 & 27.80237 & 13.368 $\pm$ 0.002 & 12.079 $\pm$ 0.002 & 9.71 $\pm$ 0.03 & 6673 & 2.784 & 57.7 \\ 
-- & 8317 & 198.59835 & 30.48399 & 14.790 $\pm$ 0.003 & 14.021 $\pm$ 0.003 & 11.94 $\pm$ 0.09 & 6287 & 2.473 & 66.4 \\ 
-- & 8328 & 198.85662 & 27.30323 & 16.044 $\pm$ 0.005 & 15.333 $\pm$ 0.006 & \multicolumn{1}{c}{--} & 6740 & 2.405 & 73.8 \\ 
-- & 8366 & 199.78432 & 28.50692 & 13.720 $\pm$ 0.002 & 12.710 $\pm$ 0.002 & 10.07 $\pm$ 0.06 & 6900 & 2.729 & 74.0 \\
\hline
\end{tabular}

* Two further galaxies are Coma cluster members in \citet{spr07-599}, but removed from our spiral sample: 1) AGC8076, at a cluster centric radius $>2$ Mpc and by far the lowest $cz=2788$ km s$^{-1}$, is almost certainly a foreground galaxy; 2) GMP1582, at a radius of $\sim$1 Mpc and $cz=9214$ km s$^{-1}$, is likely to be in the background. GMP1582 is also $\sim0.5$ mag fainter than any other source in our sample, and exhibits a clumpy, irregular morphology, causing large uncertainties in the photometry.
\end{table*} 

\begin{table*}
\centering
\caption{Tully--Fisher relation parameters for the Coma spiral galaxies sample introduced in Section \ref{sec:spirals}. Columns as in Table \ref{tab:tfr}. $\delta M_X$ includes 0.12 mag uncertainty from the assumption that every galaxy is at the same line-of-sight distance as the cluster mean.} 
\label{tab:tfr_sp} 
\begin{tabular}{@{}crcccrrrc}
\\ 
\hline
\multicolumn{1}{c}{GMP} & \multicolumn{1}{c}{U/AGC} & \multicolumn{1}{c}{$\Delta d_{CC}$} & \multicolumn{1}{c}{$\Sigma$} & \multicolumn{1}{c}{$E(B-V)$} & \multicolumn{1}{c}{$M_g$} & \multicolumn{1}{c}{$M_i$} & \multicolumn{1}{c}{$M_{K_s}$} & \multicolumn{1}{c}{$V_{\rm c}$} \\
\multicolumn{1}{c}{ID} & \multicolumn{1}{c}{ID} & \multicolumn{1}{c}{Mpc} & \multicolumn{1}{c}{Mpc$^{-2}$} & & \multicolumn{1}{c}{mag} & \multicolumn{1}{c}{mag} & \multicolumn{1}{c}{mag} & \multicolumn{1}{c}{km s$^{-1}$} \\
\hline
-- & 7845 & 6.81 & 14 & 0.015 & --20.32 $\pm$ 0.12 & --20.89 $\pm$ 0.12 & --22.84 $\pm$ 0.17 & 135 $\pm$ 1 \\ 
-- & 7877 & 6.39 & 11 & 0.015 & --19.93 $\pm$ 0.12 & --20.82 $\pm$ 0.12 & --23.26 $\pm$ 0.17 & 158 $\pm$ 1 \\ 
-- & 7890 & 6.16 & 3 & 0.017 & --20.49 $\pm$ 0.12 & --21.06 $\pm$ 0.12 & --23.12 $\pm$ 0.17 & 156 $\pm$ 6 \\ 
-- & 7955 & 5.13 & 5 & 0.015 & --20.55 $\pm$ 0.12 & --21.53 $\pm$ 0.12 & --23.96 $\pm$ 0.16 & 180 $\pm$ 2 \\ 
-- & 221022 & 4.63 & 9 & 0.013 & --20.55 $\pm$ 0.12 & --21.59 $\pm$ 0.12 & --24.41 $\pm$ 0.15 & 174 $\pm$ 26 \\ 
-- & 221033 & 4.87 & 2 & 0.011 & --20.39 $\pm$ 0.12 & --21.31 $\pm$ 0.12 & --23.92 $\pm$ 0.16 & 158 $\pm$ 5 \\ 
-- & 8004 & 6.31 & 4 & 0.014 & --20.61 $\pm$ 0.12 & --21.16 $\pm$ 0.12 & --22.67 $\pm$ 0.16 & 151 $\pm$ 1 \\ 
-- & 8013 & 3.36 & 7 & 0.012 & --20.67 $\pm$ 0.12 & --21.24 $\pm$ 0.12 & --23.08 $\pm$ 0.18 & 177 $\pm$ 1 \\ 
-- & 8025 & 3.42 & 2 & 0.019 & --21.59 $\pm$ 0.12 & --22.53 $\pm$ 0.12 & --25.03 $\pm$ 0.15 & 247 $\pm$ 1 \\ 
gmp5422 & 221130 & 1.68 & 28 & 0.010 & --19.89 $\pm$ 0.12 & --20.74 $\pm$ 0.12 & --22.71 $\pm$ 0.20 & 108 $\pm$ 4 \\ 
gmp5234 & 221147 & 1.83 & 46 & 0.010 & --20.02 $\pm$ 0.12 & --21.11 $\pm$ 0.12 & --23.61 $\pm$ 0.16 & 164 $\pm$ 6 \\ 
gmp5197 & 221149 & 1.93 & 8 & 0.010 & --20.32 $\pm$ 0.12 & --21.20 $\pm$ 0.12 & --23.44 $\pm$ 0.16 & 151 $\pm$ 8 \\ 
gmp5006 & 8069 & 2.03 & 17 & 0.012 & --21.03 $\pm$ 0.12 & --22.03 $\pm$ 0.12 & --24.44 $\pm$ 0.15 & 194 $\pm$ 13 \\ 
-- & 221174 & 2.56 & 15 & 0.016 & --20.40 $\pm$ 0.12 & --21.24 $\pm$ 0.12 & --23.70 $\pm$ 0.16 & 167 $\pm$ 7 \\ 
gmp4437 & 221206 & 1.37 & 11 & 0.013 & --20.21 $\pm$ 0.12 & --21.16 $\pm$ 0.12 & --23.59 $\pm$ 0.16 & 131 $\pm$ 1 \\ 
gmp3896 & 8096 & 0.41 & 284 & 0.014 & --20.99 $\pm$ 0.12 & --21.64 $\pm$ 0.12 & --24.23 $\pm$ 0.15 & 182 $\pm$ 8 \\ 
gmp2987 & 8108 & 1.77 & 21 & 0.010 & --21.49 $\pm$ 0.12 & --22.38 $\pm$ 0.12 & --24.60 $\pm$ 0.15 & 206 $\pm$ 7 \\ 
gmp2582 & 221402 & 0.71 & 120 & 0.010 & --19.57 $\pm$ 0.12 & --20.50 $\pm$ 0.12 & --23.19 $\pm$ 0.16 & 132 $\pm$ 7 \\ 
gmp2559 & 221406 & 0.32 & 171 & 0.011 & --19.89 $\pm$ 0.12 & --20.54 $\pm$ 0.12 & --23.36 $\pm$ 0.16 & 134 $\pm$ 8 \\ 
gmp2544 & 8118 & 1.76 & 12 & 0.014 & --21.05 $\pm$ 0.12 & --21.96 $\pm$ 0.12 & --24.42 $\pm$ 0.16 & 195 $\pm$ 3 \\ 
gmp2374 & 8128 & 0.49 & 169 & 0.008 & --21.59 $\pm$ 0.12 & --22.76 $\pm$ 0.12 & --25.27 $\pm$ 0.15 & 270 $\pm$ 13 \\ 
gmp1900 & 8140 & 1.90 & 19 & 0.017 & --21.47 $\pm$ 0.12 & --22.17 $\pm$ 0.12 & --24.52 $\pm$ 0.15 & 232 $\pm$ 1 \\ 
gmp1657 & 221460 & 2.27 & 14 & 0.012 & --21.19 $\pm$ 0.12 & --22.17 $\pm$ 0.12 & --24.65 $\pm$ 0.15 & 239 $\pm$ 12 \\ 
-- & 8161 & 2.70 & 2 & 0.013 & --20.69 $\pm$ 0.12 & --21.64 $\pm$ 0.12 & --24.05 $\pm$ 0.16 & 192 $\pm$ 1 \\ 
gmp0455 & 230051 & 2.01 & 7 & 0.013 & --19.84 $\pm$ 0.12 & --20.18 $\pm$ 0.12 & --21.60 $\pm$ 0.22 & 110 $\pm$ 9 \\ 
-- & 8194 & 2.95 & 19 & 0.011 & --21.50 $\pm$ 0.12 & --22.55 $\pm$ 0.12 & --24.92 $\pm$ 0.15 & 234 $\pm$ 5 \\ 
-- & 8195 & 3.65 & 7 & 0.012 & --18.91 $\pm$ 0.12 & --19.50 $\pm$ 0.12 & --22.58 $\pm$ 0.19 & 123 $\pm$ 4 \\ 
-- & 8209 & 6.00 & 1 & 0.017 & --20.88 $\pm$ 0.12 & --21.76 $\pm$ 0.12 & --23.71 $\pm$ 0.16 & 151 $\pm$ 6 \\ 
-- & 8220 & 6.32 & 2 & 0.022 & --21.41 $\pm$ 0.12 & --22.35 $\pm$ 0.12 & --24.83 $\pm$ 0.16 & 258 $\pm$ 1 \\ 
-- & 8229 & 3.32 & 18 & 0.015 & --21.15 $\pm$ 0.12 & --22.06 $\pm$ 0.12 & --24.69 $\pm$ 0.17 & 195 $\pm$ 8 \\ 
-- & 230117 & 3.35 & 36 & 0.010 & --19.42 $\pm$ 0.12 & --19.99 $\pm$ 0.12 & --22.24 $\pm$ 0.24 & 107 $\pm$ 7 \\ 
-- & 8244 & 3.71 & 27 & 0.009 & --19.79 $\pm$ 0.12 & --20.46 $\pm$ 0.12 & --22.47 $\pm$ 0.20 & 129 $\pm$ 1 \\ 
-- & 230139 & 4.87 & 3 & 0.013 & --20.31 $\pm$ 0.12 & --21.19 $\pm$ 0.12 & --23.31 $\pm$ 0.19 & 145 $\pm$ 5 \\ 
-- & 8294 & 7.16 & 2 & 0.012 & --20.31 $\pm$ 0.12 & --20.87 $\pm$ 0.12 & \multicolumn{1}{c}{--} & 122 $\pm$ 5 \\ 
-- & 8300 & 4.98 & 5 & 0.017 & --22.13 $\pm$ 0.12 & --23.25 $\pm$ 0.12 & --25.44 $\pm$ 0.15 & 304 $\pm$ 5 \\ 
-- & 8317 & 6.65 & 8 & 0.014 & --20.57 $\pm$ 0.12 & --21.21 $\pm$ 0.12 & --23.24 $\pm$ 0.17 & 149 $\pm$ 2 \\ 
-- & 8328 & 5.83 & 2 & 0.011 & --19.31 $\pm$ 0.12 & --19.89 $\pm$ 0.12 & \multicolumn{1}{c}{--} & 127 $\pm$ 1 \\ 
-- & 8366 & 7.07 & 2 & 0.019 & --22.04 $\pm$ 0.12 & --22.79 $\pm$ 0.12 & --25.14 $\pm$ 0.16 & 268 $\pm$ 1 \\
\hline
\end{tabular}
\end{table*} 

\label{lastpage}

\end{document}